\documentclass[phd,bottom,nosig]{usbthesis}
\usepackage{graphicx}
\usepackage{color}
\usepackage{bm}
\usepackage{amsmath}
\usepackage{amssymb}
\usepackage{mathtools}
\usepackage{upgreek}
\usepackage{setspace}
\usepackage{array}
\usepackage{tabularx}
\usepackage{multirow}
\usepackage{afterpage}
\usepackage{hhline}
\usepackage{url}
\usepackage{slashed}
\usepackage{subfig}
\usepackage[linktocpage=true]{hyperref} 
\usepackage{footmisc} 
\usepackage{xspace}
\usepackage[numbers,sort&compress]{natbib}
\usepackage{pdfpages}
\usepackage{scalerel}

\newcommand{\para}{\scalerel*{//}{\perp}}

\def\simge{%
    \mathrel{\rlap{\raise 0.511ex
    \hbox{$>$}}{\lower 0.511ex \hbox{$\sim$}}}}
\def\simle{%
    \mathrel{\rlap{\raise 0.511ex
    \hbox{$<$}}{\lower 0.511ex \hbox{$\sim$}}}}

\DeclareMathAccent{\ring}{\mathalpha}{operators}{"17}

\providecommand{\renewoperator}[3]{\renewcommand*{#1}{\mathop{#2}#3}}
\renewoperator{\Re}{\mathrm{Re}}{\nolimits}
\renewoperator{\Im}{\mathrm{Im}}{\nolimits}

\makeatletter
\providecommand*{\diff}{\@ifnextchar^{\DIfF}{\DIfF^{}}}
\def\DIfF^#1{\mathop{\mathrm{\mathstrut d}}\nolimits^{#1}\gobblespace}
\def\gobblespace{\futurelet\diffarg\opspace}
\def\opspace{%
    \let\DiffSpace\!%
    \ifx\diffarg(%
        \let\DiffSpace\relax
    \else
        \ifx\diffarg[%
            \let\DiffSpace\relax
        \else
            \ifx\diffarg\{%
                \let\DiffSpace\relax
            \fi\fi\fi\DiffSpace}

\usepackage{bibunits}

\definecolor{SCGPblue}{rgb}{0.2196, 0.3255, 0.6431}

\hypersetup{colorlinks=true, urlcolor=SCGPblue,linkcolor=SCGPblue,citecolor=SCGPblue}

\title{Universalities of Defects\\in Quantum Field Theories}
\author{Siwei Zhong}

\month{May}
\year{2026}
\program{Physics}

\director{Zohar Komargodski}{Professor, Simons Center for Geometry and Physics}

\chairman{Leonardo Rastelli}{Professor,  Yang Institute of Theoretical Physics}

\fstmember{Clark McGrew}{Associate Professor, Department of Physics and Astronomy}

\outmember{Yifan Wang}{Assistant Professor, Center for Cosmology and Particle Physics}{New York University}

\dean{Celia Marshik}%

\begin{document}

\singlespacing %
\pagenumbering{roman} %
\maketitle %

\begin{abstract}
Defects are both physically rich objects and powerful tools in modern quantum field theory. They are extended operators, such as boundaries, impurities, and probe particles, embedded in many-body systems. In this dissertation, we study the universal aspects of defect dynamics from the perspective of symmetry principles. We bring together several themes, including defect renormalization group flows, effective string theory, and impurities in atomic quantum gases.
\end{abstract}

\tableofcontents %

\listofpublications

\begin{enumerate}
\raggedright
\item A.~Raviv-Moshe and S.~Zhong,
\textit{Phases of surface defects in Scalar Field Theories},
\textit{JHEP} \textbf{08} (2023) 143,
[\href{https://arxiv.org/abs/2305.11370}{\texttt{arXiv:2305.11370}}].

\item G.~Cuomo, Z.~Komargodski, and S.~Zhong,
\textit{Chiral modes of giant superfluid vortices},
\textit{Phys. Rev. B} \textbf{110} (2024), no.~14 144514,
[\href{https://arxiv.org/abs/2312.06095}{\texttt{arXiv:2312.06095}}].

\item Z.~Komargodski and S.~Zhong,
\textit{Baryon junction and string interactions},
\textit{Phys. Rev. D} \textbf{110} (2024), no.~5 056018,
[\href{https://arxiv.org/abs/2405.12005}{\texttt{arXiv:2405.12005}}].

\item A.~Raviv-Moshe and S.~Zhong,
\textit{Impurities in Schr\"odinger field theories and s-wave resonance},
\textit{JHEP} \textbf{09} (2025) 134,
[\href{https://arxiv.org/abs/2411.04040}{\texttt{arXiv:2411.04040}}].

\item M.~Barkeshli, C.~Fechisin, Z.~Komargodski, and S.~Zhong,
\textit{Disclinations, Dislocations, and Emanant Flux at Dirac Criticality},
\textit{Phys. Rev. X} \textbf{16} (2026), no.~1 011017,
[\href{https://arxiv.org/abs/2501.13866}{\texttt{arXiv:2501.13866}}].

\item S.-H.~Shao and S.~Zhong,
\textit{Where non-invertible symmetries end: twist defects for electromagnetic duality},
\textit{JHEP} \textbf{01} (2026) 118,
[\href{https://arxiv.org/abs/2509.21279}{\texttt{arXiv:2509.21279}}].

\item X.~Lou and S.~Zhong,
\textit{Baryon Junction and String Interactions: Part II},
\href{https://arxiv.org/abs/2602.17771}{\texttt{arXiv:2602.17771}}.

\end{enumerate}
    
\listoffigures

\listoftables 

\conventions

{
\let\clearpage\relax
\let\cleardoublepage\relax
\makeatletter
\setlength{\jot}{10pt}
\begin{equation*}
\begin{aligned}
    & \updelta X  &&\text{infinitesimal variation of $X$;}\\
    & \delta_{\mu\nu}~,~\delta_{\mu}{}^{\nu}~\quad \quad \quad \quad &&\text{Kronecker delta symbol;}\\
    & \delta^p(x) &&\text{$p$-dimensional Dirac delta function;}\\
    & \star F &&\text{Hodge dual of the differential form $F$;}\\
    & \big(x\big)_n &&\text{Pochhammer symbol, i.e., }\big(x\big)_n\equiv \Gamma(x+n)/\Gamma(x);\\
    &\textbf{q}~,~\textbf{q}_x~,\dots &&\text{Modular parameters};\\
    & \eta(\textbf{q}) && \text{Dedekind eta function}; \\
    & E_{2k}(\textbf{q}) && \text{Eisenstein series;}\\
    & K_{\alpha}(x) && \text{Modified Bessel function of the second type;}\\
    &\Theta(\tau) && \text{Heaviside step function;}\\
\end{aligned}
\end{equation*}
}

\begin{acknowledgements}
    
\hspace{\parindent}This dissertation would not have been possible without the guidance and support of my advisor, Zohar Komargodski. To me, he has been an inspiring teacher, a generous collaborator, and an exceptionally visionary scientist. Beyond all the physics lessons, he has taught me by setting an example to always stay curious about nature, to always be kind to others, and to always pursue excellence in research. My appreciation is beyond words. 
\\

I am deeply grateful to Gabriel Cuomo, Leonardo Rastelli, and Shu-Heng Shao. I got to know Gabriel just as I was starting my first research project. His comments have always been kind, thoughtful, and full of physical insight. During my first years of graduate school, I benefited greatly from Leonardo's lectures on several advanced topics. They were among the most memorable lectures of my student years. As I began to find my footing in research, I learned a great deal from many discussions with Shu-Heng. His broad perspective and enthusiasm for physics have greatly influenced the way I think about my own research.
\\

I would like to acknowledge my collaborators Avia Raviv-Moshe, Maissam Barkeshli, and Christopher Fechisin. I am especially thankful to Avia, who was my very first coauthor. Her kind support and stimulating ideas meant a lot to me as a beginning researcher. \\

I am grateful to Paul Goldbart, Yin-Chen He, Mengkun Liu, Martin Roček, George Sterman, Tzu-Chieh Wei, and the entire Physics Department. It has been quite a journey studying and working at Stony Brook. Special thanks go to Clark McGrew and Yifan Wang for kindly serving on my dissertation committee.\\

I would like to thank Mykola Dedushenko, Diego Delmastro, Justin Kulp, Fedor K. Popov, Brandon C. Rayhaun, Fei Yan, Yunqing Zheng, and the entire SCGP community. It has been a wonderful experience exploring research topics and developing new ideas together with them. I would also like to thank Jan Albert, Yichul Choi, and Ho Tat Lam for many productive discussions and enjoyable conversations.\\

I am grateful to Heidi Ciolfi, Dawn Huether, Brigid O'Connor, and Donald Sheehan for rescuing me from endless administrative paperwork. I would also like to thank the SCGP café staff for their delicious and healthy food. Because of them, SCGP will always hold a special place in my heart.\\

I would like to thank Ling-Yan Hung, Yidun Wan, Xinan Zhou, and Yang Zhou for kindly hosting my visits and for their support since my undergraduate years. Before this list becomes infinitely long, I would also like to thank Jinwei Chu, Xuanjing Chu, Anirudh Deb, Wenhan Guo, Shenyang Shi, Xuejing Huang, Xinyi Sun, Jialun Wang, Meng Ye, Haolan Xu, and all my dear friends. My life would not have been the same without them.\\

Finally, I would like to thank my mom. She may not know much about physics, but she certainly knows how love can help a son go farther than he ever thought possible.
\end{acknowledgements}
\pagestyle{thesis}
\newpage
\pagenumbering{arabic}

\chapter{Introduction and Summary}

Quantum field theory provides a universal language for describing physical systems across vastly different scales. Prominent examples include the Landau--Ginzburg theory of phase transitions in ferromagnets and Quantum Chromodynamics (QCD), which describes the strong interactions between elementary particles. One of the central lessons of modern quantum field theory is that the long-distance, infrared (IR) properties of a system are often insensitive to its microscopic ultraviolet (UV) details. This idea underlies the concept of universality classes. Different systems can exhibit the same universal IR behavior, which is largely determined by their symmetries.

This paradigm is enriched by defects in a fundamental way. Defects arise naturally in a wide range of physical contexts, including boundaries and interfaces in statistical-mechanics models, Wilson lines and ’t Hooft operators in gauge theories, crystalline impurities in lattice systems, and vortices in superfluids. Although the UV realizations of these defects can differ significantly from one model to another, one may ask whether they exhibit universal behavior in the IR. In this dissertation, we address this question using symmetry principles.

While defects are rich subjects of study in their own right, they also serve as powerful tools to understand the quantum field theories in which they are embedded. In particular, we note that certain defects can be continuously deformed without affecting any physical observables. These are topological defects, which generalize the very notion of symmetry in modern physics \cite{Gaiotto:2014kfa, Freed:2022qnc}. This perspective has shed new light on many profound phenomena in quantum field theories, and we will apply it extensively throughout this dissertation.

\section{Outline}

The chapters of this dissertation are devoted to different themes. Below, we outline the main topics covered by each chapter and summarize the key results.

In Chapter \ref{cha_1}, we discuss Renormalization Group (RG) flows governing the evolution of defect dynamics under changes of scale. Defects with enhanced conformal symmetry lie at special fixed points of these RG flows. We show that the dynamics of conformal defects are subject to strong constraints. In certain cases \cite{Shao:2025qvf}, they are universally determined by conformal symmetry and the dynamics of the bulk theory, independent of microscopic details. We also examine how different fixed points are connected by defect RG flows. In particular, we investigate the phase diagram of defects embedded in the $O(N)$ Wilson--Fisher theory \cite{Raviv-Moshe:2023yvq}. Finally, we identify impurities in microscopic lattice models with defects in the continuum field theory. The RG flows connecting these two descriptions are subject to novel symmetry constraints, which we explain through examples in \cite{Barkeshli:2025cjs}.

In Chapter \ref{cha_2}, we discuss defects that play important roles in understanding the strongly coupled regime of QCD. These include chromoelectric flux tubes and their junctions \cite{Komargodski:2024swh, Lou:2026xqr}. In the long-distance limit, we show that the dynamics of these defects are universally fixed by the symmetry-breaking pattern. Using the open-closed duality, we further determine the leading contributions to the coupling constants governing the $s$-wave scattering of large glueball excitations. We comment on selection rules associated with a novel accidental symmetry of the defects.

In Chapter \ref{cha_3}, we turn to defects immersed in atomic quantum gases. We develop a general framework for scale-invariant defects in nonrelativistic quantum field theories \cite{Raviv-Moshe:2024yzt}. For point-like impurities, we show a correspondence between defect operators and many-body states in a harmonic trap with possible $s$-wave resonance. As a representative example, we apply this framework to impurities in dilute Fermi gases. We also analyze the giant vortex \cite{Cuomo:2023vvd}, a macroscopic defect that arises in rotating superfluids. We show that rapidly rotating superfluids confined to a two-dimensional trap exhibit new universal behavior, including chiral modes with emergent warped conformal symmetry.  

Taken together, these chapters survey defects across a variety of physical settings. We aim to deepen our understanding of their universal properties and to demonstrate how defects offer new perspectives on exotic physical phenomena.

\section{Future directions}

We close this chapter with comments on several future directions and open questions.

Defects respond to both intrinsic and extrinsic geometry. For conformal defects, one class of universal responses is given by conformal anomalies. These anomalies have long played an important role in the study of defect RG flows, as they govern monotonicity theorems for defect RG flows in many cases \cite{Jensen:2015swa, Cuomo:2021rkm, Shachar:2022fqk}. This motivates the question of whether defect conformal anomalies can be extracted directly from physical observables. In particular, we would like to explore this question for parity-odd anomalies. An example is provided by the chiral edge modes of three-dimensional Chern-Simons theories \cite{Witten:1988hf, Elitzur:1989nr, Wen:1992uk, Wen:1992vi}, where the edge gravitational anomaly is encoded in the Casimir momentum. It would be interesting to understand whether an analogous statement can be made for gapless systems.

Another promising future direction is to develop effective field theories for special defect configurations. For instance, the effective description of two parallel conformal boundaries was studied in \cite{Diatlyk:2024qpr}. More generally, defects can support global symmetries that are absent from the ambient quantum field theory. This perspective becomes particularly interesting in gauge theories. See \cite{Verresen:2022mcr, Chung:2024hsq} for some recent works. By studying the symmetries, anomalies, and dynamics of defect effective field theories, we aim to sharpen our understanding of exotic quantum phases and the universal structures that organize them.

Finally, it is intriguing to realize and test defect models in experimental settings. Our analyses in \cite{Cuomo:2023vvd, Barkeshli:2025cjs} were strongly motivated by experimental studies and led to testable predictions. It would be valuable to explore the phenomenological implications of other defect models.

\chapter{Defect Renormalization Group Flows}
\label{cha_1}

Historically, the idea of renormalization was introduced to address the divergences arising from loop corrections in quantum field theories. For early developments of RG flows, see \cite{Gell-Mann:1954yli, Wilson:1971bg, Wilson:1971dh, Gross:1973id, Politzer:1973fx, Polchinski:1983gv, Zamolodchikov:1986gt}, among many others.

From a modern perspective, the RG flow describes how a physical system evolves under scale transformations. A generic quantum field theory can be viewed as a point in the space of theories, interpolating between the fixed points of RG flows. In most cases\footnote{For Lorentz-invariant theories, examples of scale-invariant but not conformally invariant theories include \cite{Riva:2005gd, El-Showk:2011xbs, Jackiw:2011vz, Gimenez-Grau:2023lpz}. In Chapter \ref{cha_3}, we discuss Galilean-invariant theories, where scaling symmetry is enhanced to Schrödinger symmetry instead of the conformal symmetry.}, these fixed points are strongly constrained by the enhanced conformal symmetry, which renders their dynamics more tractable. In recent years, substantial effort has been devoted to understanding the dynamics of these Conformal Field Theories (CFTs) through their symmetries and consistency conditions. See \cite{Simmons-Duffin:2016gjk, Rychkov:2016iqz} for pedagogical reviews of the modern CFT framework, and \cite{Polyakov:1974gs, Rattazzi:2008pe, El-Showk:2012cjh, Komargodski:2012ek, Fitzpatrick:2012yx} for a partial list of relevant studies. Throughout this dissertation, we make extensive use of standard concepts of CFT, such as state-operator correspondence, conformal primaries and descendants, and Operator Product Expansion (OPE). 

The concept of RG flow extends naturally to defects in quantum field theory. A defect is usually characterized by a set of parameters describing its dynamics and its couplings to the ambient field theory, both of which evolve under scale transformations. As in ordinary quantum field theories, defect RG flows can also admit fixed points with enhanced conformal symmetry. Defect Conformal Field Theory (DCFT) \cite{McAvity:1995zd, Liendo:2012hy, Gliozzi:2015qsa, Billo:2016cpy, Lauria:2017wav, Lauria:2018klo, Liendo:2019jpu, Lauria:2020emq, Herzog:2020bqw, Chalabi:2021jud, Cuomo:2021rkm} provides a framework for analyzing universalities of defects at these fixed points.

In this chapter, we discuss defect RG flows through examples drawn from \cite{Shao:2025qvf, Raviv-Moshe:2023yvq, Barkeshli:2025cjs}. We first study general aspects of defects with extended conformal symmetry and introduce the basics of DCFT in Section \ref{sec_conformal defects}. Conformal symmetry imposes strong constraints on defect dynamics. As an example, in Section \ref{sec_Constraints from bulk dynamics} we show that a chiral conformal defect in Maxwell theory is universally determined by symmetry principles, regardless of the microscopic details. In Section \ref{sec_RG flows in the $O(N)$ models}, we examine how different fixed points are connected via defect RG flows. In particular, we explore the phase diagram of surface defects embedded in the $O(N)$ Wilson--Fisher conformal field theory. Finally, in Section \ref{sec_Matching UV and IR defects}, we identify impurities in UV discrete lattice models with defects in IR continuum field theory. The RG flows connecting these two descriptions are subject to novel symmetry constraints, which we discuss in detail.

\section{Conformal defects}
\label{sec_conformal defects}

We begin by discussing the general properties of defects at the RG flow fixed points. To set the stage, we consider unitary CFTs defined in $d$-dimensional Euclidean space $x_\mu \in \mathbb{R}^d$. Local operators in the CFT furnish representations of the $d$-dimensional conformal algebra $\mathfrak{so}(d+1,1)$, and are characterized by their scaling dimension $\Delta\geq 0$ and spin $j\geq 0$. Schematically, we denote local primaries in the CFT by $\mathcal{O}_{j}$. These operators transform as tensors for integer $j$ and as spinors for half-integer $j$ under $\mathfrak{so}(d)\subset \mathfrak{so}(d+1,1)$.

We next consider a $p$-dimensional defect supported on the hyperplane $\mathbb{R}^p \subseteq \mathbb{R}^d$, with $x^{\para}_\nu$ denoting the coordinates parallel to the defect plane and $x^{\perp}_\sigma$ denoting the coordinates perpendicular to it.\footnote{Point-like defects $(p=0)$ essentially coincide with local operators, which have been extensively studied in CFT. We will not discuss them further in this dissertation.} As noted in the preface, we focus on the cases where the translation symmetries along the perpendicular directions are explicitly broken by the defect. The maximal subalgebra of $\mathfrak{so}(d+1,1)$ that can be preserved by the planar defect is therefore  
\begin{equation}
\label{eq_DCFT symmetry}
   \mathfrak{so}(p+1,1)\times \mathfrak{so}(q)~,
\end{equation}
where $q=d-p$ is the codimension of the defect. In what follows, we assume that scaling invariance is enhanced to the residual conformal algebra $\mathfrak{so}(p+1,1)$ at the fixed points of defect RG flows.\footnote{This enhancement for CFT boundaries ($q=1$) is discussed in \cite{Nakayama:2012ed}. In particular, there exist non-unitary boundary theories that are scale-invariant but not conformally invariant.} We further assume the transverse rotation algebra $\mathfrak{so}(q)$ is also preserved, so that the defect is isotropic in the perpendicular directions when $q\geq 2$. The dynamics of defects preserving the symmetry \eqref{eq_DCFT symmetry} are encoded in the DCFT data, which we detail in the remainder of this section.

Local operators confined to the $p$-dimensional defect are characterized by their scaling dimension $\Delta\geq 0$, parallel spin $\ell\geq 0$, and transverse spin $s \geq 0$. In analogy with the bulk CFT, we denote local primaries in the DCFT by $\hat{\mathcal{O}}^s_{\ell}$. The parallel spin $\ell$ labels the tensor and spinor representations of $\mathfrak{so}(p)\subset \mathfrak{so}(p+1,1)$ and takes either integer or half-integer values. The selection rule for the transverse spin $s$, on the other hand, depends on the possible topological volumes associated with the defect; see Sections \ref{sec_Constraints from bulk dynamics} and \ref{sec_Matching UV and IR defects}. 

\subsection{Bulk-to-defect OPE}

DCFT concerns the relation between bulk operators $\mathcal{O}_j$ and defect operators $\hat{\mathcal{O}}_\ell^s$. One possibility is that the correlation functions between bulk and defect operators completely factorize. For example, one can stack a decoupled $p$-dimensional CFT onto the given DCFT without affecting its dynamics, so that correlation functions of operators in the decoupled sector factorize from those of bulk operators. A non-trivial example of such factorization is discussed in Section \ref{sec_Constraints from bulk dynamics} below.

More interestingly, we bring a bulk operator $\mathcal{O}_j$ close to the defect at $x^\perp_{\sigma}=0$. It follows from the branching rule of $\mathfrak{so}(d+1,1)\to \mathfrak{so}(p+1,1)\times \mathfrak{so}(q)$ that bulk operators can be expanded in an infinite sum of defect operators:
\begin{equation}
\label{eq_formal bulk-to-defect OPE}
    \mathcal{O}_j(x^\perp,x^{\para})\to \sum C_{\mathcal{O}_j}{}^{\hat{\mathcal{O}}_\ell^s}\big(\hat{\mathcal{O}}_\ell^s(x^{\para})+\text{descendants}\big)~.
\end{equation}
This is known as the bulk-to-defect OPE \cite{McAvity:1995zd, Liendo:2012hy, Billo:2016cpy, Lauria:2017wav}, where $C_{\mathcal{O}_j}{}^{\hat{\mathcal{O}}_\ell^s}$ denote the corresponding OPE coefficients. At the level of representations, a spin-$j$ representation of $\mathfrak{so}(d)$ can be decomposed into representations of $\mathfrak{so}(p)\times \mathfrak{so}(q)$ with spin $(\ell,s)$ satisfying $\ell+s\leq j$. Notably, one can use the transverse vector $x^\perp_\sigma$ to construct tensors that carry the $\mathfrak{so}(q)$ spin on the right-hand side of \eqref{eq_formal bulk-to-defect OPE}, while tensors built from $x^{\para}_\nu$ are forbidden by translation symmetry. We therefore find the selection rule for the OPE coefficients
\begin{equation}
\label{eq_bulk-to-defect OPE selection rule}
    C_{\mathcal{O}_j}{}^{\hat{\mathcal{O}}_\ell^s}=0~,~~\text{for}~~\ell> j~.
\end{equation}
For instance, a bulk scalar primary $\mathcal{O}$ with $j=0$ can be expanded in terms of defect scalar primaries $\hat{\mathcal{O}}^s$ with $\ell=0$. The explicit bulk-to-defect OPE for $\mathcal{O}$ is fixed by the algebra \eqref{eq_DCFT symmetry} and takes the form \cite{Billo:2016cpy, Lauria:2020emq}
\begin{equation}
\label{eq_explicit scalar OPE}
\mathcal{O}(x^\perp,x^{\para})=\sum_{\hat{\mathcal{O}}^s}\frac{Y_{s}(x^{\perp}_\sigma/|x^{\perp}|)}{|x^{\perp}|^{\Delta(\mathcal{O})-\Delta(\hat{\mathcal{O}}^s)}}C_{\mathcal{O}}{}^{\hat{\mathcal{O}}^s}\sum_{n\geq 0}\frac{(-|x^{\perp}|^2\partial^\nu\partial_\nu/4)^n}{n!\big(\Delta(\hat{\mathcal{O}}^s)+1-\frac{p}{2}\big)_n}\hat{\mathcal{O}}^{s}(x^{\para})~,
\end{equation}
where $Y_{s}$ denote the spin $s$ spherical harmonic on $S^{q-1}$ with its $\mathfrak{so}(q)$ indices suppressed, as those of $\hat{\mathcal{O}}^{s}$.

The OPE \eqref{eq_formal bulk-to-defect OPE} encodes how the presence of the defect affects bulk observables. This can be seen by considering bulk–defect two-point functions. One can choose a normalization for the two-point function of the defect scalar primary $\hat{\mathcal{O}}^s$ as follows
\begin{equation}
\label{eq_defect 2pt normalization}
    \langle \hat{\mathcal{O}}^{s}(x^{\para})\hat{\mathcal{O}}^{s}(\Tilde{x}^{\para})\rangle=|x^{\para}-\Tilde{x}^{\para}|^{-2\Delta(\hat{\mathcal{O}}^s)}~,
\end{equation}
such that it is non-degenerate and diagonal in the $\mathfrak{so}(q)$ indices. Using \eqref{eq_explicit scalar OPE} and \eqref{eq_defect 2pt normalization}, one finds the correlation function between the bulk scalar $\mathcal{O}$ and the defect scalar $\hat{\mathcal{O}}^{s}$:
\begin{equation}
\label{eq_bulk scalar to defect scalar}
    \langle \mathcal{O}(x^\perp,x^{\para})\hat{\mathcal{O}}^{s}(\Tilde{x}^{\para})\rangle=\frac{Y_{s}(x^{\perp}_\sigma/|x^{\perp}|)C_{\mathcal{O}}{}^{\hat{\mathcal{O}}^s}}{|x^{\perp}|^{\Delta(\mathcal{O})-\Delta(\hat{\mathcal{O}}^s)}(|x^{\perp}|^2+|x^{\para}-\Tilde{x}^{\para}|^2)^{\Delta(\hat{\mathcal{O}}^s)}}~.
\end{equation}
The two-point function \eqref{eq_bulk scalar to defect scalar} probes the location of the defect when $\Delta(\mathcal{O})\neq \Delta(\hat{\mathcal{O}}^s)$ and $C_{\mathcal{O}}{}^{\hat{\mathcal{O}}^s}\neq 0$, indicating that the defect does not factorize from the bulk.

Our analysis extends to general bulk and defect operators, despite the more involved tensor structures in \eqref{eq_formal bulk-to-defect OPE}. See, e.g., \cite{Billo:2016cpy, Lauria:2018klo, Herzog:2020bqw} for a systematic enumeration of these tensors. In what follows, we discuss the defect operators that play a special role in the DCFT, in particular the identity operator, the displacement operator, and the tilt operator. 

\subsection{Defect identity operator}

In this subsection, we consider the defect identity operator\footnote{We have assumed that the defect identity operator is unique, which in turn implies that the CFT with the defect inserted has a unique vacuum.} $\hat{I}$, for which $\ell=0$, $s=0$, and $\Delta(\hat{I})=0$. Given a bulk CFT operator $\mathcal{O}_j$, it is natural to ask whether $\hat{I}$ appears on the right-hand side of the expansion \eqref{eq_formal bulk-to-defect OPE} with a non-zero OPE coefficient. Such coefficients determine the one-point functions of bulk operators in the presence of the defect. We denote them by
\begin{equation}
\label{eq_1pt OPE coeff}
    C_{\mathcal{O}_j}\equiv C_{\mathcal{O}_j}{}^{\hat{I}}~.
\end{equation}
For example, the one-point function of a bulk scalar operator $\mathcal{O}$ simply follows from the dimensional analysis, and is given by
\begin{equation}
\label{eq_scalar bulk 1pt function}
    \langle \mathcal{O}(x^\perp,x^{\para})\rangle=\frac{C_{\mathcal{O}}}{|x^\perp|^{\Delta(\mathcal{O})}}~,
\end{equation}
which is also consistent with \eqref{eq_bulk scalar to defect scalar} upon taking the defect operator to be $\hat{I}$.

The OPE coefficients \eqref{eq_1pt OPE coeff} depend on both the bulk and defect dynamics, while also being constrained by the conformal algebra \eqref{eq_DCFT symmetry}. In the following discussion, we focus on bulk operators transforming in the symmetric traceless tensor representation of $\mathfrak{so}(d)$. For a spinning operator $\mathcal{O}_j$, it is convenient to define a degree-$j$ polynomial \cite{Costa:2011mg}
\begin{equation}
\label{eq_bulk polarization polynomial}
    \mathcal{P}_{\mathcal{O}_{j}}\equiv v^{\mu_1} \dots v^{\mu_j}(\mathcal{O}_j)_{\mu_1 \dots\mu_j}~,
\end{equation}
where $v_\mu$ is a null vector satisfying $v^\mu v_\mu=0$. The operator $\mathcal{P}_{\mathcal{O}_{j}}$ transforms as a scalar under the conformal algebra \eqref{eq_DCFT symmetry}. Moreover, the one-point function of $\mathcal{P}_{\mathcal{O}_{j}}$ must also be degree-$j$ polynomial in $v_\mu$. We thereby find \cite{Billo:2016cpy,Lauria:2018klo}
\begin{equation}
\label{eq_bulk polarization 1pt}
    \langle \mathcal{P}_{\mathcal{O}_{j}}(x^\perp,x^{\para})\rangle=\frac{C_{\mathcal{O}_{j}}}{|x^\perp|^{\Delta(\mathcal{O}_{j})}}\left(v^\sigma v_\sigma-\frac{(v^\sigma x^\perp_\sigma)^2}{|x^\perp|^2}\right)^{\frac{j}{2}}~.
\end{equation}
For conformal boundaries and interfaces, the one-point function \eqref{eq_bulk polarization 1pt} implies that \cite{Liendo:2012hy}
\begin{equation}
    C_{\mathcal{O}_{j}}=0~,~~\text{for integer}~j\geq 1~\text{when}~q=1~.
\end{equation}
Meanwhile, the polynomial in parentheses in \eqref{eq_bulk polarization 1pt} only becomes a perfect square for defect codimension $q=2$. We note that
\begin{equation}
    C_{\mathcal{O}_{j}}=0~,~~\text{for odd integer}~j~\text{when}~q\geq 3~.
\end{equation}
For example, the bulk conserved current $J_\mu$ admits a non-zero one-point function only for defects of codimension $q=2$. The one-point function of bulk stress-energy tensor $T_{\mu_1\mu_2}$, by contrast, can take non-zero values for defects of any codimension $q\geq 2$. In Section \ref{sec_Matching UV and IR defects}, we study the OPE coefficients \eqref{eq_1pt OPE coeff} associated with $J_\mu$ and $T_{\mu_1\mu_2}$ in concrete physical settings.

\subsection{Displacement and tilt operators}

In this subsection, we turn to other defect operators whose quantum numbers, like that of the identity operator $\hat{I}$, are fixed across different DCFTs. This includes the displacement operator $\hat{D}$ \cite{McAvity:1993ue,McAvity:1995zd,Liendo:2012hy,Jensen:2015swa}, which is a universal defect primary for local DCFTs. 

To define the $\hat{D}$ operator, it is helpful to generalize the background metric on the $d$-dimensional space to $(ds)^2=g_{\mu_1\mu_2}dx^{\mu_1}dx^{\mu_2}~.$
We denote the embedding of the $p$-dimensional defect by $X_\mu(x^{\para})$, and the induced metric on the defect is given by
\begin{equation}
    \hat{g}_{\nu_1\nu_2}\equiv g_{\mu_1\mu_2}\frac{\partial X^{\mu_1}}{\partial x^{\para}_{\nu_1}}\frac{\partial X^{\mu_2}}{\partial x^{\para}_{\nu_2}}~.
\end{equation}
The defect dynamics depends both on the induced metric $\hat{g}_{\nu_1\nu_2}$ and on the embedding $X_\mu$ into the ambient space. In particular, the displacement operator $\hat{D}$ is defined through the leading linear response of the DCFT action to variations of $X_\mu$ as follows
\begin{equation}
\label{eq_displacement def}
\mathtoolsset{multlined-width=0.9\displaywidth}
\begin{multlined}
    \updelta S_{\text{DCFT}}=-\frac{1}{2}\int d^dx\sqrt{g}\ \updelta g_{\mu_1\mu_2}T^{\mu_1\mu_2}\\
    -\frac{1}{2}\int d^p x^{\para}\sqrt{\hat{g}}\ \updelta g_{\mu_1\mu_2}\hat{T}^{\mu_1\mu_2}-\int d^p x^{\para}\sqrt{\hat{g}}\ \updelta X_\mu (\hat{D})^\mu+\text{higher-order terms}~,
\end{multlined}
\end{equation}
where $T_{\mu_1\mu_2}$ is the bulk stress-energy tensor. For planar defects embedded in the Euclidean space $\mathbb{R}^d$, we have $g_{\mu_1\mu_2}=\delta_{\mu_1\mu_2}$ and $X_\mu=\delta_\mu{}^{\nu}x^{\para}_\nu$.\footnote{The definition analogously applies to defects in curved spacetimes. However, the quantum numbers of the corresponding operator may depend on the background metric. We discuss such examples in Section \ref{sec_Matching UV and IR defects}.} Clearly, the embedding function $X_\mu$ in this case has the dimension $-1$, which in turn fixes the scaling dimension of the displacement operator to be $\Delta(\hat{D})=p+1$.

The DCFT action is expected to be invariant under bulk and defect reparametrizations. Requiring $\updelta S_{\text{DCFT}}=0$ under the defect coordinate variation $\updelta g_{\mu_1\mu_2}=0$ and $\updelta X_\mu =\delta_\mu{}^\nu\updelta x^{\para}_{\nu}$, we find from \eqref{eq_displacement def} that the displacement operator has $\ell=0$ and $s=1$. We also consider the bulk coordinate variation:
\begin{equation}
    \updelta g_{\mu_1\mu_2}=\partial_{\mu_1}\updelta x_{\mu_2}+\partial_{\mu_2}\updelta x_{\mu_1}~,~~\updelta X_\mu=-\updelta x_\mu~.
\end{equation}
The Ward identities following from $\updelta S_{\text{DCFT}}=0$ are then given by
\begin{equation}
\label{eq_displacement Ward identities}
    \partial_{\mu} T^{\mu \sigma}_{\text{tot}}=\delta^q(x^\perp)(\hat{D})^\sigma~,~~\text{and}~~\partial_{\mu} T^{\mu \nu}_{\text{tot}}=0~,
\end{equation}
where $T^{\mu_1\mu_2}_{\text{tot}}\equiv T^{\mu_1\mu_2}+\delta^q(x^\perp)\hat{T}^{\mu_1\mu_2}$. In the $x^{\para}_\nu$-directions, \eqref{eq_displacement Ward identities} implies that momentum can flow between the bulk and the defect and remain conserved, while the explicit breaking of translation symmetries in the $x^{\perp}_\sigma$-directions leads to the displacement operator $\hat{D}$. Notably, the Ward identities fix the norm of the two-point function of the $\hat{D}$ operator, which takes the form
\begin{equation}
\label{eq_displacement 2pt}
    \langle (\hat{D})_{\sigma_1}(x^{\para})(\hat{D})_{\sigma_2}(\Tilde{x}^{\para})\rangle=\delta_{\sigma_1\sigma_2}C_{\hat{D}\hat{D}}|x^{\para}-\Tilde{x}^{\para}|^{-2(p+1)}~.
\end{equation}
The coefficient $C_{\hat{D}\hat{D}}$ encodes information about the defect Weyl anomaly; see \cite{Chalabi:2021jud} for details.

Each continuous global symmetry of the bulk CFT is associated with a conserved current operator. For simplicity, we consider a $0$-form $U(1)$ symmetry and the corresponding bulk current $J_\mu$ in the following. Generalizations to non-abelian groups and higher-form symmetries are straightforward \cite{Antinucci:2024izg}.

The Ward identity for $J_\mu$ is generally modified in the presence of the defect. As in \eqref{eq_displacement Ward identities}, it takes the form
\begin{equation}
\label{eq_tilt Ward identity}
    \partial_\mu J^\mu=\delta^q(x^\perp) \hat{\tau}~.
\end{equation}
This implies that the $\hat{\tau}$ operator has quantum numbers $\ell=0$, $s=0$, and $\Delta(\hat{\tau})=p$. 

The $\hat{\tau}$ operator encodes the DCFT response to the symmetry transformation. To see that, we consider the topological operator $\exp(i\alpha\int \star J)$ and sweep it across the defect at $x^\perp_\sigma=0$. The Ward identity \eqref{eq_tilt Ward identity} indicates that 
\begin{equation}
\label{eq_tilt deformation}
    \updelta S_{\text{DCFT}}=i\alpha \int d^px^{\para}\ \hat{\tau}~.
\end{equation}
We note that $\updelta S_{\text{DCFT}}$ becomes trivial when $\hat{\tau}=\partial_\nu \hat{J}^\nu$ is a total derivative. In such cases, the total $U(1)$ charge of the bulk and the defect remains conserved, and the defect is symmetric. By contrast, the $\hat{\tau}$ operator in \eqref{eq_tilt deformation} generates a non-trivial marginal deformation when the defect breaks the symmetry. The defect primary $\hat{\tau}$ associated with broken global symmetries is referred to as the tilt operator \cite{Metlitski:2020cqy,Cuomo:2021kfm,Drukker:2022pxk,Antinucci:2024izg}.

Deforming the DCFT with the tilt operator $\hat{\tau}$ as in \eqref{eq_tilt deformation} generates an exact defect conformal manifold, which is isomorphic to the coset space associated with the broken symmetries. The Zamolodchikov metric on this conformal manifold is determined by the two-point function of the $\hat{\tau}$ operator, which takes the form 
\begin{equation}
\label{eq_tilt 2pt}
    \langle \hat{\tau}(x^{\para})\hat{\tau}(\Tilde{x}^{\para})\rangle=C_{\hat{\tau}\hat{\tau}}|x^{\para}-\Tilde{x}^{\para}|^{-2p}~.
\end{equation}
Similar to \eqref{eq_displacement 2pt}, the local metric $C_{\hat{\tau}\hat{\tau}}$ is fixed by the Ward identity \eqref{eq_tilt Ward identity}.

\section{Constraints on DCFTs in Maxwell theory}

\label{sec_Constraints from bulk dynamics}

What is the space of possible DCFTs, up to decoupled sectors, for a given bulk theory? This is a profound and challenging problem that, in general, can only be attacked using numerical bootstrap methods \cite{Liendo:2012hy}. One may expect that the bulk-to-defect OPE \eqref{eq_formal bulk-to-defect OPE} becomes stringent for certain representations of the bulk conformal algebra $\mathfrak{so}(d+1,1)$, in particular for free theories. Indeed, the spaces of DCFTs in free scalar field theory \cite{Lauria:2020emq, Bartlett-Tisdall:2025iqx} and in free Maxwell theory \cite{Herzog:2022jqv, Fraser-Taliente:2024lea} are considerably more constrained.

In this section, we discuss the $p=2$ chiral surface defects in the $d=4$ Maxwell theory that are defined as the boundaries of topological duality defects. We show that the constraints from bulk dynamics in such cases completely determine the DCFT spectrum. Correlation functions of operators confined to these defects factorize into two sectors \cite{Shao:2025qvf}: a universal generalized free-field sector, and a chiral current sector analogous to edge modes in Chern-Simons theory.

\subsection{Topological duality defect}

We first review Maxwell theory and its global symmetries. The Maxwell action for a $U(1)$ abelian gauge field $A$ on a 4-dimensional Euclidean manifold $\mathcal{M}_4$ is 
\begin{equation}
\label{eq_review maxwell action}
    S_{\text{Maxwell}}=\frac{1}{2e^2}\int_{\mathcal{M}_{4}}F\wedge \star F~,
\end{equation}
where $e>0$ denotes the gauge coupling, and $F=dA$ is the field strength\footnote{More generally, one can include a theta-angle term in the Maxwell action \eqref{eq_review maxwell action}. 
Even though this topological term doesn't affect the equations of motion, it changes the global properties of the theory and leads to new non-invertible symmetries \cite{Choi:2022zal,Choi:2022rfe, Niro:2022ctq}. 
However, we will not study this generalization here.  }. The action \eqref{eq_review maxwell action} has an ordinary $\mathbb{Z}_2$ symmetry, which acts on the gauge field by charge conjugation $C:~ A\to -A$. Furthermore, it has a continuous 1-form global symmetry $U(1)_\text{e}^{(1)}\times U(1)_\text{m}^{(1)}$ generated by the electric current $J_\text{e}$ and the magnetic current $J_\text{m}$ \cite{Gaiotto:2014kfa}. Explicitly, the 1-form symmetry currents $J_{\text{e}/\text{m}}$ and their defects $\eta_{\text{e}/\text{m}}$ are defined as follows:
\begin{equation}
\label{eq_maxwell U1xU1}
\begin{aligned}
&U(1)_\text{e}^{(1)}~:~~~J_\text{e}=\frac{i}{ e^2} F~, &&\eta_\text{e}(\alpha)=\exp{(i\alpha \int_{\mathcal{M}_2}\star J_\text{e})}~;\\
    &U(1)_\text{m}^{(1)}~:~~~J_\text{m}=\frac{1}{2\pi}\star F~,&&\eta_\text{m}(\alpha)=\exp{(i\alpha \int_{\mathcal{M}_2}\star J_\text{m})}~,
\end{aligned}
\end{equation}
where $\mathcal{M}_2 \subset \mathcal{M}_4$ is a closed 2-dimensional surface in spacetime. They are genuine 2d surface defects in spacetime, and are special examples of the Gukov-Witten defects \cite{Gukov:2006jk,Gukov:2008sn}. 
These defects are topological, which follows from the equation of motion  $d\star J_\text{e}=id\star F/e^2=0$ and the Bianchi identity $d\star J_\text{m}=dF/2\pi=0$. 

These two 1-form global symmetries in \eqref{eq_maxwell U1xU1} have a mixed 't Hooft anomaly. We couple the symmetry currents to two 2-form background gauge fields, which we denote as $B_\text{e}$ and $B_\text{m}$, respectively. 
The covariant Maxwell action reads
\begin{equation}
\label{eq_covariant maxwell action}
    S_\text{Maxwell}=\frac{1}{2e^2}\int_{\mathcal{M}_{4}}(F-B_\text{e})\wedge \star (F-B_\text{e})+\frac{i}{2\pi}\int_{\mathcal{M}_{4}}B_\text{m}\wedge (F-B_\text{e})~.
\end{equation}
The background  gauge transformations for $U(1)^{(1)}_\text{e}\times U(1)^{(1)}_\text{m}$ are:
\begin{equation}
\label{eq_1form gauge transformation}
    A\to A+a_\text{e}~,~~~B_\text{e}\to B_\text{e}+da_\text{e}~,~~~B_\text{m}\to B_\text{m}+da_\text{m}~,
\end{equation}
where $a_\text{e}$ and $a_\text{m}$ are two 1-form gauge parameters. 
Under these gauge transformations, the Maxwell action \eqref{eq_covariant maxwell action} is not invariant and transforms as
\begin{equation}
\label{eq_1form thooft anomaly}
    S_\text{Maxwell} \to  S_\text{Maxwell}-\frac{i}{2\pi}\int_{\mathcal{M}_4} a_\text{m}\wedge dB_\text{e}~.
\end{equation}
The variation term in \eqref{eq_1form thooft anomaly} signals the mixed 't Hooft anomaly between $U(1)_\text{e}^{(1)}$ and $U(1)_\text{m}^{(1)}$, which can be canceled by a $5$-dimensional anomaly inflow action.

For concreteness, we define the duality defect on the manifold $\mathcal{M}_4=T^3\times [0,2\pi]$, with coordinates $x_i\sim x_i+2\pi$ for $i=1,2,3$, and $0\leq x_4\leq 2\pi$. We impose the Neumann boundary condition $F_{14}=F_{24}=F_{34}=0$ on the two boundaries at $x_4=0$ and $x_4=2\pi$ for the Maxwell theory \eqref{eq_review maxwell action}. Furthermore, we denote the 1-form gauge fields on the boundaries as
\begin{equation}
\label{eq_boundary gauge fields}
    A^+(x_i)\equiv A(x_i,x_4=0)~, ~\text{and}~A^-(x_i)\equiv A(x_i,x_4=2\pi)~.
\end{equation}
The duality defect $\mathcal{D}_N$ is defined by a gluing action that couples the gauge fields on the two boundaries \cite{Gaiotto:2008ak,Kapustin:2009av,Choi:2021kmx, Niro:2022ctq, Cordova:2023ent,Kim:2025zdy}:
\begin{equation}
\label{eq_maxwell duality defect action}
\mathcal{D}_N~:~~~S_\text{duality}=\frac{iN}{2\pi}\int_{T^3} A^-\wedge dA^+~.
\end{equation}
Similar to the standard argument in Chern-Simons gauge theories, gauge invariance requires $N\in \mathbb{Z}$. For simplicity, we focus on the $N>0$ case in the following, noting that $\mathcal{D}_{-N}=C\times \mathcal{D}_{N}=\mathcal{D}_{N}\times C$, where $C$ is the charge conjugation defect. 

It follows from the equation of motion that the stress-energy tensor is conserved across the duality defect \eqref{eq_maxwell duality defect action} if 
\begin{equation}
\label{eq_defect topo cond}
    e^2=\frac{2\pi }{N}~.
\end{equation}
We will always assume this condition from now on, so that $\mathcal{D}_N$ is a topological defect\footnote{More generally, the topological duality defect can be defined for $e^2/2\pi \in \mathbb{Q}^+$; See \cite{Niro:2022ctq, Cordova:2023ent, Shao:2025qvf}.}.

We note the variation of the Maxwell action \eqref{eq_review maxwell action} on the open manifold $\mathcal{M}_4=T^3\times [0,2\pi]$ is given by
\begin{align}\label{dSbulk}
\updelta S_\text{Maxwell} =  {N\over 2\pi} \int_{T^3\times [0,2\pi]} \updelta A\wedge d\star F
-{N\over 2\pi} \int_{T^3} \updelta A^+ \wedge \star F^+
+{N\over 2\pi} \int_{T^3} \updelta A^{-} \wedge \star F^-\,.
\end{align}
Clearly, the first term vanishes when the equations of motion are imposed, while the second and the third terms have to be canceled against the variation of the defect action \eqref{eq_maxwell duality defect action}:
\begin{align}
\updelta S_\text{duality}= {iN\over 2\pi} \int_{T^3} (\updelta A^{-} \wedge dA^+ +A^- \wedge d\updelta A^+)
\end{align}
We therefore find
\begin{equation}
\label{eq_4d modified Neumann}
    F^+=i\star F^-~.
\end{equation}
This gluing condition across $\mathcal{D}_N$ implements the electromagnetic duality (or S-duality). In particular, \eqref{eq_4d modified Neumann} implies the fusion rules
\begin{equation}
\label{eq_maxwell fusion rule 1}
\begin{aligned}
    \eta_\text{e}(\alpha)\times \mathcal{D}_N=&\mathcal{D}_N\times \eta_\text{m}(N\alpha)~,\\
     \mathcal{D}_N\times \eta_\text{e}(\alpha)=&\eta_\text{m}(-N\alpha)\times\mathcal{D}_N ~.
\end{aligned}
\end{equation}
We note that the duality defect can absorb the electric 1-form symmetry defects associated with the subgroup $\mathbb{Z}^{(1)}_N\subset U(1)^{(1)}_\text{e}$:
\begin{equation}
\label{eq_maxwell fusion corollary}
    \eta_\text{e}(\alpha)\times \mathcal{D}_N=\mathcal{D}_N\times \eta_\text{e}(\alpha)=\mathcal{D}_N~,~~~\text{if } \alpha\in \frac{2\pi \mathbb{Z}}{N}~.
\end{equation}
This indicates that $\mathcal{D}_N$ is non-invertible when $N>1$. At $N=1$, \eqref{eq_defect topo cond} reduces to the self-dual radius, and the duality defect $\mathcal{D}_{N=1}$ is invertible. 

\subsection{Twist defects for electromagnetic duality}

\begin{figure}[thb]
\centering
\includegraphics[width=.95\textwidth]{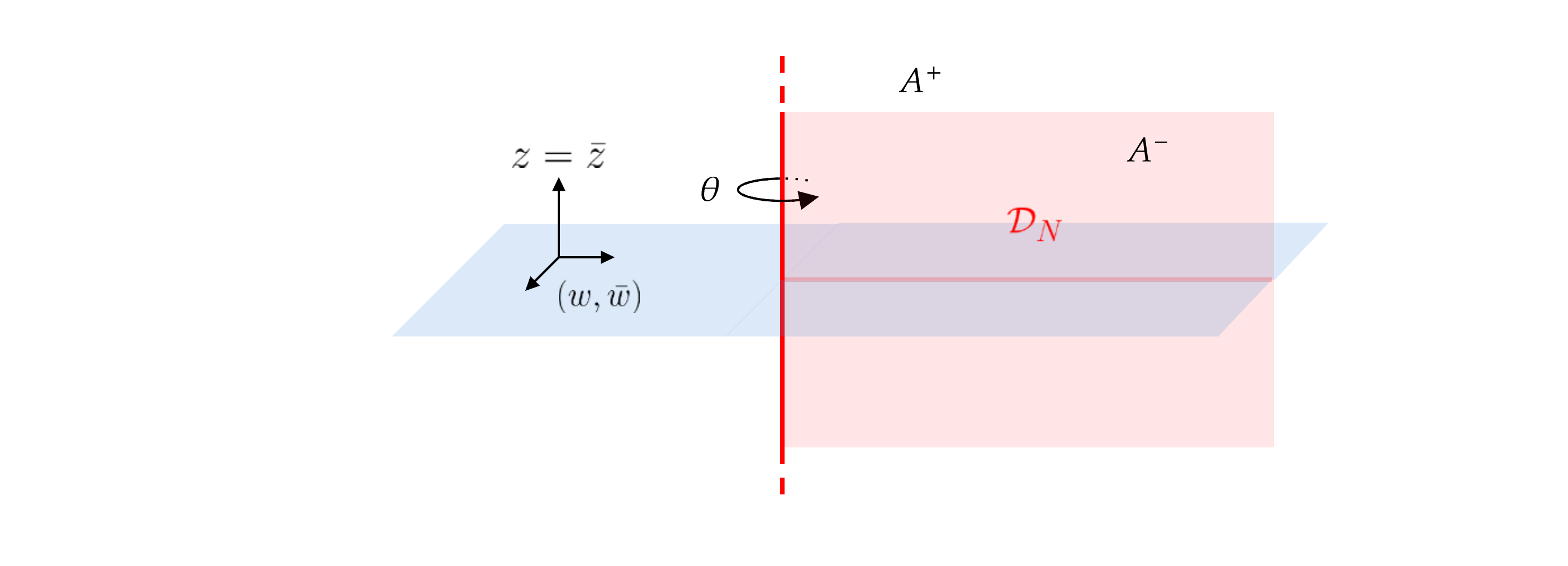}
  \caption[The configuration of the conformal twist defect in $\mathbb{R}^4$.]{\label{pic_complex coordinate} The configuration of the conformal twist defect and the topological duality defect in $\mathbb{R}^4$.  
  The vertical red line is the 2-dimensional twist defect, which is localized at $w=\bar{w}=0$ and extends along the $(z,\bar{z})$ plane. The twist defect is attached to the 2-dimensional topological defect $\mathcal{D}_N$ shown as the half-infinite red plane. This figure only displays the $\mathbb{R}^3$ slice defined by $z=\bar{z}$.} 
\end{figure}

With preliminaries in place, we now introduce the setup for the $p=2$ chiral surface defect. We consider the Euclidean spacetime $\mathcal{M}_4=\mathbb{R}^4$, with complex coordinates $z,w\in \mathbb{C}$ and the metric
\begin{equation}
\label{eq_Euclidean metric}
    ds^2=2dzd\bar{z}+2dwd\bar{w}~.
\end{equation}
We implement the topological duality defect $\mathcal{D}_N$ so that it covers the complex $z$-plane and extends along the positive real axis $w>0$ of the complex $w$-plane, terminating at the origin $w=\bar{w}=0$. See Figure \ref{pic_complex coordinate}. More concretely, the action for the duality defect reads
\begin{equation}
\label{eq_maxwell twist defect attach}
\mathcal{D}_N~:~~~S_\text{duality}=\frac{iN}{2\pi}\int_{w\in \mathbb{R}^+} A^-\wedge dA^+~.
\end{equation}
where $A_\pm$ are the gauge fields on the two sides of $\mathcal{D}_N$:
\begin{equation}
\label{eq_gauge fields across defect}
A^\pm(z,\bar{z},|w|)=\lim_{\epsilon\to0^{\pm}}A(z,\bar{z},w=|w|+ i\epsilon,\bar{w}=|w|- i\epsilon)~.
\end{equation}
Since the duality defect $\mathcal{D}_N$ is topological, it preserves the $\mathfrak{so}(2)$ rotation symmetry of the complex $w$-plane. Denoting $w=|w|e^{ i \theta }$, we find that the gluing condition \eqref{eq_4d modified Neumann} implies the following monodromy condition under $\theta\to \theta+2\pi$:
 \begin{equation}
 \label{eq_Euclidean twist condition}
 F \longrightarrow -i\star F\longrightarrow -F\longrightarrow i\star F \longrightarrow F~,
\end{equation}
The $\theta\sim \theta+2\pi$ periodicity is therefore lifted to $\theta\sim \theta+8\pi$ in the presence of $\mathcal{D}_N$.

The twist defect \cite{Billo:2013jda, Gaiotto:2013nva, Giombi:2021uae,Bianchi:2021snj} is defined by the $p=q=2$ boundary of the topological defect $\mathcal{D}_N$ as in Figure \ref{pic_complex coordinate}. The Maxwell theory in 4-dimensional spacetime has the conformal symmetry $\mathfrak{so}(5,1)$. In what follows, we assume that the twist defect preserves a $\mathfrak{so}(3,1)\times \mathfrak{so}(2)$ conformal subalgebra at the defect RG fixed point.

What are the possible twist defect CFTs compatible with the duality defect \eqref{eq_maxwell twist defect attach} and the bulk monodromy condition \eqref{eq_Euclidean twist condition}? In the following subsections, we address this question using the techniques developed in \cite{Herzog:2020bqw, Herzog:2022jqv, Shao:2025qvf}.

\subsection{Bulk-defect two-point functions}

\begin{figure}[thb]
\centering
\includegraphics[width=1\textwidth]{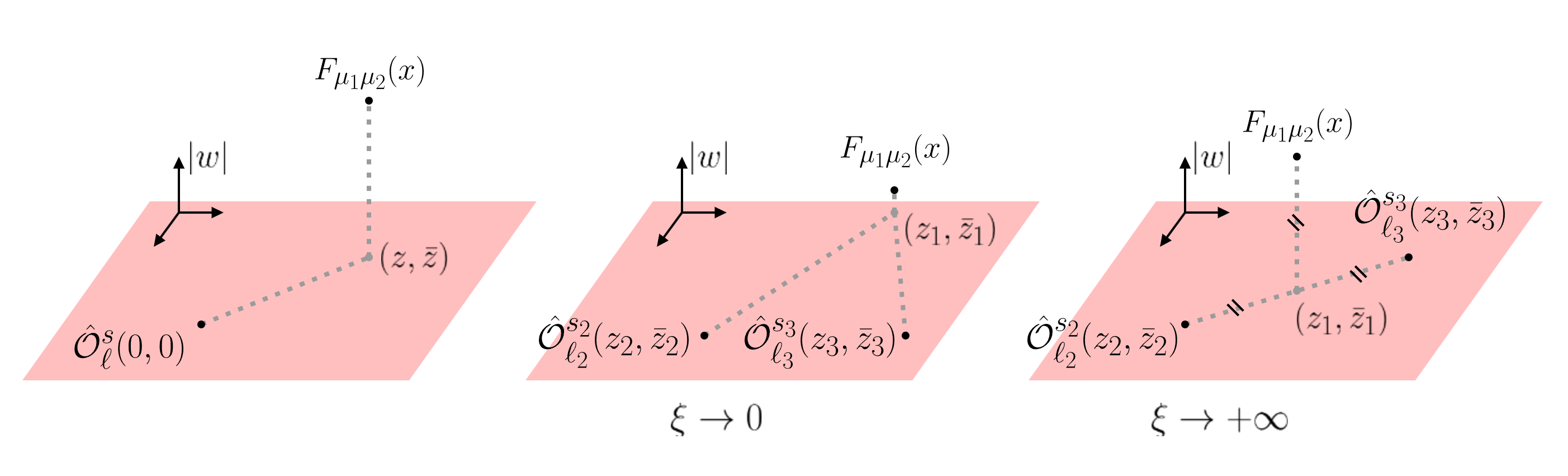}
  \caption[DCFT two-point and three-point functions.]{\label{pic_DCFT points}DCFT two-point and three-point functions. The red plane represents the twist defect at $w=\bar{w}=0$. Left: the bulk-defect two-point function $\langle F \hat{\mathcal{O}}^s_\ell\rangle$. Middle: the bulk-defect-defect three-point function $\langle F\hat{\mathcal{O}}_{\ell_2}^{s_2} \hat{\mathcal{O}}_{\ell_3}^{s_3}\rangle$ in the limit $\xi\to 0$, where the bulk point $x$ approaches the defect. Right: the three-point function $\langle F\hat{\mathcal{O}}_{\ell_2}^{s_2} \hat{\mathcal{O}}_{\ell_3}^{s_3}\rangle$ in the limit $\xi \to +\infty$, where the projection of the bulk point onto the defect plane coincides with the midpoint between $(z_2,\bar{z}_2)$ and $(z_3,\bar{z}_3)$. } 
\end{figure}

We first consider the two-point function $\langle F \hat{\mathcal{O}}_{\ell}^s\rangle$ between the bulk field strength operator $F_{\mu_1\mu_2}(x)$ and a defect primary $ \hat{\mathcal{O}}_{\ell}^s$ at $z=\bar{z}=0$. See Figure \ref{pic_DCFT points}. With slight abuse of notation, we now allow the parallel spin $\ell$ and the transverse spin $s$ to take both positive and negative values, corresponding to the doublet representations of $\mathfrak{so}(2)$. We also denote the conformal weights of $\mathfrak{so}(3,1)$ by $(h,\bar{h})$, such that $\Delta=h+\bar{h}$ and $\ell=h-\bar{h}$.

Our first hint follows from \eqref{eq_Euclidean twist condition}. For the two-point function to be nontrivial, the defect operator has to satisfy the spin selection rule 
\begin{align}\label{eq_spinselection}
\langle F\hat{\mathcal{O}}_{\ell}^s\rangle \neq 0 ~~ \Rightarrow ~~ s\in \mathbb{Z}\pm \frac 14\,.
\end{align}
The form of the non-zero function $\langle F \hat{\mathcal{O}}_{\ell}^s\rangle$ is strongly constrained by the $so(3,1)\times so(2)$ conformal symmetry. In particular, covariant tensor structures in the two-point function between a bulk primary and a defect primary are spanned by polynomials of the following basis tensors:
\begin{equation}
\label{eq_tensor building blocks}
\begin{aligned}
    \mathcal{X}_\mu=&\frac{|w|}{|z|^2+|w|^2}\left(\bar{z},z,\frac{\bar{w}}{2}(1-|z/w|^2),\frac{w}{2}(1-|z/w|^2)\right)~;\\
    \mathcal{Y}_\mu=&\frac{1}{2}\left(0,0,\frac{\bar{w}}{|w|},-\frac{w}{|w|}\right)~;\\
   \mathcal{I}_{\mu z}=&-\frac{1}{2(|z|^2+|w|^2)}\left(\bar{z}^2,-|w|^2,\bar{z} \bar{w},\bar{z}w\right)~;\\
   \mathcal{I}_{\mu \bar{z}}=&-\frac{1}{2(|z|^2+|w|^2)}\left(-|w|^2,z^2,z\bar{w},zw\right) ~.
\end{aligned}
\end{equation}
We have $\mathcal{X}^\mu \mathcal{X}_\mu=\frac{1}{2}$, $\mathcal{Y}^\mu \mathcal{Y}_\mu=-\frac{1}{2}$, and $\mathcal{I}^\mu_{\phantom{\mu} z}\mathcal{I}_{\mu \bar{z}}=\frac{1}{4}$, while the index contractions between other pairs of these tensors yield zero. Since one cannot build a tensor out of \eqref{eq_tensor building blocks} with a pair of anti-symmetrized indices $[\mu_1\mu_2]$ and parallel spin $|\ell|\geq 2$,  the two-point function $\langle F \hat{\mathcal{O}}_{\ell}^s\rangle$ must vanish unless $|\ell|<2$. To admit a non-zero two-point function, the defect primary $\hat{\mathcal{O}}_{\ell}^s$ is either a scalar ($\ell=0$) or a vector ($\ell=\pm1$). It is further shown in \cite{Herzog:2022jqv} that a defect scalar primary $\hat{\mathcal{O}}^s$ yields $\langle F \hat{\mathcal{O}}^s\rangle=0$ unless $s=0$ and $\Delta(\hat{\mathcal{O}}^s)=2$, which is obviously incompatible with the spin selection rule \eqref{eq_spinselection}. We are therefore left with the vector primaries.

We denote the vector primary of parallel spin $\ell=1$ and transverse spin $s$ on the twist defect by $\hat{V}_{z}^{s}$. The two-point function $ \langle F_{\mu_1\mu_2}\mathcal{V}^{s}_z\rangle$ admits two compatible tensor structures, namely $\mathcal{X}_{[\mu_1}\mathcal{I}_{\mu_2] z}$ and $\mathcal{Y}_{[\mu_1}\mathcal{I}_{\mu_2] z}$. Note that the linear combination $\mathcal{X}_{[\mu_1}\mathcal{I}_{\mu_2] z}+\mathcal{Y}_{[\mu_1}\mathcal{I}_{\mu_2] z}$ is self-dual under the Hodge star operation, while $\mathcal{X}_{[\mu_1}\mathcal{I}_{\mu_2] z}-\mathcal{Y}_{[\mu_1}\mathcal{I}_{\mu_2] z}$ is anti-self-dual. The conformal algebra $so(3,1)\times so(2)$ and the monodromy condition \eqref{eq_Euclidean twist condition} fix the functional form of $ \langle F_{\mu\nu}\mathcal{V}^{s}_z\rangle$ up to the OPE coefficient as follows
\begin{equation}
\label{eq_bulktodefect 2pt 1}
     \left. \langle F_{\mu_1\mu_2}(x) \hat{V}_{z}^{s}(0) \rangle\right|_{s\in \mathbb{Z}\pm \frac{1}{4}}=\frac{C_{F}{}^{\hat{V}_{z}^{s}}}{|w|^2}\left(\frac{|w|}{|w|^2+|z|^2}\right)^{\Delta(\hat{V}_{z}^{s})}\left(\mathcal{X}_{[\mu_1}\mathcal{I}_{\mu_2] z} \mp \mathcal{Y}_{[\mu_1}\mathcal{I}_{\mu_2] z}\right)\left(\frac{w}{|w|}\right)^{s}~.
\end{equation}
A similar analysis can be applied to the defect vector primary $\hat{V}_{\bar{z}}^{s}$ of parallel spin $\ell=-1$, and we find 
\begin{equation}
\label{eq_bulktodefect 2pt 2}
\left. \langle F_{\mu_1\mu_2}(x) \hat{V}_{\bar{z}}^{s}(0) \rangle\right|_{s\in \mathbb{Z}\pm \frac{1}{4}}=\frac{C_{F}{}^{\hat{V}_{\bar{z}}^{s}}}{|w|^2}\left(\frac{|w|}{|w|^2+|z|^2}\right)^{\Delta(\hat{V}_{\bar{z}}^{s})}\left(\mathcal{X}_{[\mu_1}\mathcal{I}_{\mu_2] \bar{z}} \pm \mathcal{Y}_{[\mu_2}\mathcal{I}_{\mu_1] \bar{z}}\right)\left(\frac{w}{|w|}\right)^{s}~.
\end{equation}

Imposing the Bianchi identity $dF=0$ on \eqref{eq_bulktodefect 2pt 1} and \eqref{eq_bulktodefect 2pt 2}, we find that non-zero OPE coefficients are only compatible with the following scaling dimensions 
\begin{align}
\Delta(\hat{V}_{z}^{s})|_{s\in \mathbb{Z}\pm \frac{1}{4}}=1\pm s \,,~~~~~
\Delta(\hat{V}_{\bar{z}}^{s})|_{s\in \mathbb{Z}\pm \frac{1}{4}}=1\mp s\,.
\end{align}
Note that the unitarity bound for a defect vector is $\Delta \geq 1$, which is saturated by the conserved spin-1 currents. We therefore conclude that the primaries on the twist defect that appear in the OPE of the bulk field strength $F$ are given by
\begin{equation}
\label{eq_defect vector primary}
    \begin{aligned}
\hat{V}_{z}^{s}~:&~~~(h,\bar{h})=\left(1+\frac{|s|}{2},\frac{|s|}{2}\right)~,~~~\text{where}~|s|\in \mathbb{N}+\frac{1}{4}~;\\
\hat{V}_{\bar{z}}^{s}~:&~~~(h,\bar{h})=\left(\frac{|s|}{2},1+\frac{|s|}{2}\right)~,~~~\text{where}~|s|\in \mathbb{N}+\frac{3}{4}~.
    \end{aligned}
\end{equation}
Notably, the conformal weights in \eqref{eq_defect vector primary} are completely determined by conformal symmetry and unitarity. 
They are universal and insensitive to the possible interactions localized on the twist defect. Moreover, the transverse spin $s$ in \eqref{eq_defect vector primary} is subject to the selection rule that depends on the parallel spin $\ell$. This manifests a spin-momentum lock of photons propagating along the defect.

\subsection{Bulk-defect-defect three-point functions}

Next, we consider the three-point function $\langle F\hat{\mathcal{O}}_{\ell_2}^{s_2} \hat{\mathcal{O}}_{\ell_3}^{s_3}\rangle$ between the bulk field strength operator $F$ at $x=(z_1,\bar{z}_1,w,\bar{w})$ and defect primaries $\hat{\mathcal{O}}_{\ell_2}^{s_2}$ at $(z_2, \bar{z}_2)$ and $\hat{\mathcal{O}}_{\ell_3}^{s_3}$ at $(z_3, \bar{z}_3)$. See Figure \ref{pic_DCFT points}.

It follows from the monodromy condition \eqref{eq_Euclidean twist condition} that $\langle F\hat{\mathcal{O}}_{\ell_2}^{s_2} \hat{\mathcal{O}}_{\ell_3}^{s_3}\rangle=0$ unless $s=s_2+s_3\in \mathbb{Z}\pm \frac{1}{4}$, while $s_2$ and $s_3$ themselves need not satisfy the spin selection rule \eqref{eq_spinselection}. The form of the non-zero function $\langle F\hat{\mathcal{O}}_{\ell_2}^{s_2} \hat{\mathcal{O}}_{\ell_3}^{s_3}\rangle$ is fixed by the conformal algebra $so(3,1)\times so(2)$ up to functions of the cross ratio $\xi$, defined as:
\begin{equation}
\label{eq_crossration}
    \xi=\frac{|z_{23}|^2|w|^2}{(|w|^2+z_{12}\bar{z}_{13})(|w|^2+\bar{z}_{12}z_{13})}>0~,
\end{equation}
where $z_{ij}=z_i-z_j$. 

Tensor structures in the three-point function $\langle F\hat{\mathcal{O}}_{\ell_2}^{s_2} \hat{\mathcal{O}}_{\ell_3}^{s_3}\rangle$ can be enumerated in a way similar to \eqref{eq_bulktodefect 2pt 1} and \eqref{eq_bulktodefect 2pt 2}. They are spanned by polynomial combinations of four vector building blocks: $\mathcal{X}_{12,\mu}$, $\mathcal{X}_{13,\mu}$, $\mathcal{Y}_{\mu}$, and $\mathcal{I}_{\mu}$. The pseudo-vector $\mathcal{Y}_\mu$ follows the same definition as in \eqref{eq_tensor building blocks}, while the other vectors are given by
\begin{equation}
\begin{aligned}
    \mathcal{X}_{ij,\mu}={}&\frac{|w|}{|z_{ij}|^2+|w|^2}\left(\bar{z}_{ij},z_{ij},\frac{\bar{w}}{2}(1-|z_{ij}/w|^2),\frac{w}{2}(1-|z_{ij}/w|^2)\right)~;\\
    \mathcal{I}_\mu={}&\frac{z_{12}\bar{z}_{13}+|w|^2}{2|w|z_{23}(|z_{13}|^2+|w|^2)}\left(-|w|^2,z_{13}^2,\bar{w}z_{13},wz_{13}\right)~.
\end{aligned}
\end{equation}
We consider the following combinations of these building block vectors
\begin{equation}
\label{eq_rank2 antisymmetric base}
    \mathcal{X}_{12,[\mu_1}\mathcal{X}_{13,\mu_2]}~,~~\mathcal{X}_{12,[\mu_1}\mathcal{Y}_{\mu_2]}~,~~\mathcal{X}_{12,[\mu_1}\mathcal{I}_{\mu_2]}~,~~\mathcal{X}_{13,[\mu_1}\mathcal{Y}_{\mu_2]}~,~~\mathcal{X}_{13,[\mu_1}\mathcal{I}_{\mu_2]}~,~~\mathcal{Y}_{[\mu_1}\mathcal{I}_{\mu_2]}~.
\end{equation}
These six tensors are linearly independent, and they provide a complete basis for rank-2 anti-symmetric tensors. We organize the \eqref{eq_rank2 antisymmetric base} into the combinations that are either self-dual $(+)$ and anti-self-dual $(-)$ under the Hodge star operation:
\begin{equation}
\label{eq_rank2 antisymmetric tensor def}
\mathtoolsset{multlined-width=0.9\displaywidth}
\begin{multlined}
\mathcal{Z}_{1,\mu_1\mu_2}^\pm= \mathcal{X}_{13,[\mu_1}\mathcal{I}_{\mu_2]}\mp \mathcal{Y}_{[\mu_1} \mathcal{I}_{\mu_2 ]}~;\hfill\\
\mathcal{Z}_{2,\mu_1\mu_2}^\pm=\mathcal{X}_{12,[\mu_1}\mathcal{X}_{13,\mu_2]}-2\mathcal{X}_{12,[\mu_1}\mathcal{I}_{\mu_2]}+2\mathcal{X}_{13,[\mu_1}\mathcal{I}_{\mu_2]}\pm (\mathcal{X}_{12,[\mu_1}\mathcal{Y}_{\mu_2]}+ \mathcal{X}_{13,[\mu_1 }\mathcal{Y}_{\mu_2 ]})~;\hfill\\
\mathcal{Z}_{3,\mu_1\mu_2}^\pm=\xi^{-1}(\mathcal{X}_{12,[\mu_1}\mathcal{Y}_{\mu_2 ]}-\mathcal{X}_{13,[\mu_1 }\mathcal{Y}_{\mu_2 ]})\hfill\\
\pm(2\mathcal{X}_{12,[\mu_1}\mathcal{I}_{\mu_2]}+2\mathcal{X}_{13,[\mu_1}\mathcal{I}_{\mu_2]}+\xi^{-1}\mathcal{X}_{12,[\mu_1}\mathcal{X}_{13,\mu_2]})~.
\end{multlined}
\end{equation}

With \eqref{eq_rank2 antisymmetric tensor def} at hand, we find that a general three-point function $\langle F\hat{\mathcal{O}}_{\ell_2}^{s_2} \hat{\mathcal{O}}_{\ell_3}^{s_3}\rangle$ takes the form
\begin{equation}
 \label{eq_3pt function}
\mathtoolsset{multlined-width=0.9\displaywidth}
\begin{multlined}
\left. \langle F_{\mu_1 \mu_2}(x)\hat{\mathcal{O}}_{\ell_2}^{s_2}(z_2,\bar{z}_2) \hat{\mathcal{O}}_{\ell_3}^{s_3}(z_3,\bar{z}_3)\rangle\right|_{s\in \mathbb{Z}\pm \frac{1}{4}}=\frac{(w/|w|)^{s}|w|^{\Delta(\hat{\mathcal{O}}_{\ell_2}^{s_2})+\Delta(\hat{\mathcal{O}}_{\ell_3}^{s_3})-2}}{(|z_{12}|^2+|w|^2)^{\Delta(\hat{\mathcal{O}}_{\ell_2}^{s_2})}(|z_{13}|^2+|w|^2)^{\Delta(\hat{\mathcal{O}}_{\ell_3}^{s_3})}}\hfill\\
\times \left(\frac{|w|^2+\bar{z}_{12}z_{13}}{z_{23}|w|}\right)^{\ell_2}\left(\frac{|w|^2+\bar{z}_{13}z_{12}}{z_{23}|w|}\right)^{\ell_3}\left(\sum_{a=1}^3f^\mp_a(\xi) \mathcal{Z}_{a,\mu_1\mu_2}^{\mp}\right)~,
\end{multlined}
\end{equation}
where $f^\pm_{a}(\xi)$ are real-valued functions of the cross ratio \eqref{eq_crossration}. Imposing the Bianchi identity $dF=0$ on \eqref{eq_3pt function}, we obtain an algebraic relation among the functions $f^\pm_{a}$: 
\begin{equation}
\label{eq_three point linear}
   \big(
   \frac{s}{4}\pm (h_2-h_3)
   \big)
   f_1^{\pm}+sf_2^{\pm}+(\ell_2-\ell_3)f_3^\pm=0~,\text{ for }s\in \mathbb{Z}\mp \frac{1}{4}~.
\end{equation}
This relation allows us to express $f_2^{\pm}$ in terms of $f_1^{\pm}$ and $f_3^{\pm}$. Furthermore, the Bianchi identity yields an ordinary differential equation:
\begin{equation}
\label{eq_three point closure condition}
\mathtoolsset{multlined-width=0.9\displaywidth}
\begin{multlined}
    s\xi(1+\xi)\partial_\xi f_1^\pm=2\left(s^2+\xi(\ell_2-\ell_3)^2\right)f_3^\pm\hfill\\
    \hfill+\left[\left(\frac{s}{2}(2+\ell_2+\ell_3)\pm(h_2-h_3)(\ell_2-\ell_3)\right)\xi+s(1\pm \frac{s}{2}-\bar{h}_2-\bar{h}_3)\right]f^\pm_1~,\\
    s\xi(1+\xi)\partial_\xi f_3^\pm=\frac{\xi}{8}(s^2-4(h_2-h_3)^2)f_1^\pm\hfill\\
    \hfill+\left[\left(\frac{s}{2}(2+\ell_2+\ell_3)\mp(h_2-h_3)(\ell_2-\ell_3)\right)\xi+s(1\mp\frac{s}{2}-\bar{h}_2-\bar{h}_3)\right]f^\pm_3~.
\end{multlined}
\end{equation}
The two hypergeometric solutions to \eqref{eq_three point closure condition} are discussed in \cite{Shao:2025qvf}. These solutions are subject to constraints from unitarity (at $\xi \to 0$) as well as convergence (at $\xi \to \infty$). Crucially, we find that $\langle F\hat{\mathcal{O}}_{\ell_2}^{s_2} \hat{\mathcal{O}}_{\ell_3}^{s_3}\rangle=0$ unless the following conditions are satisfied:
\begin{equation}
\label{eq_double twist condition}
\begin{aligned}
\text{if}~|s|\in \mathbb{N}+\frac{1}{4}~,~~~ \text{either}~(h_2,\bar{h}_2)-(h_3,\bar{h}_3) \in {}& (1+\frac{|s|}{2}+\mathbb{N},\frac{|s|}{2}+\mathbb{N})\\
\text{or}~(h_3,\bar{h}_3)-(h_2,\bar{h}_2) \in {}&(1+\frac{|s|}{2}+\mathbb{N},\frac{|s|}{2}+\mathbb{N})~;\\
\text{if}~|s|\in \mathbb{N}+\frac{3}{4}~,~~~ \text{either}~(h_2,\bar{h}_2)-(h_3,\bar{h}_3) \in {}& (\frac{|s|}{2}+\mathbb{N},1+\frac{|s|}{2}+\mathbb{N})\\
\text{or}~(h_3,\bar{h}_3)-(h_2,\bar{h}_2) \in {}&(\frac{|s|}{2}+\mathbb{N},1+\frac{|s|}{2}+\mathbb{N})~.\\
\end{aligned}
\end{equation}
We see that the differences between the allowed values of $(h_2,\bar h_2)$ and $(h_3,\bar h_3)$ correspond to the conformal weights of $\hat{V}_{z}^s$ and $\hat{V}_{\bar z}^s$ in \eqref{eq_defect vector primary}, or their descendants.

The selection rule \eqref{eq_double twist condition} implies that the operator spectrum in the OPE between $\hat{V}_{z}^s$, $\hat{V}_{\bar{z}}^s$, and a generic defect primary $\hat{\mathcal{O}}^s_\ell$ is of the double-twist type, with no anomalous dimension. The correlation functions of these defect operators therefore obey Wick's theorem \cite{Lauria:2020emq, Herzog:2022jqv}. We thus conclude that $\hat{V}_{z}^s$, $\hat{V}_{\bar{z}}^s$, and their descendants form a universal sector of generalized free fields on the twist defect.

Notably, the generalized free field spectrum contains a unique operator doublet with $|s|=1$, $\ell=0$, and $\Delta=3$. They are exactly the displacement operators \eqref{eq_displacement def}, whose existence marks the mobility of the twist defect. Up to overall normalizations, we identify the displacement operators as
\begin{equation}
    \hat{D}^{+ 1}=\hat{V}_z^{+ \frac{1}{4}}\hat{V}_{\bar{z}}^{+ \frac{3}{4}}~,~~~\hat{D}^{- 1}=\hat{V}_z^{- \frac{1}{4}}\hat{V}_{\bar{z}}^{- \frac{3}{4}}~.
\end{equation}

\subsection{Chiral current sector and anyonic branes}

We have determined a universal sector of the twist defect CFT using just conformal symmetry \eqref{eq_DCFT symmetry}, the monodromy condition \eqref{eq_Euclidean twist condition}, and the unitarity. In this subsection, we argue that there has to be another sector, which we call the chiral current sector, enforced by anomaly inflow and the 1-form global symmetry \eqref{eq_maxwell U1xU1}. This is intuitively clear, since the action \eqref{eq_maxwell twist defect attach} for the 3d topological duality defect $\mathcal{D}_N$ takes the form of a Chern-Simons action. 
Our discussion below follows closely the standard treatment of conformal boundary conditions of chiral 3d Chern-Simons theory in \cite{Witten:1988hf,Elitzur:1989nr}. 

Again, let us consider the defect configuration in Figure \ref{pic_complex coordinate}. Since the topological defect $\mathcal{D}_N$ is supported on a three-dimensional manifold with a boundary, variation of its action yields an extra boundary contribution:
\begin{align}
\updelta S_\text{duality}= {iN\over 2\pi} \int_{w\in \mathbb{R}^+} (\updelta A^{-} \wedge dA^+ + \updelta A^+ \wedge dA^- ) 
+{i N\over 2\pi} \int_{w=\bar w=0} A\wedge \updelta A \,.
\end{align}
With the gluing condition \eqref{eq_4d modified Neumann}, the first two terms are canceled by the variation of the bulk action in \eqref{dSbulk}. 
Following \cite{Elitzur:1989nr}, we pick a complex structure along the twist defect at $w=\bar w=0$ and introduce a local counterterm:
\begin{equation}
\label{eq_counter term}
S_{\text{counter}}=-\frac{N}{2\pi}\int_{w=\bar{w}=0} A_z A_{\bar{z}}dzd\bar{z}~. 
\end{equation}
The variation of the total action is then
\begin{equation}
\label{eq_twist defect current}
\updelta(S_\text{bulk}+S_\text{duality}+S_\text{counter})=-\frac{N}{\pi}\int_{w=\bar{w}=0} A_{\bar{z}}\updelta A_{z} dzd\bar{z}~.
\end{equation}
To ensure a well-defined variational principle, we impose a Dirichlet condition on the holomorphic component of the gauge field:
\begin{equation}
\label{eq_chiral defect condition}
    \left. A_{z}\right|_{w=\bar{w}=0}=0~.
\end{equation}
This is precisely the usual conformal boundary condition for the chiral WZW model in the chiral Chern-Simons theory \cite{Witten:1988hf, Elitzur:1989nr}. 
The standard analysis implies that there is a chiral compact boson $\hat{\varphi}$ living along the twist defect, which, on-shell, is related to the gauge field as $A_{\bar z} |_{w=\bar w=0}= \partial_{\bar z}\hat{\varphi}$. 
From \eqref{eq_twist defect current}, we define a current operator as:
\begin{equation}\label{chiralcurrent}
    \hat{J}_{\bar{z}}=-\frac{i N}{\pi} A_{\bar{z}} = - {iN\over \pi} \partial_{\bar z}\hat{\varphi}~. 
\end{equation}
It has conformal weights $(h,\bar{h})=(0,1)$ and zero transverse spin $s=0$. 
From the spin selection rule in \eqref{eq_spinselection}, the two-point function $\langle F \hat{J}_{\bar{z}}\rangle=0$ vanishes. We have also shown in \eqref{eq_double twist condition} that fusion of defect operators $\hat{V}_z^s$, $\hat{V}_{\bar{z}}^s$, and their descendants cannot produce current operators with either $h=0$ or $\bar{h}=0$. We therefore conclude that correlation functions associated with the chiral current sector and the generalized free field sector factorize.

\begin{figure}[thb]
\centering
\includegraphics[width=1\textwidth]{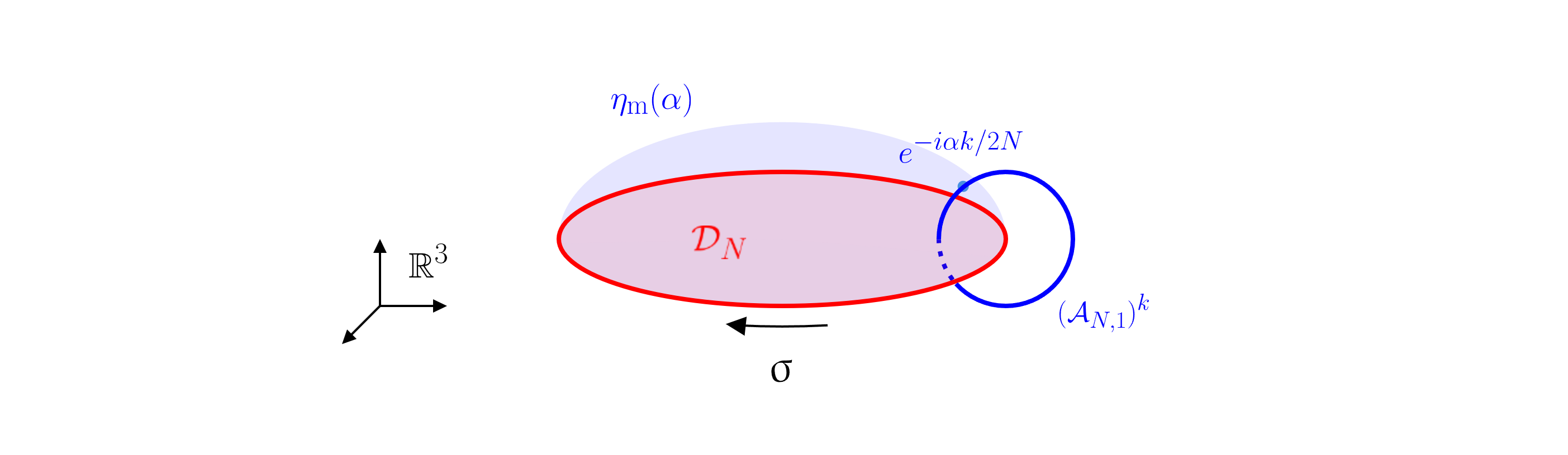}
  \caption[The configuration of the anyonic brane $(\mathcal{A}_{N})^k$ and the twist defect.]{\label{pic_linking} The configuration of the anyonic brane $(\mathcal{A}_{N})^k$ and the twist defect. The red circle denotes the twist defect defined on a Lorentzian cylinder. The red disk denotes the topological duality defect $\mathcal{D}_N$. The anyonic brane is shown as a blue circle, which links with the twist defect. The magnetic 1-form symmetry operator $\eta_m(\alpha)$ is shown as the blue surface.  This figure only shows the spatial $\mathbb{R}^3$, and every defect except for $\eta_\text{m}(\alpha)$ extends in the time direction.} 
\end{figure}

How does the chiral current localized on the twist defect couple to the bulk Maxwell theory? To see this, we place the twist defect on a Lorentzian cylinder parameterized by $(t,\upsigma)$ and embed it in the $(3+1)$-dimensional spacetime $\mathbb{R}^{3,1}$. See Figure \ref{pic_linking}. The effective action of the right-moving chiral current is given by\footnote{For manifestly Lorentz invariant formulations of the chiral boson action, see, e.g. \cite{Floreanini:1987as, Sen:2019qit, Mkrtchyan:2019opf}.}
\begin{equation}
\label{eq_chiral compact boson action}
    S_{\text{chiral}}=-\frac{N}{2\pi}\int dtd\upsigma [(\partial_{\upsigma}\hat{\varphi})^2+\partial_t \hat{\varphi}\partial_{\upsigma} \hat{\varphi}]~,
\end{equation}
where the compact boson field $\hat{\varphi} \sim \hat{\varphi}+2\pi$. This theory has a chiral $U(1)_{2N}$ global symmetry that shifts the chiral boson, which is generated by the current $\hat{J}=\frac{N}{\pi}\partial_\upsigma\hat{\varphi}(dt-d\upsigma)$.

Next, we discuss the charge quantization for this chiral $U(1)_{2N}$ symmetry. The total $U(1)_{2N}$ charge on the spatial circle of the twist defect is
\begin{equation}
\label{eq_chiral current charge}
    Q_{\text{R}}=\int_{S^1}d\upsigma \hat{J}_t 
    = {2N\over 2\pi} \int_{S^1} d\upsigma \partial_\upsigma \hat{\varphi}={2N\over 2\pi}\int_{S^1}d\upsigma A_\upsigma~.
\end{equation}
The charge $Q_{\text{R}}$ can be understood in two complementary ways: In the second equation of \eqref{eq_chiral current charge}, it is the winding number of $\hat{\varphi}$ multiplied by $2N$. In the third equation of \eqref{eq_chiral current charge}, we interpret it as the gauge field holonomy, which equals the magnetic 1-form symmetry defect ending on the twist defect. See also figure \ref{pic_linking}. Note that a fractional winding number of $\hat{\varphi}$ would result in $Q_\text{R} \notin 2N\mathbb{Z}$. 
This signals a discontinuity in the chiral boson field and can be interpreted as the insertion of a 1-form symmetry defect, which we now explain.

Consider the following discrete $\mathbb{Z}_{2N}^{(1)}$ subgroup of the $U(1)^{(1)}_\text{e} \times U(1)^{(1)}_\text{m}$ 1-form global symmetry \eqref{eq_maxwell U1xU1} generated by the operator $\mathcal{A}_{N}$ supported on a 2d  surface $\mathcal{M}_2$: 
\begin{equation}\label{anyonic}
\mathcal{A}_{N}=\eta_{\text{e}}(\frac{\pi}{N})\eta_{\text{m}}(\pi)=\exp\left(\frac{i}{2}\int_{\mathcal{M}_2}(F-\star F)\right)~.
\end{equation}
It is a certain linear combination of the electric and magnetic fluxes and is of order $2N$, i.e., $(\mathcal{A}_{N})^{2N}=1$. 
It follows from the fusion rule \eqref{eq_maxwell fusion rule 1} that 
$\mathcal{A}_{N}$ commutes with $\mathcal{D}_{N}$:
\begin{equation}
\label{eq_maxwell commutation relation 1}
    \mathcal{A}_{N}\times \mathcal{D}_N=\mathcal{D}_N\times \mathcal{A}_{N}~.
\end{equation}
In other words, $\mathbb{Z}_{2N}^{(1)}$ is the 1-form symmetry subgroup preserved by the duality symmetry defect $\mathcal{D}_{N}$. 
We can therefore consider the defect configuration of $\mathcal{D}_N$ intersects topologically with the $k$-th power of $\mathcal{A}_{N}$ at a point.

Let us consider the magnetic 1-form symmetry operator $\eta_\text{m}(\alpha)$, defined on an open surface that terminates on the twist defect as in figure \ref{pic_linking}. The mixed 't Hooft anomaly between $\mathcal{A}_{N}$ and $\eta_\text{m}(\alpha)$, which follows from \eqref{eq_1form thooft anomaly}, modifies the latter operator from its original form in \eqref{eq_maxwell U1xU1} to:
\begin{equation}
    \eta_\text{m}(\alpha)=\exp{\left(\frac{i\alpha}{2\pi}\int F-\frac{i\alpha k}{2N}\right)}=\exp\left(\frac{i\alpha}{2N}(Q_\text{R}-k)\right)~.
\end{equation}
Since the $\eta_\text{m}(2\pi)$ surface can be annihilated with $\mathcal{D}_N$, we demand $\eta_\text{m}(2\pi)=1$, leading to the charge quantization condition $Q_\text{R}\in 2N\mathbb{Z}+k$. 
The insertions of $(\mathcal{A}_{N,1})^k$ correspond to the $2N$ primaries of the compact chiral boson \eqref{eq_chiral compact boson action}, whose conformal weights are
\begin{equation}
\label{eq_chiral primaries}
    (h,\bar{h})=(0,\frac{Q_\text{R}^2}{4N})~,~~~\text{where}~ Q_\text{R}\in 2N\mathbb{Z}+k~.
\end{equation}
In the context of 3d chiral Chern-Simons theories, these primaries of the chiral boson correspond to the insertions of the bulk anyon lines in a similar way. 
For this reason, we will refer to $\mathcal{A}_{N}$ as an anyonic brane. 

Finally, we note that the twist defect CFT is stable against perturbations from the generalized free field sector and the chiral current sector. The lowest-lying primary operator that preserves both the defect Lorentz symmetry and the transverse rotation symmetry is $\hat{J}_{\bar{z}}\hat{J}_{\bar{z}}\hat{V}^{+1/4}_z\hat{V}^{-1/4}_z$, which has scaling dimension $\Delta=4.5>2$ and is therefore irrelevant. Even if we relax constraints from the transverse rotation symmetry, the lowest-lying primaries $\hat{J}_{\bar{z}}\hat{V}^{+1/4}_z$ and $\hat{J}_{\bar{z}}\hat{V}^{-1/4}_z$ have $\Delta=2.25>2$, and the twist defect CFT remains stable against anitropic perturbations.

\section{Defect RG flows in the $O(N)$ models}
\label{sec_RG flows in the $O(N)$ models}

In Sections \ref{sec_conformal defects} and \ref{sec_Constraints from bulk dynamics}, we have discussed the fixed points of the defect RG flows. In particular, we show that the surface defect attached to the topological duality operator in Maxwell theory admits a universal chiral DCFT. 

Away from the fixed points, the parameters governing the defect dynamics generally evolve under scale transformations. In this section, we study the defect RG flow itself, with particular emphasis on the monotonicity theorem and defect phase diagrams. For concreteness, we discuss surface defects embedded in a free scalar field theory and in the $O(N)$ Wilson--Fisher theory \cite{Raviv-Moshe:2023yvq, Giombi:2023dqs, Trepanier:2023tvb}. These models are closely related to boundary and defect criticalities in the 3-dimensional Ising model and Heisenberg model. See, e.g., \cite{Deng:2005dh, Gliozzi:2015qsa, Toldin:2020wbn, Metlitski:2020cqy, Padayasi:2021sik, Zhou:2024dbt} for both theoretical and numerical studies of this subject.

\subsection{Bilinear defect deformation}

We begin with a Gaussian defect model, defined by adding a localized mass term to a free scalar field $\phi \in \mathbb{R}$ in $d$-dimensional Euclidean spacetime $\mathbb{R}^d$. We consider the bilinear mass deformation $\phi^2$ on a 2-dimensional plane $\mathbb{R}^2 \subset \mathbb{R}^d$, such that the action reads
\begin{equation}
\label{eq_exactly solvable}
S_\text{free}=\frac{1}{2}\int_{\mathbb{R}^{d}}d^{d}x\,(\partial_\mu \phi)^2+\frac{\gamma_{\text{b}}}{2}\int_{\mathbb{R}^2}d^2x^{\para}\,\phi^2~,
\end{equation}
where $\gamma_\text{b}$ is the bare defect coupling constant. The free theory Hamiltonian is unbounded from below when $\gamma_\text{b}<0$ \footnote{The defect RG flows associated with $\gamma_\text{b}<0$ are conjectured to exhibit runaway behavior \cite{Cuomo:2021rkm, Cuomo:2022xgw, Aharony:2022ntz, Raviv-Moshe:2023yvq}.}, and we therefore restrict to unitary defects with $\gamma_\text{b}\geq 0$. A simple power-counting also shows that the defect deformation is irrelevant when $d>4$, classically marginal when $d=4$, and relevant when $d<4$. In this and the following subsections, we will also denote $d=4-\epsilon$ with $\epsilon\geq 0$.

The defect RG flow of the model \eqref{eq_exactly solvable} is exactly solvable. Following the Wilsonian analysis \cite{Wilson:1971bg,Wilson:1971dh,Wilson:1973jj}, we introduce a UV momentum cutoff $\Lambda$ to the model \eqref{eq_exactly solvable}. In terms of Fourier modes, the scalar field $\phi$ can be written as 
\begin{equation}
    \phi(x)=\int_{k^2\leq \Lambda^2} \frac{d^dk}{(2\pi)^d} e^{ikx}\phi_{k}~.
\end{equation}
Correspondingly, the Gaussian action \eqref{eq_exactly solvable} takes the form
\begin{equation}
\label{eq_free theory fourier modes}
S_\text{free}(\Lambda)=\frac{1}{2}\int_{k_1^2\leq \Lambda^2} \frac{d^dk_1}{(2\pi)^d}\int_{k_2^2\leq \Lambda^2} \frac{d^dk_2}{(2\pi)^d}\big(k_1^2\delta^{d-2}({k_1^\perp+k_2^\perp})+\gamma_{\text{b}}\big) \delta^2(k_1^{\para}+k_2^{\para})\phi_{k_1}\phi_{k_2}.
\end{equation}
We consider the coarse-graining procedure by lowering the cutoff from $\Lambda$ to $\tilde{\Lambda}$ and integrating the Fourier modes between $\Lambda$ and $\tilde{\Lambda}$. Clearly, the UV modes with $\tilde{\Lambda}\leq |k| \leq \Lambda$ couple the IR modes with $|k|\leq \tilde \Lambda $ due to the explicit breaking of translation symmetry along the $x^\perp_\sigma$-directions. This coarse-graining procedure gives the following correction term:
\begin{equation}
    S_\text{free}(\tilde\Lambda)+\updelta S_\text{free}(\tilde\Lambda,\Lambda)=-\ln{\Big(\int \left. (d\phi _{k})\right|_{\tilde{\Lambda}\leq |k|\leq \Lambda}\, e^{-S_\text{free}(\Lambda)}\Big)}~.
\end{equation}
A straightforward linear algebra computation yields
\begin{equation}
\label{eq_Wilsonian term}
\updelta S_\text{free}(\tilde\Lambda,\Lambda)=-\frac{\gamma_{\text{b}}}{2}\int_{k_1^2\leq \tilde\Lambda^2} \frac{d^dk_1}{(2\pi)^d}\int_{k_2^2\leq \tilde\Lambda^2} \frac{d^dk_2}{(2\pi)^d}\frac{\gamma_{\text{b}}W(k_1^{\para};\tilde\Lambda,\Lambda)}{1+\gamma_{\text{b}}W(k_1^{\para};\tilde\Lambda,\Lambda)}\delta^2(k_1^{\para}+k_2^{\para})\phi_{k_1}\phi_{k_2}~,
\end{equation}
where we have introduced the shorthand
\begin{equation}
\begin{aligned}
W(k^{\para};\tilde\Lambda,\Lambda)\equiv{}&\int_{\tilde\Lambda^2-(k^{\para})^2\leq (k^{\perp})^2\leq \Lambda^2-(k^{\para})^2}\frac{d^{2-\epsilon}k^{\perp}}{(2\pi)^{2-\epsilon}}\frac{1}{(k^{\para})^2+(k^{\perp})^2}\\
={}&\frac{(4\pi)^{\frac{\epsilon}{2}} }{ \pi\epsilon ^2 \Gamma (-\frac{\epsilon }{2})}\big(\Lambda^{-\epsilon
   }-\tilde\Lambda^{-\epsilon
   }\big)+O\big((k^{\para})^2\big)~.    
\end{aligned}
\end{equation}

Next, we define the renormalized defect coupling $\gamma$ and the coordinate $\uptau$ along the RG flow as follows
\begin{equation}
\gamma\equiv{\Lambda}^{-\epsilon}\gamma_{\text{b}}  ~,~~\text{and}~~\uptau \equiv-\ln(\tilde{\Lambda}/\Lambda)\geq 0~.
\end{equation}
Using the Wilsonian term \eqref{eq_Wilsonian term}, we find the exact RG running of the dimensionless defect coupling $\gamma$, which reads
\begin{equation}
\label{eq_exact rg}
\begin{aligned}
\frac{1}{\gamma(\uptau )}=\frac{(4\pi)^{\frac{\epsilon}{2}}(1-e^{-\epsilon \uptau})}{2\pi \Gamma(1-\frac{\epsilon}{2})}+\frac{e^{-\epsilon \uptau}}{\gamma(0)}~.
    \end{aligned}
\end{equation}
Let us first consider two special cases of \eqref{eq_exact rg}:
\begin{equation}
    \begin{aligned}
        \gamma(\uptau)={}&\frac{2\pi \gamma(0) }{2\pi +\uptau \gamma(0)}~,~~&&\text{for}~~d=4~;\\
        \gamma(\uptau)={}&e^{2\uptau}\gamma(0)~,~~&&\text{for}~~d=2~.
    \end{aligned}
\end{equation}
In both cases, the defect RG flow admits only the trivial fixed point at $\gamma=0$. When $d=4$, one-loop effects render the bilinear defect deformation in \eqref{eq_exactly solvable} marginally irrelevant. The surface defect fills the entire spacetime when $d=2$, in which case \eqref{eq_exact rg} simply describes the trivial scaling of the free field mass term.

When $2<d<4$, the defect RG flow \eqref{eq_exact rg} admits a non-trivial IR-stable fixed point at
\begin{equation}
\label{eq_fixed point}
\gamma_{\text{fixed}}=\frac{2\pi \epsilon}{(4\pi)^{\frac{\epsilon}{2}}}\Gamma(1-\frac{\epsilon}{2})~.
\end{equation}
In particular, we find that $\gamma_{\text{fixed}}=\pi$ when $d=3$. To see the physical meaning of this fixed point, we examine the propagator derived from \eqref{eq_free theory fourier modes}:
\begin{equation}
    \left.\langle \phi_{k_1}\phi_{k_2}\rangle\right|_{\gamma_\text{b}=\pi \Lambda}=\frac{1}{(k_1)^2}\Big[\delta(k_1^{\perp}+k_2^{\perp})-\frac{2|k_2^{\para}|}{(k_2)^2}+O(\Lambda^{-3})\Big]\delta^2(k_1^{\para}+k_2^{\para})~,
\end{equation}
where the subleading terms are suppressed by the cutoff $\Lambda$ and are scheme-dependent. It is useful to evaluate the layer susceptibility \cite{Shpot:2019iwk,Dey:2020lwp}, given by
\begin{equation}
\begin{aligned}
\chi(x_1^{\perp},x_2^{\perp})\equiv {}&\lim_{|k^{\para}|\to 0}\int \frac{d k_1^{\perp}}{2\pi} \int\frac{d k_2^{\perp}}{2\pi} \frac{e^{i (k_1^{\perp} x_1^{\perp}+k_2^{\perp} x_2^{\perp})}}{(k^{\para})^2+(k_1^{\perp})^2}\Big[\delta(k_1^{\perp}+k_2^{\perp})-\frac{2|k^{\para}|}{(k^{\para})^2+(k_2^{\perp})^2}\Big]\\
={}&\frac{1}{2}(|x^\perp_1|+|x^\perp_2|-|x^\perp_1-x^\perp_2|). 
    \end{aligned}
\end{equation}
The susceptibility vanishes when $x^\perp_1$ and $x^\perp_2$ are on different sides of the defect plane, 
so that the defect effectively splits $\mathbb{R}^3$ into two decoupled half-spaces in the IR. We conclude that, when $d=3$, the fixed point \eqref{eq_fixed point} corresponds to two Dirichlet boundaries of the free field. Another perspective follows from the defect central charge, as we discuss in Section \ref{subsec_defect central charge}. 
\begin{figure}[!h]
\centering
\includegraphics[width=0.6\textwidth ]{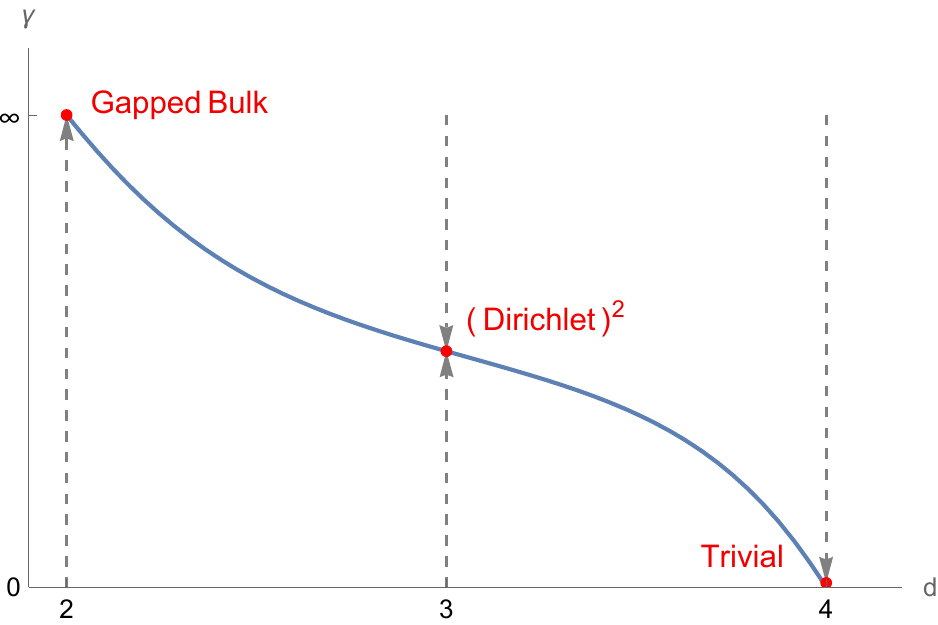}
  \caption[IR-stable fixed points in the free theory.]{\label{pic_Free FP}IR-stable fixed points associated with bilinear defect deformations in the free theory. The blue curve shows how the fixed point depends on the spacetime dimension $d$, while the grey dashed arrow indicates the defect RG flow from the UV to the IR. Fixed points with physical interpretations in integer dimensions are marked in red. }
\end{figure}

The fixed point \eqref{eq_fixed point} interpolates between the trivial defect at $d=4$, the Dirichlet boundaries at $d=3$, and the gapped bulk theory at $d=2$. See Figure \ref{pic_Free FP}. The DCFT at this fixed point is governed by the conformal algebra $\mathfrak{so}(3,1)\times \mathfrak{so}(d-2)$. As a technical side note, the spin-$s$ representation of $\mathfrak{so}(q)$ is of dimension
\begin{equation}
\label{eq_harmonics_degeneracy}
\text{deg}(q,s)\equiv (2s+q-2)\frac{\Gamma (s+q -3)}{\Gamma (s+1)\Gamma (q-1)}~,
\end{equation}
which is non-integer for $s>0$ and generic $q$. Nevertheless, the spin-0 representation is unambiguous, as $\text{deg}(q,0)=1$. We denote by $\hat{\phi}$ the defect primary with $s=0$ and $\ell=0$ that appears in the bulk-to-defect OPE of the scalar field $\phi$. At the trivial fixed point $\gamma=0$, the defect operator $\hat{\phi}$ coincides with the bulk operator $\phi$. Once the bilinear defect deformation is turned on, $\hat{\phi}$ deviates from $\phi$ and becomes distinct. It follows from the RG equation \eqref{eq_exact rg} that the $\hat{\phi}$ operator at the fixed point $\gamma=\gamma_{\text{fixed}}$ has scaling dimension
\begin{equation}
\Delta(\hat{\phi})=3-\frac{d}{2}~,
\end{equation}
while the bulk free scalar field has $\Delta(\phi)=\frac{d}{2}-1$.

The same conclusion can also be reached by studying a free scalar field conformally coupled to the curved spacetime $\text{AdS}_3\times S^{d-3}$, whose isometry is precisely $\mathfrak{so}(3,1)\times \mathfrak{so}(d-2)$. We focus on the Kaluza-Klein mode that is independent of the coordinates on $S^{d-3}$, which transforms in the spin-0 representation of $\mathfrak{so}(d-2)$. In this setup, the two defect RG fixed points $\gamma=0$ and $\gamma=\gamma_{\text{fixed}}$ correspond to the different conformal boundary conditions of the scalar field on $\text{AdS}_3$, namely the alternative and standard quantizations \cite{Burges:1985qq, DHoker:2002nbb, Giombi:2021uae}.

\subsection{Monotonicity and the defect central charge}
\label{subsec_defect central charge}

The monotonicity of the surface defect RG flow is associated with the defect central charge, also known as the $b$-coefficient \cite{Jensen:2015swa, Shachar:2022fqk}. To facilitate the definition, we consider a spherical surface defect supported on $S^2 \subset \mathbb{R}^d$ with radius $R$. At RG fixed points, the defect contribution to the free energy is given by
\begin{equation}
\label{eq_defect_partition_func_contr}
\mathcal{F}\equiv-\ln \frac{Z_{\text{DCFT}}}{Z_{\text{CFT}}} ~,
\end{equation}
where $Z_{\text{CFT}}$ is the partition function of the ambinet CFT on $\mathbb{R}^d$, and $Z_{\text{DCFT}}$ is the corresponding partition function in the presence of the spherical defect.

Next, we consider the large-size limit of the spherical defect. By locality, the leading terms of the free energy $\mathcal{F}$ in this limit take the form
\begin{equation}
\label{eq_b-def}
\mathcal{F}= a^{(-2)}({\Lambda}' R)^2+a^{(0)}+ \frac{b}{3}\log ({\Lambda}' R)+O\big(({\Lambda}' R)^{-2}\big)~, 
\end{equation}
where ${\Lambda}'$ s the UV cutoff scale. The scale ${\Lambda}'$ is clearly scheme-dependent, and so are the coefficients $a^{(-2)}$ and $a^{(0)}$ appearing in the expansion \eqref{eq_b-def}. By contrast, the coefficient $b$ extracted from the logarithmic divergence of \eqref{eq_b-def} is scheme-independent and corresponds to the defect Weyl anomaly.

The monotonicity theorem \cite{Jensen:2015swa, Shachar:2022fqk} for surface defects states  that if a UV DCFT and an IR DCFT are connected by a defect RG flow, then their defect central charges satisfy
\begin{equation}
\label{eq_b-theorem}
    b_{\text{UV}}\geq b_{\text{IR}}~.
\end{equation}
Conversely, if $b_{\text{UV}}< b_{\text{IR}}$, then the UV DCFT cannot admit a deformation that takes it back to the IR DCFT. This establishes the irreversibility of defect RG flows. 

Let us now examine the theorem \eqref{eq_b-theorem} for the defect RG fixed point \eqref{eq_fixed point}. Since the spherical surface defect preserves the rotational symmetry $\mathfrak{so}(3)$, the problem reduces to finding the Laplacian eigenvalues associated with spin-$l$ spherical harmonics. We find 
\begin{equation}
\label{eq_F-function}
\begin{aligned}
\mathcal{F}=\frac{1}{2}\sum_{l\in \mathbb{N}}(2l+1)\ln{\left(1+ \frac{\pi \epsilon(\Lambda R)^{\epsilon}}{2\sin(\pi \epsilon/2)} \frac{ \Gamma \left(l-\frac{\epsilon }{2}+1\right)}{\Gamma \left(l+\frac{\epsilon
   }{2}+1\right)} \right)}~.
    \end{aligned}
\end{equation}
To extract the defect central charge from \eqref{eq_F-function}, we apply the dimensional regularization to the defect dimension $p$ and then take the limit $ R\to +\infty$. In this limit, \eqref{eq_F-function} can be simplified as 
\begin{equation}
\label{eq_F-function IR limit}
\begin{aligned}
\mathcal{F}=-\frac{1}{2}\lim_{p\to 2}\,\sum_{l\in \mathbb{N}}\text{deg}(p+1,l)\ln{\frac{ \Gamma \left(l+\frac{p}{2} +\frac{\epsilon }{2}\right)}{ \Gamma \left(l+\frac{p}{2}-\frac{\epsilon }{2}\right)}} ~.
    \end{aligned}
\end{equation}
We identify the $1/(p-2)$-pole term in \eqref{eq_F-function IR limit} with the logarithmic IR-divergence in the cutoff regularization scheme \cite{Graham:1999jg, Diaz:2007an}. For the free scalar field theory in $d=4-\epsilon$ dimensions, we obtain
\begin{equation}\label{eq_defect_central_charge_free}
b_{\text{IR}}=-\frac{\epsilon^3}{8}~.
\end{equation}

For the trivial fixed point $\gamma=0$, it follows directly from the definition \eqref{eq_defect_partition_func_contr} that $b_{\text{UV}}=0$. The first consistency check for \eqref{eq_defect_central_charge_free} is $b_{\text{IR}}<b_{\text{UV}}=0$ when $d<4$, hence the inequality \eqref{eq_b-theorem} is indeed satisfied. We also observe that
\begin{equation}
    \left. b_{\text{IR}}\right|_{d=3}=2b_{\text{D}}=-\frac{1}{8}~,
\end{equation}
so that the defect central charge agrees with that of two free field Dirichlet boundaries \cite{Jensen:2015swa}. Finally, when d=2, one can no longer perform a Weyl transformation on the defect independently, since the defect is then indistinguishable from the ambient theory. It is well known that a two-dimensional free scalar has central charge $c=1$. Therefore, the physical Weyl anomaly coefficient is $b_{\text{IR}}+c=0$. This is consistent with \eqref{eq_exact rg} and the phase diagram in Figure \ref{pic_Free FP}, according to which the theory flows to a trivially gapped phase.

\subsection{Phases of defects in interacting theories} 
\label{subsec_Phases of defects in interacting theories}

In this section, we study bilinear defect deformations in the $O(N)$ model at the Wilson--Fisher fixed point \cite{Wilson:1973jj}. This bulk theory consists of $N$ scalar fields $\phi_i$, with $1\leq i \leq N$, interacting through quartic couplings in Euclidean spacetime $\mathbb{R}^d$. We denote the dimensionless, fully symmetric coupling tensor by $\lambda_{i_1i_2i_3i_4}$, so that the bulk action takes the form
\begin{equation}
\label{bulk model}
\begin{aligned}
S_{\text{WF}}=\int_{\mathbb{R}^d}d^dx\left(\frac{1}{2}(\partial \phi^i)^2+\frac{\Lambda^{4-d}}{4!}\lambda_{i_1i_2i_3i_4}\phi^{i_1}\phi^{i_2}\phi^{i_3}\phi^{i_4}\right)~,
    \end{aligned}
\end{equation}
where $\Lambda$ is the UV cutoff scale. This theory admits perturbative interacting fixed points when $\epsilon=4-d\ll 1$. At the one-loop level, the $O(N)$ symmetric fixed point is given by
\begin{equation}
\label{eq_phi 4 interaction}
(\lambda_{\text{fixed}})_{i_1i_2i_3i_4}= \frac{16\pi^2\epsilon}{N+8}(\delta_{i_1i_2}\delta_{i_3i_4}+\delta_{i_1i_3}\delta_{i_2i_4}+\delta_{i_1i_4}\delta_{i_2i_3})+O\left(\epsilon^2\right)~,
\end{equation}
which we adopt in the remainder of this section.

There are two types of bilinear operators in this weakly-interacting CFT, distinguished by their transformation properties under $O(N)$. We introduce the singlet operator \cite{Kehrein:1995ia,Henriksson:2022rnm}:
\begin{equation}
\label{eq_ON singlet operator}
    O\equiv \frac{1}{2N} (\phi_i)^2~,~~\Delta(O)=2-\frac{6}{N+8}\epsilon+O\left (\epsilon^2\right)~,
\end{equation}
and the symmetric traceless operator
\begin{equation}
    \Upsilon_{i_1i_2}\equiv \phi_{i_1}\phi_{i_2}-\frac{\delta_{i_1i_2}}{N}(\phi_i)^2~,~~\Delta(\Upsilon)=2-\frac{N+6}{N+8}\epsilon+O\left (\epsilon^2\right)~.
\end{equation}
Both operators are even under the $\mathbb{Z}_2 \subset O(N):~\phi_i\to -\phi_i$. In analogy with the defect model \eqref{eq_exactly solvable}, we consider the bilinear deformations localized on a 2-dimensional plane: 
\begin{equation}\label{eq_action_with_def}
S_{\text{dWF}}=S_{\text{WF}}+\int_{\mathbb{R}^2}d^2x^{\para} \left( \Lambda^{2-\Delta(O)} \gamma^O O+\Lambda^{2-\Delta(\Upsilon)}\gamma^{\Upsilon}_{i_1i_2}\Upsilon^{i_1i_2}\right)~,
\end{equation}
where $\gamma^O$ and $\gamma^\Upsilon$ denotes the dimensionless defect coupling constants. The defect RG flow of the model \eqref{eq_action_with_def} can be analyzed perturbatively in $\epsilon=4-d\ll 1$, assuming that both $\gamma^O$ and $\gamma^\Upsilon$ are of order $O(\epsilon)$.

In what follows, we compute the beta functions for $\gamma^O$ and $\gamma^\Upsilon$ and determine the associated defect RG fixed points at one-loop order. Using the conformal perturbation theory \cite{Komargodski:2016auf}, we find the bulk one-point function of the singlet operator as follows
\begin{equation}
\mathtoolsset{multlined-width=0.9\displaywidth}
\begin{multlined}
\langle O(x^{\perp},0)\rangle=-\Lambda^{\frac{6\epsilon}{N+8}}\gamma^O\int d^2x^{\para} \langle O(x^{\perp},0) O(0,x^{\para})\rangle \hfill\\
\hfill+\frac{1}{2}\Lambda^{2\frac{N+6}{N+8}\epsilon}\gamma^{\Upsilon}_{i_1i_2}\gamma^{\Upsilon}_{i_3i_4}\int d^2z_1d^2z_2 \langle O(x^{\perp},0) \Upsilon^{i_1i_2}(0,x^{\para}_1)\Upsilon^{i_3i_4}(0,x^{\para}_2)\rangle\\
\hfill + \frac{1}{2}\Lambda^{\frac{12\epsilon}{N+8}}(\gamma^O)^2\int d^2x_1^{\para}d^2x_2^{\para} \langle O(x^{\perp},0) O(0,x^{\para}_1)O(0,x^{\para}_2)\rangle+O\left (\epsilon^3\right)\\
\phantom{\langle O(x^{\perp},0)\rangle}=-\frac{|x^\perp|^{-\Delta(O)}}{32 \pi ^3 N }\left( \gamma^O -\frac{N+8}{12 \pi  N \epsilon }(\gamma^O)^2-\frac{(N+8)}{\pi  (N+3) \epsilon }\text{Tr}\big((\gamma^\Upsilon)^2\big)+O\left(\epsilon^2\right)\right).\hfill \\
\end{multlined}
\end{equation}
It then follows from the minimal subtraction scheme that the beta function for $\gamma^O$ reads
\begin{equation}
\label{eq_scalar_beta}
\begin{aligned}
-\beta(\gamma^O)=\frac{6\epsilon}{N+8}\gamma^O-\frac{1}{2\pi N}(\gamma^O)^2-\frac{2}{\pi}\text{Tr}\big((\gamma^\Upsilon)^2\big)+O\left(\epsilon^3\right)\,. 
    \end{aligned}
\end{equation}
Similarly, we consider the bulk one-point function of the symmetric traceless operator
\begin{equation}
\mathtoolsset{multlined-width=0.9\displaywidth}
\begin{multlined}
\langle \Upsilon_{i_1i_2}(x^{\perp},0)\rangle=-\Lambda^{\frac{N+6}{N+8}\epsilon}\gamma^\Upsilon_{i_3i_4}\int d^2z \langle \Upsilon_{i_1i_2}(x^{\perp},0)\Upsilon^{i_3i_4}(0,x^{\para})\rangle \hfill\\
\hfill +\frac{1}{2}\Lambda^{2\frac{N+6}{N+8}\epsilon}\gamma^\Upsilon_{i_3i_4}\gamma^\Upsilon_{i_5i_6}\int d^2x^{\para}_1d^2x^{\para}_2 \langle \Upsilon_{i_1i_2}(x^{\perp},0)\Upsilon^{i_3i_4}(0,x^{\para}_1)\Upsilon^{i_5i_6}(0,x^{\para}_2)\rangle\\
\hfill+ \Lambda^{\frac{N+12}{N+8}\epsilon}\gamma^O \gamma^\Upsilon_{i_3i_4}\int d^2x^{\para}_1d^2x^{\para}_2 \langle \Upsilon_{i_1i_2}(x^{\perp},0)O(0,x^{\para}_1)\Upsilon^{i_3i_4}(0,x^{\para}_2)\rangle+O\left (\epsilon^3\right)\\
\phantom{\langle \Upsilon_{i_1i_2}(x^{\perp},0)\rangle}=-\frac{|x^\perp|^{-\Delta(\Upsilon)}}{8 \pi ^3} \left(\gamma^\Upsilon_{i_1i_2}-\frac{N+8}{6 \pi  N \epsilon } \gamma^\Upsilon_{i_1i_2}\gamma^O-\frac{N+8}{\pi (N+6) \epsilon }((\gamma^\Upsilon)^2)_{i_1i_2}\right.\hfill\\
\hfill\left.+\frac{N+8}{\pi N (N+6) \epsilon }\delta_{i_1i_2}\text{Tr}\big((\gamma^\Upsilon)^2\big)\right)~. \\
\end{multlined}
\end{equation}
With the minimal subtraction scheme, we find
\begin{equation}
\label{eq_stt_beta}
\begin{aligned}
-\beta(\gamma^{\Upsilon}_{i_1i_2})=\frac{N+6}{N+8}\epsilon \gamma^{\Upsilon}_{i_1i_2}-\frac{1}{\pi N}\gamma^{\Upsilon}_{i_1i_2}\gamma^O-\frac{1}{\pi  }\big((\gamma^{\Upsilon})^2\big)_{i_1i_2}+\frac{\delta_{i_1i_2}}{\pi N}\text{Tr}\big((\gamma^\Upsilon)^2\big)+O\left(\epsilon^3\right).
    \end{aligned}
\end{equation}

The beta functions \eqref{eq_scalar_beta} and \eqref{eq_stt_beta} govern the defect RG flows near the trivial fixed point $\gamma^O=\gamma^{\Upsilon}_{i_1i_2}=0$. For every $N$, there is also a nontrivial $O(N)$-symmetric fixed point:
\begin{equation}
\label{eq_O(N) FP}
    (\gamma^O)_{\text{fixed}}\equiv\frac{12\pi N}{N+8}\epsilon+O\left(\epsilon^2\right)~;~~(\gamma^{\Upsilon}_{i_1i_2})_{\text{fixed}}=0~.
\end{equation}
At this fixed point, the lowest-lying defect primaries with $s=0$ and $\ell=0$ include the $O(N)$ vector $\hat{\phi}_i$, the singlet $\hat{O}$, and the symmetric traceless tensor $\hat{\Upsilon}_{i_1i_2}$. Their scaling dimensions are given by \cite{Raviv-Moshe:2023yvq}:
\begin{equation}
    \begin{aligned}
        \Delta(\hat{\phi})={}&1-\frac{N-4  }{2N+16}\epsilon+O\left(\epsilon^2\right)~,\\
        \Delta(\hat{O})={}&2+\frac{6}{N+8}\epsilon+O\left(\epsilon^2\right)~,\\
        \Delta(\hat{\Upsilon})={}&2-\frac{N-6 }{N+8}\epsilon+O\left(\epsilon^2\right).
    \end{aligned}
\end{equation}
The defect RG fixed point \eqref{eq_O(N) FP} is clearly stable against the $O(N)$ symmetric deformations. By contrast, the deformations triggered by the $\hat{\Upsilon}_{i_1i_2}$ operators break $O(N)$ down to subgroups containing the $\mathbb{Z}_2$ symmetry. These deformations are relevant for $N>6$ and irrelevant for $N<6$. When $N=6$, the relevance of these deformations depends on the subgroup they preserve, as we explain below.

We consider the bilinear defect deformations that preserve the $O(M)\times O(N-M)$ subgroup of the $O(N)$ symmetry. In particular, we take $\gamma^\Upsilon_{i_1i_2}\Upsilon^{i_1i_2}=\gamma_M^{\Upsilon}\Upsilon_M$, where the projection onto the symmetric traceless tensor, up to field redefinitions, is given by
\begin{equation}
     \Upsilon_M\equiv -\frac{1}{M}\sum_{1\leq i\leq M}(\phi_i)^2+\frac{1}{N-M}\sum_{M<i\leq N}(\phi_i)^2~.
\end{equation}
Since $\Upsilon_M$ and $\Upsilon_{N-M}$ preserve the same subgroup symmetry, the analyses for $\gamma^\Upsilon_M\geq 0$ proceed in the same way as that of $\gamma^\Upsilon_{N-M}\leq 0$. To avoid overcounting of equivalent cases, we adopt the convention $\gamma_M^{\Upsilon}\geq 0$. 

\begin{figure}[!h]
\centering
  \includegraphics[width=\textwidth ]{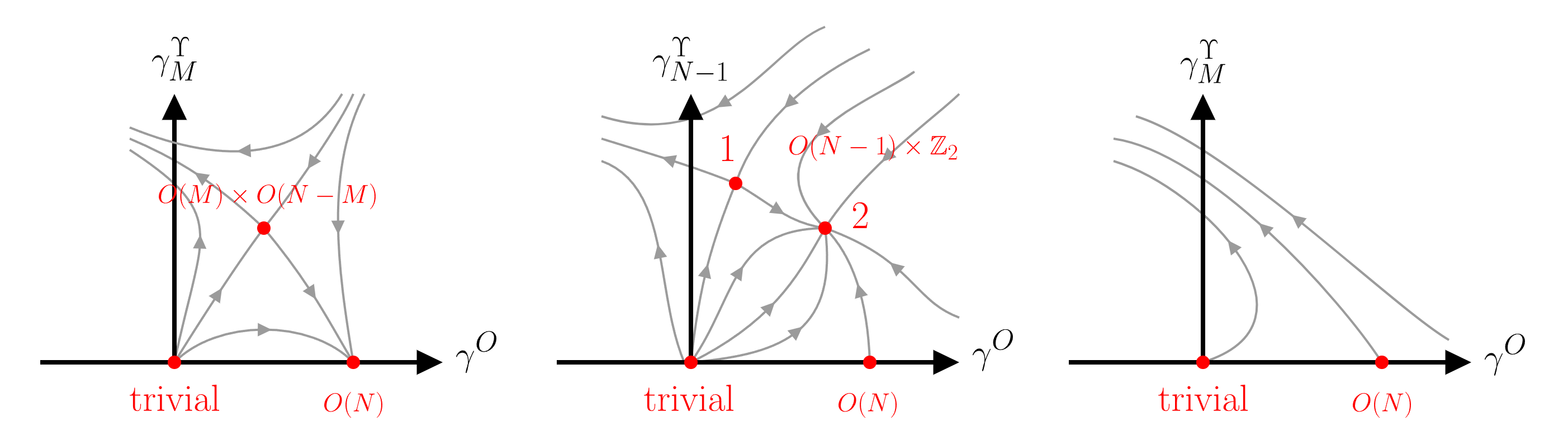}
  \caption[Perturbative phase diagram for surface defects in the Wilson--Fisher CFT.]{\label{pic_Phase Diagram}Perturbative phase diagram for $O(M)\times O(N-M)$ surface defects in the Wilson--Fisher CFT. Left: cases listed in table \ref{tab_N<=6 defect fixed points}. Middle: cases listed in table \ref{tab_N>6 defect fixed points}. Right: cases of no fixed point other than the trivial one and the $O(N)$-symmetric one. }
\end{figure}

The beta function for the defect coupling constant $\gamma^\Upsilon_M$ follows directly from \eqref{eq_stt_beta}:
\begin{equation}
\label{eq_g_M_beta}
-\beta(\gamma^\Upsilon_M)=\frac{N+6}{N+8} \epsilon \gamma^\Upsilon_M-\frac{1}{\pi  N}\gamma^\Upsilon_M\gamma^O-\frac{2M-N}{\pi M(N-M)}(\gamma^\Upsilon_M)^2+O\left(\epsilon^3\right).
\end{equation}
Using \eqref{eq_scalar_beta} and \eqref{eq_g_M_beta}, we identify new defect RG fixed points that preserve the $O(M)\times O(N-M)$ subgroup. We find that when the $\hat{\Upsilon}_{M}$ operators are irrelevant at \eqref{eq_O(N) FP}, they connect the $O(N)$-symmetric stable fixed point to metastable fixed points. When $N<6$, each choice of $1\leq M\leq N-1$ admits a single perturbative fixed point with $(\gamma_M^{\Upsilon})_{\text{fixed}}>0$. When $N=6$, we find the beta function at the $O(N)$-symmetric fixed point \eqref{eq_O(N) FP} as follows
\begin{equation}
    \left. -\beta(\gamma^\Upsilon_M) \right|_{\gamma^O=(\gamma^O)_{\text{fixed}},~N=6}=\frac{2(M-3)}{\pi M(M-6)} (\gamma^\Upsilon_M)^2+O\left(\epsilon^3\right)~.
\end{equation}
We conclude that, for $N=6$, the defect bilinear operators $\hat{\Upsilon}_{M=1,2}$ are marginally relevant, whereas $\hat{\Upsilon}_{M=4,5}$ are marginally irrelevant, and $\hat{\Upsilon}_{M=3}$ remains marginal at the one-loop level.  See Table \ref{tab_N<=6 defect fixed points} for the fixed point solutions and the left panel in Figure \ref{pic_Phase Diagram} for schematic defect RG flows.
{\renewcommand{\arraystretch}{1.5}
\begin{table}[!h]
\begin{center}
\begin{tabular}{ |c|c|c|c|c|c| } 
 \hline
& $N=2$ & $N=3$ & $N=4$ & $N=5$ & $N=6$\\ 

  \hline
$M=1$ & $\left(\frac{\sqrt{2}}{5},\frac{8}{5}\right)$ & $\left( \frac{4 \sqrt{7}-2}{33},\frac{26+2 \sqrt{7}}{11} \right)$ & $\left( \frac{\sqrt{6}-1}{8}, \frac{9+\sqrt{6}}{3}\right)$ & $\left(\frac{8 \sqrt{5}-12}{65},\frac{46 + 6 \sqrt{5}}{13} \right) $ & no fixed point \\ 

  \hline
$ M=2$ &  & $\left(\frac{4 \sqrt{7}+2}{33},\frac{26-2 \sqrt{7}}{11} \right)$ & $\left(\frac{\sqrt{5}}{6},\frac{10}{3}\right)$ & $\left(\frac{12 \sqrt{3}-6}{65},\frac{54 + 2 \sqrt{3}}{13} \right)$ & no fixed point\\ 
\hline

$M=3$ &  &  & $\left( \frac{\sqrt{6}+1}{8}, \frac{9-\sqrt{6}}{3}\right)$ & $\left(\frac{12 \sqrt{3}+6}{65},\frac{54 - 2 \sqrt{3}}{13} \right)$ & undetermined \\ 
\hline
$M=4$ &  &  & & $\left(\frac{8 \sqrt{5}+12}{65},\frac{46 - 6 \sqrt{5}}{13} \right)$ & $\left(\frac{8}{21},\frac{32}{7} \right)$  \\ 
\hline
$M=5$ &  &  & &  & $\left(\frac{10}{21},\frac{20}{7} \right)$ \\ 
\hline
\end{tabular}
\caption[$O(M)\times O(N-M)$-symmetric defect RG fixed points for $2\leq N\leq 6$.]{\label{tab_N<=6 defect fixed points}
$O(M)\times O(N-M)$-symmetric defect RG fixed points for $2\leq N\leq 6$. This table presents $((\gamma^{\Upsilon}_M)_{\text{fixed}},(\gamma^O)_{\text{fixed}})$ in units of $\pi \epsilon$, with terms of order $O(\epsilon^2)$ omitted. For $N=6$ and $M=3$, the existence of the fixed point needs to be examined at higher loop orders.}
\end{center}
\end{table}}

When $N>6$, most values of $N$ and $M$ admit no perturbative fixed point other than the trivial fixed point and the $O(N)$-symmetric fixed point \eqref{eq_O(N) FP}. See also the right panel of Figure \ref{pic_Phase Diagram}. The exceptions arise for $M=N-1$ and $7\leq N\leq 9$, in which case the beta function admits two $O(N-1)\times \mathbb{Z}_2$-symmetric fixed points. For convenience, we denote the metastable one by fixed point 1 and the stable one by fixed point 2. When $N=10$, fixed point 1 and fixed point 2 collide at the one-loop order, and higher-loop calculations are needed to determine the existence of the $O(N-1)\times \mathbb{Z}_2$ fixed points. See Table \ref{tab_N>6 defect fixed points} for these fixed-point solutions and the middle panel of Figure \ref{pic_Phase Diagram} for the corresponding defect RG flows.
{\renewcommand{\arraystretch}{1.5}
\begin{table}[!h]
\begin{center}
\begin{tabular}{ |c|c|c|c|c|c| } 
 \hline
 & $N=7$ & $N=8$ & $N=9$ & $N=10$ \\ 

  \hline
fixed point 1 & $\left(\frac{10+4 \sqrt{3}}{35}, \frac{66-10 \sqrt{3}}{15} \right)$ & $\left(\frac{21+7 \sqrt{2}}{64}, \frac{19-3 \sqrt{2}}{4} \right)$ & $\left( \frac{8}{17}, \frac{72}{17}\right)$ & undetermined  \\ 

  \hline
fixed point 2 & $\left(\frac{10-4 \sqrt{3}}{35}, \frac{66+10 \sqrt{3}}{15} \right)$ & $\left(\frac{21-7 \sqrt{2}}{64}, \frac{19+3 \sqrt{2}}{4} \right)$ & $\left( \frac{40}{153}, \frac{100}{17}\right)$ & undetermined \\ 
\hline
\end{tabular}
\caption[$O(N-1)\times \mathbb{Z}_2$-symmetric defect RG fixed points for $7\leq N\leq 10$.]{\label{tab_N>6 defect fixed points}
$O(N-1)\times \mathbb{Z}_2$-symmetric defect RG fixed points for $7\leq N\leq 10$. This table presents $((\gamma^{\Upsilon}_M)_{\text{fixed}},(\gamma^O)_{\text{fixed}})$ in units of $\pi \epsilon$, with terms of order $O(\epsilon^2)$ omitted. For $N=10$ and $M=9$, the two defect RG fixed points collide at the one-loop level, and higher loop corrections are needed to
distinguish them.}
\end{center}
\end{table}}

\subsection{IR effective theory of defects}

In addition to the perturbative fixed points, the beta functions \eqref{eq_scalar_beta} and \eqref{eq_g_M_beta} also admit regimes in which both $\gamma^O$ and $\gamma^{\Upsilon}_M$ flow to large values. See also Figure \ref{pic_Phase Diagram}. What is the IR theory governing the field dynamics near the surface defect in these regimes? In this section, we follow the approach in \cite{Metlitski:2020cqy, Krishnan:2023cff, Raviv-Moshe:2023yvq} and analyze the effective theory of such defects.

Classically, the defect tends to acquire localized degrees of freedom for large negative $\gamma^O$ and large positive $\gamma^{\Upsilon}_M$ \footnote{For studies on the subject of spontaneous symmetry breaking on surface defects, see \cite{Cuomo:2023qvp}.}. We adopt the assumption that the IR effective action of the defect consists of three parts: a DCFT governing the dynamics of bulk scalar fields near the defect, the action of the effective degrees of freedom localized on the defect, and weak interactions between the DCFT and the effective degrees of freedom. In what follows, we first identify this DCFT and then examine the stability of the proposed effective action in the IR.

We begin by considering a DCFT in which the bulk order parameter $\phi_i$ develops a vacuum expectation value near the defect, thereby breaking the $O(N)$ symmetry down to the $O(N-1)$ subgroup. This is analogous to the pinning defect in the Wilson--Fisher fixed point \cite{Cuomo:2021kfm}. Specifically, we consider the bulk one-point functions:
\begin{equation}
\label{eq_ON breaking 1pt}
    \langle \phi_i(x)\rangle=\frac{\delta_{i1}C_{\phi_1}}{|x^{\perp}|^{\Delta(\phi)}}~,~~\text{where}~~\Delta(\phi)=1-\frac{\epsilon}{2}+O(\epsilon^2)~.
\end{equation}
At leading order, the OPE coefficient $C_{\phi_1}$ can be determined by a mean-field calculation. Let us consider a classical field profile $\phi^{\text{cl}}_1(x)$ obeying the equation of motion from \eqref{bulk model} as follows
\begin{equation}
\label{eq_mean field eom}-\partial^2_{\mu}\phi_1^{\text{cl}}+\Lambda^\epsilon\big(\frac{8\pi^2\epsilon}{N+8}+O(\epsilon)\big)(\phi_1^{\text{cl}})^3=0~.
\end{equation}
In addition to the trivial solution $\phi^{\text{cl}}_1(x)=0$, equation \eqref{eq_mean field eom} also admits a solution that is singular near the defect at $x^\perp=0$: 
\begin{equation}
\label{eq_classical profile}
    \phi^{\text{cl}}_1(x)=\frac{1}{\Lambda^{\frac{\epsilon}{2}}|x^\perp|}\sqrt{\frac{N+8}{8\pi^2\epsilon}}\left(1+O(\epsilon)\right)~.
\end{equation}
By matching \eqref{eq_ON breaking 1pt} with \eqref{eq_classical profile} to the leading order in $\epsilon$, we obtain
\begin{equation}
    \begin{aligned}\label{asigma}
        C_{\phi_1}=\sqrt{\frac{N+8}{8\pi^2\epsilon}}+O\left(\epsilon^{1/2}\right)~.
    \end{aligned}
\end{equation}

The symmetry-breaking DCFT associated with \eqref{eq_ON breaking 1pt} contains tilt operators $\hat{\tau}_j$ with $2\leq j\leq N$. In general, the two-point function between the bulk scalar $\phi_j$ and the tilt operator $\hat{\tau}_j$ takes the form
 \begin{equation}
    \langle \phi_{j_1}(x)\hat{\tau}_{j_2}(\tilde{x}^{\para})\rangle=\delta_{j_1j_2} \frac{C_{\phi_j}^{\hat{\tau}}C_{\hat{\tau}\hat{\tau}}}{|x^\perp|^{\Delta(\phi)}}\left(\frac{|x^\perp|}{|x^\perp|^2+|x^{\para}-\tilde{x}^{\para}|^2}\right)^2~,
\end{equation}
where $C_{\phi_j}^{\hat{\tau}}$ is the OPE coefficent and $C_{\hat{\tau}\hat{\tau}}$ is fixed by \eqref{eq_tilt 2pt}. While computing $C_{\phi_j}^{\hat{\tau}}$ and $C_{\hat{\tau}\hat{\tau}}$ separately is rather involved, we can determine the product $(C_{\phi_j}^{\hat{\tau}})^2C_{\hat{\tau}\hat{\tau}}$ at leading order in $\epsilon$ by mapping the system \eqref{bulk model} onto $\text{AdS}_{3}\times S^{1-\epsilon}$ through a Weyl transformation \cite{Giombi:2021uae, Raviv-Moshe:2023yvq}. The key observation is that the tilt operator is the lowest-lying defect primary appearing on the right-hand side of the OPE of $\phi_j$, with scaling dimension $\Delta(\hat{\tau})=2$. We consider the two-point function $\langle \phi_{j_1}\phi_{j_2}\rangle $ on $\text{AdS}_{3}\times S^{1-\epsilon}$, with the leading order result in $\epsilon$ dominated by the free field propagator. See appendix in \cite{Barkeshli:2025cjs} for a compact review of these propagators. Notably, the product $(C_{\phi_j}^{\hat{\tau}})^2C_{\hat{\tau}\hat{\tau}}$ can be extracted by taking the two points close to the $\text{AdS}_{3}$ boundary, and it is given by
\begin{equation}
(C_{\phi_j}^{\hat{\tau}})^2C_{\hat{\tau}\hat{\tau}}=\frac{1}{2\pi}+O(\epsilon)~.
\end{equation}

Next, we turn to the effective degrees of freedom localized on the defect and their couplings to the DCFT. We consider a Non-Linear Sigma Model (NL$\Sigma$M) whose target space is the $M$-dimensional unit sphere $S^M$. The kinetic action of this model is given by
\begin{equation}
\label{eq_NLSM action}
\begin{aligned}
S_{\text{NL$\Sigma$M}}=\frac{1}{2g_{\Sigma}}\sum_{i=1}^M\int_{\mathbb{R}^2}d^2x^{\para}(\partial_\sigma\Omega_i)^2,
    \end{aligned}
\end{equation}
where $\sum_{i=1}^M(\Omega_i)^2=1$ and $g_{\Sigma}> 0$ is the coupling constant. It is convenient to parametrize $S^M$ by $\Omega_j$, with $2\leq j\leq M$, so that
\begin{equation}
    \Omega_1=\big(1-\sum_{j=2}^M(\Omega_j)^2\big)^{\frac{1}{2}}~.
\end{equation}
The interactions between these $\Omega_j$ fields are governed by the NL$\Sigma$M coupling constant $g_{\Sigma}$. As implied by Coleman's theorem \cite{Coleman:1973ci}, for a standalone two-dimensional NL$\Sigma$M \eqref{eq_NLSM action}, the coupling $g_{\Sigma}$ flows to large values at long distances, and the $\Omega_j$ fields become unstable. Nevertheless, we will take $g_{\Sigma}\ll 1$ and show that the coupling between the NL$\Sigma$M and the DCFT indeed stabilizes the $\Omega_j$ fields.

As in \cite{Metlitski:2020cqy}, we take the complete IR effective action for the surface defect in the regime of large negative $\gamma^O$ and large positive $\gamma^{\Upsilon}_M$ to be
\begin{equation}
\label{eq_IR_effective}
S_{\text{IR}}=S_{\text{DCFT}}+S_{\text{NL$\Sigma$M}}-\frac{\gamma_{\Sigma}}{\sqrt{C_{\hat{\tau}\hat{\tau}}}}\sum_{j=2}^M\int_{\mathbb{R}^2}d^2 x^{\para}\  \Omega_j\hat{\tau}_j~,
\end{equation}
where $\gamma_{\Sigma}$ couples the $\Omega_j$ fields with the tilt operators in the DCFT. The coupling constant $\gamma_{\Sigma}$ is uniquely fixed by requiring that the effective action \eqref{eq_IR_effective} preserves the symmetry $O(M)\times O(N-M)$. To see that, we consider an infinitesimal rotation in the NL$\Sigma$M target space, generated by $\Omega_2\to \Omega_2+\updelta\Omega_2$. In the limit $g_\Sigma \ll 1$, we find that the $\phi_2$ scalar field acquires a vacuum expectation value under this rotation:
\begin{equation}
\label{eq_gamma coupling 1}
    \updelta \langle \phi_2(x)\rangle=\frac{\updelta\Omega_2\gamma_{\Sigma}}{\sqrt{C_{\hat{\tau}\hat{\tau}}}}\int d^2\tilde{x}^{\para} \langle \phi_{2}(x)\hat{\tau}_{2}(\tilde{x}^{\para})\rangle=\updelta\Omega_2\pi\gamma_{\Sigma}C_{\phi_j}^{\hat{\tau}}\sqrt{C_{\hat{\tau}\hat{\tau}}}|x^{\perp}|^{-\Delta(\phi)}~.
\end{equation}
For the $O(M)$ subgroup not explicitly broken, we require that the result be reproduced by rotating the vacuum expectation value of the $\phi_1$ by the same angle:
\begin{equation}
\label{eq_gamma coupling 2}
    \updelta \langle \phi_2(x)\rangle=\updelta\Omega_2\langle \phi_1(x)\rangle=\updelta\Omega_2C_{\phi_1}|x^{\perp}|^{-\Delta(\phi)}~.
\end{equation}
From \eqref{eq_gamma coupling 1} and \eqref{eq_gamma coupling 2}, we find
\begin{equation}
\gamma_{\Sigma}=\frac{C_{\phi_1}}{\pi C^{\hat{\tau}}_{\phi_{j}}\sqrt{C_{\hat{\tau}\hat{\tau}}}}=\sqrt{\frac{N+8}{4\pi^3 \epsilon}}+O\left(\epsilon^{1/2}\right).
\end{equation}
Crucially, the coupling constant $\gamma_{\Sigma}$ does not run under the defect RG and is parametrically large when $\epsilon\ll 1$.

At the one-loop level, we obtain the beta function for the defect RG flow of the NL$\Sigma$M coupling constant $g_\Sigma$ as follows \cite{Raviv-Moshe:2023yvq}
\begin{equation}
\begin{aligned}
-\beta(g_{\Sigma})={}&\left(M-2-(\pi \gamma_{\Sigma})^2\right)\frac{(g_{\Sigma})^2}{2\pi}+O\left((g_{\Sigma})^3\right)\\
={}&-\frac{N+8}{8\pi^2\epsilon}(g_{\Sigma})^2+O\left((g_{\Sigma})^3,\epsilon^0\right)~.
    \end{aligned}
\end{equation}
When $\epsilon\ll 1$, interactions from the DCFT drive $g_{\Sigma}$ to small values at long distances, thereby stabilizing the $\Omega_j$ fields. We have thus established the self-consistency of the IR effective action \eqref{eq_IR_effective}. This completes the phase diagrams shown in Figure \ref{pic_Phase Diagram}.

\section{Matching UV and IR defects}
\label{sec_Matching UV and IR defects}

So far, we have discussed defect RG flows within the framework of continuum quantum field theory. In many cases, however, the UV description of a physical system is given by a discrete lattice model. The RG flow connecting a UV lattice model to an IR field theory is often subtle. For example, one needs to carefully distinguish accidental symmetries and emanant global symmetries \cite{Lieb:1961fr, Cheng:2022sgb, Seiberg:2023cdc}.

In this section, we match the crystalline impurities in UV lattice models with defects in IR field theories \cite{Barkeshli:2025cjs}. Our analysis is motivated by two considerations. First, these crystalline impurities flow to DCFTs that, on their own, exhibit intriguing physical properties. Second, such defects serve as probes of exotic topological invariants that constrain the RG flow from the UV to the IR, as we elaborate below. 

\subsection{UV and IR symmetries}

We first discuss how UV symmetries are matched to IR symmetries. We denote the spatial and internal symmetry groups of the UV lattice model by $G_{\text{UV}}$, and those of the IR field theory by $G_{\text{IR}}$. In general, there exists a group homomorphism
\begin{align}
\label{eq_def homomorphism}
    \rho~:~~ G_{\text{UV}} \to  G_{\text{IR}}~.
\end{align}
Let us comment on a few basic properties of the homomorphism $\rho$. When $\rho$ has a non-trivial kernel, some UV internal symmetries become trivial in the IR. This can happen when massive degrees of freedom decouple from the spectrum, whose associated symmetries become invisible to the field theory at long distances. When $\rho$ has a non-trivial cokernel, the IR field theory is endowed with accidental symmetries, whose corresponding symmetry-breaking operators are irrelevant and therefore suppressed at low energies.

In particular, emanant symmetries arise when elements $g_{\text{UV}}\in G_{\text{UV}}$ and their images $\rho(g_{\text{UV}})\in G_{\text{IR}}$ generate distinct symmetry groups \cite{Cheng:2022sgb}. The symmetry generated by $\rho(g_{\text{UV}})$ is said to emanate from the group generated by $g_{\text{UV}}$; such symmetries are exact in the IR field theory, even though they may not be a subgroup of the $G_{\text{UV}}$. Such examples are common when $g_{\text{UV}}$ generates a lattice translation.

In this section, we primarily focus on cases where the UV free-fermion lattice models are defined in $(2+1)$-dimensional spacetime, and the IR field theory consists of $N$ free massless Dirac fermions\footnote{We apologize for repeatedly using $M$ and $N$ to denote integers with different physical meanings across sections.}. The symmetry group $G_{\text{UV}}$ that acts faithfully on fermionic lattice operators takes the form
\begin{equation}
\label{eq_GUV}
G_{\text{UV}} = U(1)^\text{f} \times (\mathbb{Z}^2 \rtimes \mathbb{Z}_M)~,
\end{equation}
where $\mathbb{Z}^2$ is the lattice translation group, and $\mathbb{Z}_M$ is the $M$-fold lattice rotation group. Here, $U(1)^\text{f}$ denotes a central extension of the bosonic particle-number symmetry $U(1)$ by the fermion parity $\mathbb{Z}_2^\text{f}$. We also assume that the particle-number symmetry $U(1)$ in the UV acts as a $U(1)$ symmetry in the IR, where each Dirac fermion field carries charge $+1$. If the UV model is a free-fermion lattice model, this is the only possibility. If the model includes interactions, then each IR Dirac fermion in principle could carry any odd integer charge. For simplicity, we have omitted the lattice reflection group from \eqref{eq_GUV}. The action of lattice reflection symmetries in the IR field theory involves several subtleties that lie beyond the scope of this section. See, e.g., \cite{Manjunath:2022hbm, Pace:2024tgk} for further discussion of lattice reflections.

The IR CFT of $N$ massless Dirac fermions in $(2+1)$-dimensional spacetime is endowed with the symmetry
\begin{equation}
    G_\text{IR} = \big(U(N)^\text{f} \times Spin(3,2)\big)/\mathbb{Z}_2~,
\end{equation}
where $U(N)^\text{f}$ dentoes the $U(N)$ flavor symmetry group, with the $-1$ element corresponding to the fermion parity. Note that the IR theory also may contain charge conjugation, reflection, and time-reversal symmetries, which we ignore in this discussion.

We denote the Dirac fermion fields by $\Phi_i(t,\vec{x})$, where $1\leq i\leq N$ labels the fermion species, $t$ denotes time, and $\vec{x}$ is the 2-dimensional spatial coordinate. For a given lattice model, the fermion fields describe the excitations near the Dirac cones located at momenta $\vec{k}_{i}$ in the Brillouin zone. We thus expect a UV translation $ P_{\updelta \vec{x}}$ by the lattice vector $\updelta \vec{x}$ to act on the IR fields as follows
\begin{equation}
\label{eq_lattice translation}
    \rho(P_{\updelta \vec{x}})~:~~\Psi_i(t,\vec{x}) \to  e^{i \vec{k}_{i}\cdot\vec{v}}\Psi_i(t,\vec{x}+\updelta \vec{x})~.
\end{equation}
The dislocation defect is defined by a locus in 2-dimensional space such that, upon encircling it, the fermion fields transform according to \eqref{eq_lattice translation}. The vector $\updelta \vec{x}$ is of order the lattice spacing and is negligible in the continuum limit, whereas the phase $\vec{k}_{i}\cdot \updelta \vec{x}$ remains finite and determines the long-distance physical properties of the dislocation defect.

The action of UV rotations on the IR fields is more intricate. Let us consider the rotation in the 2-dimensional lattice plane around a high symmetry point $\vec{o}$ by the angle $2\pi/M$. We denote the operator implementing this rotation by $R_{M;\vec{o}}$, with $(R_{M;\vec{o}})^M = 1$ \footnote{For fermionic lattice models, we can also consider a possible lattice rotation operators that act as $(R_{M;\vec{o}})^M =-1$ on states with odd fermion parity.}. In general, the $R_{M;\vec{o}}$ operator acts the fermion field as follows
\begin{equation}
\label{eq_lattice rotation}
    \rho(R_{M;\vec{o}})~:~~\Psi_{i}(t,\vec{x}) \rightarrow (U_{M;\vec{o}})_{i{i}'} R_{\text{IR}}(2\pi/M) \Psi_{{i}'}(t,{\vec{x}}')~,
\end{equation}
where ${\vec{x}}'$ is obtained from $\vec{x}$ by a $2\pi/M$ rotation around the fixed point $\vec{o}$. Here $R_{\text{IR}}$ is the Lorentzian rotation matrix acting on the Dirac spinor indices, and $U_{M,\vec{o}}$ denotes a flavor symmetry transformation. Since the Dirac fermion falls in the spin-$1/2$ representation of the Lorentz group, we have $(R_{\text{IR}}(2\pi/M))^M=-1$. It then follows from $(R_{M;\vec{o}})^M = 1$ that $(U_{M,\vec{o}})^M=-1$. Since the lattice translation generally does not commute with the lattice rotation, the unitary matrix $U_{M;\vec{o}}$ need not be diagonal in the flavor basis set by \eqref{eq_lattice translation}. We denote the eigenvalues of $U_{M;\vec{o}}$ by
\begin{equation}
\label{eq_quantum number s def}
    U_{M;\vec{o}}\sim \text{diag}(e^{\frac{2\pi i}{M}\textbf{s}_{1;\vec{o}}},e^{\frac{2\pi i}{M}\textbf{s}_{2;\vec{o}}},\dots, e^{\frac{2\pi i}{M}\textbf{s}_{N;\vec{o}}})~.
\end{equation}
where the quantum numbers $\textbf{s}_{i;\vec{o}}$ are defined modulo $M$. For a UV lattice rotation symmetry operator satisfying $(R_{M;\vec{o}})^M = 1$, we find the selection rule
\begin{equation}
    \textbf{s}_{i;\vec{o}}\in \mathbb{Z}+\frac{1}{2}~.
\end{equation}
Such $U_{M;\vec{o}}$ matrices exist only when the IR global symmetry group contains an exact $\mathbb{Z}_{2M}^{\text{f}}$ subgroup. Conversely, the IR field theories without such a $\mathbb{Z}_{2M}^{\text{f}}$ symmetry are incompatible with a lattice rotation operator satisfying $(R_{M;\vec{o}})^M = 1$.

Similar to the dislocation, the disclination defect is defined by a locus in the 2-dimensional space, around which the fermion fields transform according to \eqref{eq_lattice rotation}. Clearly, both the rotation angle and the phase $2\pi \textbf{s}_{i;\vec{o}}/M$ remain finite in the continuum limit, and together they determine the long-distance response of the IR fields to the disclination defect.

\subsection{Free-fermion lattice model example}

For concreteness, we now turn to the Qi--Wu--Zhang model \cite{Qi:2006xub} as an example of free-fermion lattice models. This model describes fermions hopping on a 2-dimensional square lattice with sites labeled by $(n_1,n_2)\in \mathbb{Z}^2$. We denote the fermion creation operator at the site $(n_1,n_2)$ by 
\begin{equation}
c_{(n_1,n_2)}^{\dagger}=\big(c_{(n_1,n_2),\uparrow}^{\dagger},c_{(n_1,n_2),\downarrow}^{\dagger}\big)~.
\end{equation}
The Hamiltonian of the Qi--Wu--Zhang model is controlled by the on-site potential $m$ and takes the form
\begin{equation}
\label{eq_QWZ Hamiltonian}
\mathtoolsset{multlined-width=0.89\displaywidth}
\begin{multlined}
H=m\sum_{(n_1,n_2)\in \mathbb{Z}^2} c_{(n_1,n_2)}^{\dagger} \sigma_z c_{(n_1,n_2)}+\frac{1}{2}\sum_{(n_1,n_2)\in \mathbb{Z}^2}\Big(c_{(n_1+1,n_2)}^{\dagger} (\sigma_z + i \sigma_x ) c_{(n_1,n_2)}\hfill\\
\hfill +c_{(n_1,n_2+1)}^{\dagger} (\sigma_z + i \sigma_y ) c_{(n_1,n_2)}+\text{h.c.}\Big)~,
\end{multlined}
\end{equation}
where $\sigma_x$, $\sigma_y$, and $\sigma_z$ are the Pauli matrices. This Hamiltonian can be readily diagonalized by the Fourier transformation. Let $\vec{k}=(\textbf{k}_1,\textbf{k}_2)$ denote a vector in the Brillouin zone, with $-\pi< \textbf{k}_1, \textbf{k}_2 \leq \pi$. The corresponding momentum-space Hamiltonian is
\begin{equation}
    H(\vec{k})=(\sin{\textbf{k}_1})\sigma_x+(\sin{\textbf{k}_2})\sigma_y+(m+\cos{\textbf{k}_1}+\cos{\textbf{k}_2})\sigma_z~.
\end{equation}
In particular, the Qi--Wu--Zhang model is gapless at $m=0$ and $m=\pm 1$, with the corresponding IR field theory described by free Dirac fermions.

The lattice translation operator acts on local fermion operators as follows
\begin{equation}
    P_{(\updelta n_1,\updelta n_2)} c_{(n_1,n_2)}P_{(\updelta n_1,\updelta n_2)}^\dagger=c_{(n_1+\updelta n_1,n_2+\updelta n_2)}~.
\end{equation}
For lattice rotations, we need to distinguish between the vertex center $\vec{o}_{\text{v}}=(0,0)$ and the plaquette center $\vec{o}_{\text{p}}=(\frac{1}{2},\frac{1}{2})$. Their actions on the local fermion operators are correspondingly
\begin{equation}
\label{eq_QWZ rotation action}
\begin{aligned}
     R_{4;\vec{o}_{\text{v}}} c_{(n_1,n_2)} R_{4;\vec{o}_{\text{v}}}^\dagger  ={}& \big(i c_{(-n_2,n_1),\uparrow},c_{(-n_2,n_1),\downarrow}\big)~;\\
     R_{4;\vec{o}_{\text{p}}} c_{(n_1,n_2)} R_{4;\vec{o}_{\text{p}}}^\dagger  ={}& \big(i c_{(1-n_2,n_1),\uparrow},c_{(1-n_2,n_1),\downarrow}\big)~. 
\end{aligned}
\end{equation}
The crystalline quantum numbers \eqref{eq_quantum number s def} of the Qi--Wu--Zhang model can be determined by diagonalizing \eqref{eq_QWZ rotation action} near the Dirac cones when $m=0,\pm 1$. For a more detailed treatment, see \cite{Barkeshli:2025cjs}. We summarize the quantum numbers associated with each Dirac cone in Table \ref{tab_QZW quantum numbers}.

{\renewcommand{\arraystretch}{1.5}
\begin{table}[!h]
\begin{center}
\begin{tabular}{ |c|c|c|c|c| } 
 \hline
 & Dirac cone $\vec{k}$ & $\textbf{s}_{\vec{o}_{\text{v}}}$ (vertex) & $\textbf{s}_{\vec{o}_{\text{p}}}$ (plaquette) \\ 

  \hline
$m=-1$& $(0,0)$ & $\frac{1}{2}$ & $\frac{1}{2}$  \\ 

  \hline
$m=0$, fermion 1& $(\pi,0)$ & $\frac{3}{2}$ & $\frac{1}{2}$ \\ 
\hline
$m=0$, fermion 2 & $(0,\pi)$ & $\frac{7}{2}$ & $\frac{5}{2}$\\
\hline
$m=+1$ &$(\pi,\pi)$ & $\frac{1}{2}$ & $\frac{5}{2}$\\
\hline
\end{tabular}
\caption[Crystalline quantum numbers of Dirac cones in the Qi--Wu--Zhang model]{\label{tab_QZW quantum numbers}Crystalline quantum numbers of Dirac cones in the Qi--Wu--Zhang model. We note that the band structure contains a single Dirac cone at $m=\pm 1$, whereas it contains two Dirac cones at $m=0$.}
\end{center}
\end{table}}

\subsection{Crystalline impurities and localized emanant fluxes}

\begin{figure}[thb]
\centering
\includegraphics[width=0.25\textwidth]{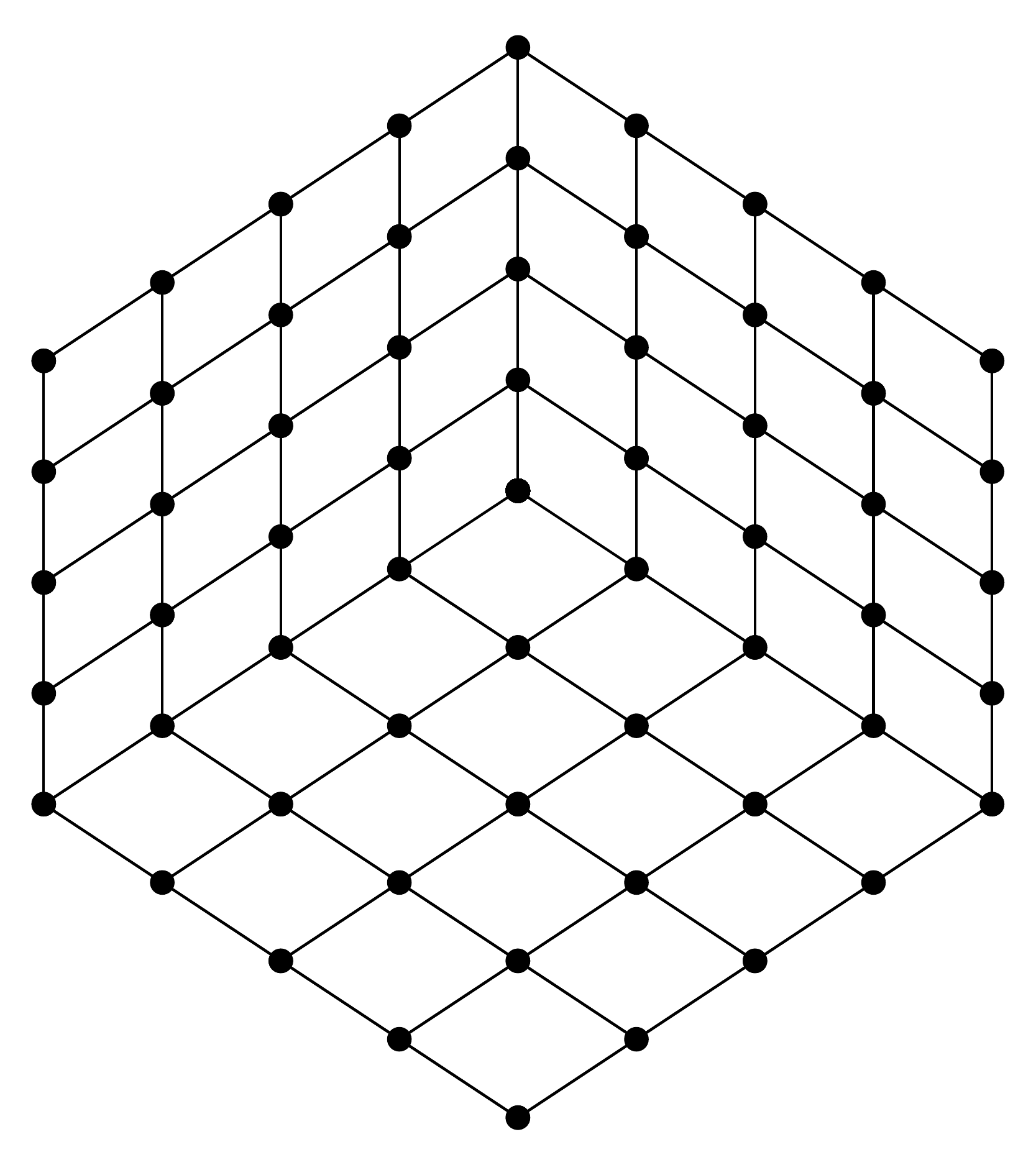}
\hspace{0.05\textwidth}
\includegraphics[width=0.26\textwidth]{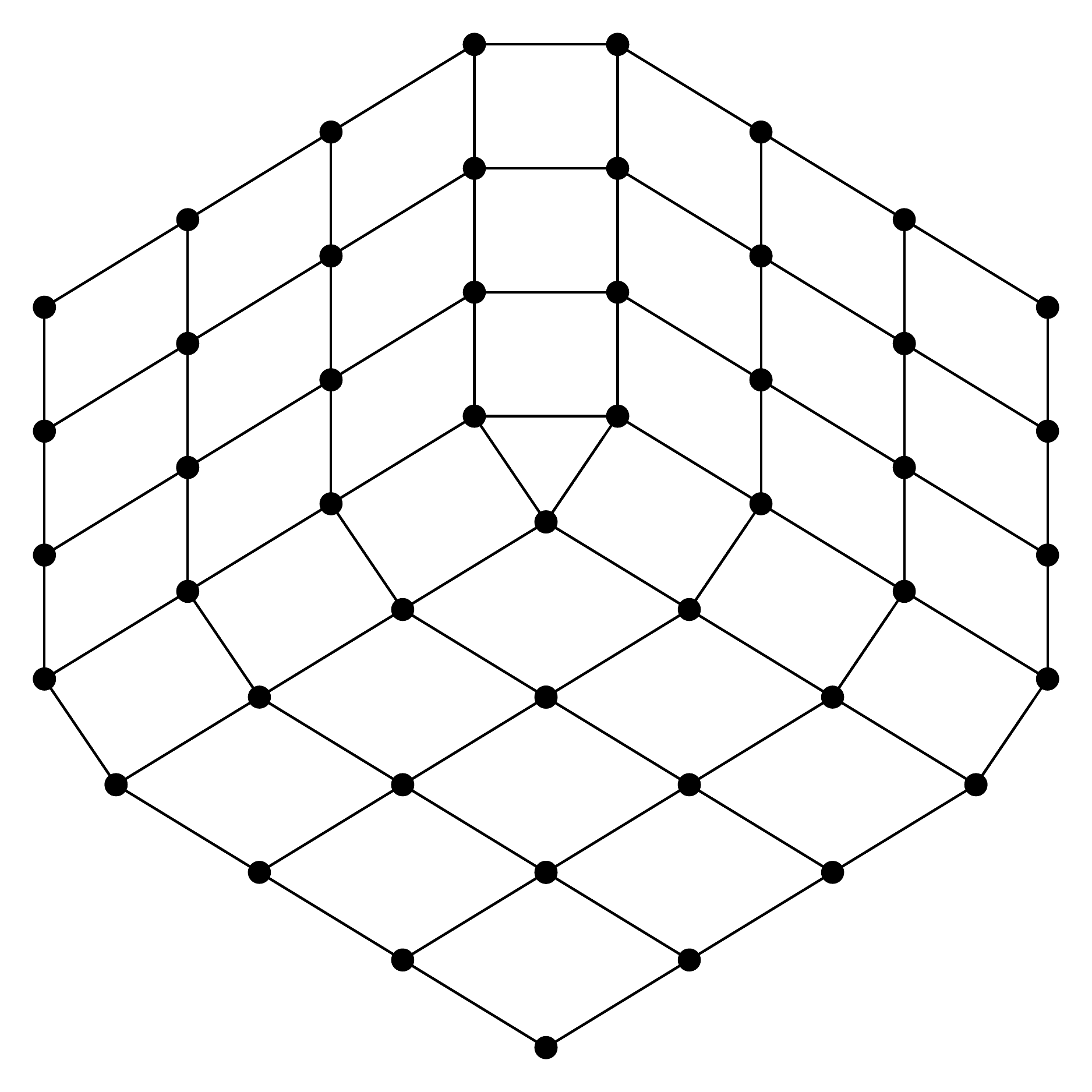}
\hspace{0.05\textwidth}
\includegraphics[width=0.3\textwidth]{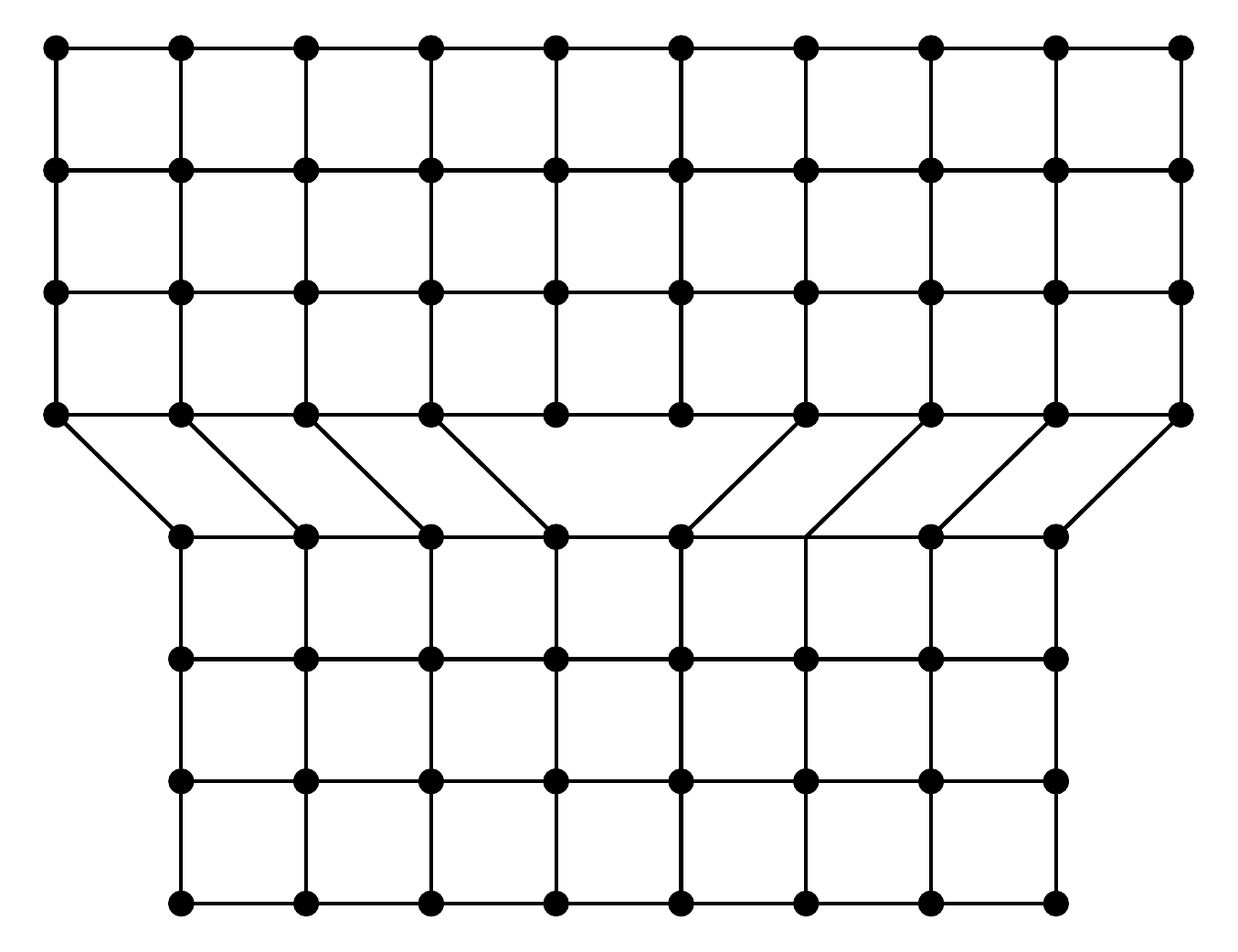}
  \caption[Crystalline impurities in the square lattice.]{\label{pic_lattice_impurities}Crystalline impurities in the square lattice. Left: vertex-centered disclination with $\beta=3/4$; Middle: plaquette-centered disclination with $\beta=3/4$; Right: dislocation with unit Burgers vector.}
\end{figure}

How can one measure the quantum numbers listed in Table \ref{tab_QZW quantum numbers} at long distances? We propose that these topological data are encoded in the DCFT one-point functions that are associated with crystalline impurities, including the disclinations and dislocations shown in Figure \ref{pic_lattice_impurities}.\footnote{We thank C. Fechisin for preparing Figures \ref{pic_lattice_impurities} and \ref{fig_QWZ_currents}, as well as for carrying out the numerical simulation.} In this and the following section, we carry out the DCFT analysis and present results from numerical simulations.

Intuitively, disclinations and dislocations can be viewed as localized fluxes of the lattice symmetry group. The operators \eqref{eq_lattice translation} and \eqref{eq_lattice rotation} act on the IR fermion fields through a combination of Lorentz and global-symmetry transformations. We therefore assume that the IR description of these crystalline impurities is given by conical defects carrying magnetic fluxes of the flavor symmetry. We refer to such global-symmetry fluxes localized at crystalline impurities as emanant fluxes.

In general, a conical singularity on the two-dimensional lattice plane is characterized by the Riemann curvature
\begin{equation}
\label{eq_ricci scalar}
R=4\pi(1-\beta)\delta^2(x)+(\vec{b}\cdot \partial) \delta^2(x)+O(\partial^2)~.
\end{equation}
Here, the first term with $\beta>0$ represents the angle deficit or excess arising from lattice disclinations. The second term, proportional to the Burgers vector $\vec{b}$, could arise from lattice dislocations \cite{kupferman2015metric}.\footnote{We note that having two opposite disclinations next to each other can also lead to the $(\vec{b}\cdot \partial) \delta^2(x)$ term in the IR description of metric.} At long distances, the IR fermion fields respond only to the angle deficit or excess. It therefore suffices to consider the effective metric of the Lorentzian spacetime 
\begin{equation}
\label{eq_cone metric}
(ds)^2=-(dt)^2+(dr)^2+\beta^2 r^2 (d\theta)^2~,
\end{equation}
where $t$ denotes the time,  $r\geq 0$ is the distance to the defect location and $\theta\sim \theta+2\pi$ is the angular polar coordinate. Crucially, the metric \eqref{eq_cone metric} admits Killing vectors leading to the conformal symmetry algebra  $so(1,2)\times so(2)\subset so(3,2)$. This is precisely the symmetry algebra \eqref{eq_DCFT symmetry} of conformal line defects in $(2+1)$-dimensional spacetime. We can thus apply the DCFT formalism introduced in Section \ref{sec_conformal defects} to this problem.

Let us now focus on a single Dirac cone in the effective metric \eqref{eq_cone metric}:
\begin{equation}
\label{eq_fermion action}
S_\text{Dirac}=\int \beta r\, dtdr  d\theta\, \Bar{\Psi}\slashed{\nabla}\Psi~,
\end{equation}
where $\slashed{\nabla}$ is the covariant Dirac operator. We also activate a static connection $A_\mu=\alpha \delta_{\mu \theta}$, with $0\leq \alpha<1$ that corresponds to the emanant Aharonov-Bohm flux at $r=0$. As the fermion field is parallel transported around the defect, the emanant flux induces the phase $\Psi\to e^{2\pi \alpha i}\Psi$. The explicit form of the Dirac operator is given by\footnote{Our convention for gamma matrices is $\upgamma_t=i \sigma_z$, $\upgamma_r=\cos{(\theta)} \sigma_x+\sin{(\theta)}\sigma_y$, and $\upgamma_\theta=\beta r(-\sin{(\theta)}\sigma_x+\cos{(\theta)}\sigma_y
)$.}
\begin{equation}
\slashed \nabla= \begin{pmatrix}
-i\partial_t & e^{-i\theta}(\partial_r+\frac{-i\partial_\theta+\alpha-\frac{\beta}{2}}{r\beta})\\
e^{i\theta}(\partial_r+\frac{i\partial_\theta-\alpha-\frac{\beta}{2}}{r\beta }) &  i\partial_t\\
\end{pmatrix}~.
\end{equation}

This DCFT can be solved by mapping the system to $\mathrm{AdS}_2\times S^1$. 
See \cite{Barkeshli:2025cjs} for a detailed treatment. 
Schematically, the bulk-to-defect OPE \eqref{eq_formal bulk-to-defect OPE} of the Dirac fermion field $\Psi$ takes the form 
\begin{equation}
\Psi(t,r,\theta)=\sum_{s}\frac{e^{i \left(s-\frac{\sigma_z}{2} \right)\theta}}{r^{1-\Delta(\hat{\Psi}^s)}}e^{\frac{i \pi}{4}\sigma_y} C_{\Psi}{}^{\hat{\Psi}^s}(\hat{\Psi}^s(t)+\text{descendants)}~,
\end{equation}
where $\hat{\Psi}^s$ are the defect spinor primaries with parallel spin $\ell=1/2$. Here, $s$ denotes the orbital angular momentum of the fermion harmonic waves, and is also identified with the transverse spin of the defect operators. With the presence of the emanant flux, the spin selection rule reads
\begin{equation}
    s\in \mathbb{Z}+\alpha+\frac{1}{2}~.
\end{equation}
In the conical space \eqref{eq_cone metric}, the allowed scaling dimensions of the defect spinor primaries are
$\Delta(\hat{\Psi}^s)=\frac{1}{2}\pm\frac{|s|}{\beta}$, whereas the unitarity of the DCFT requires $\Delta(\hat{\Psi}^s)\geq 0$. Therefore, for sufficiently large $|s|$, the defect spinors must have $\Delta(\hat{\Psi}^s)=\frac{1}{2}+\frac{|s|}{\beta}$. This analysis parallels that of Section \ref{sec_Constraints from bulk dynamics}. More generally, we denote
\begin{equation}
\label{eq_std and alt fixed points}
\begin{aligned}
        \text{standard fixed point}~:{}&~~  \forall ~s\in \mathbb{Z}+\alpha+\frac{1}{2}~,~~ \Delta(\hat{\Psi}^s)=\frac{1}{2}+ \frac{|s|}{\beta}~,\\
        \text{alternative fixed points}~:{}&~~ \exists ~s\in \mathbb{Z}+\alpha+\frac{1}{2}~,~~ \Delta(\hat{\Psi}^s)=\frac{1}{2}-\frac{|s|}{\beta}\geq0~.
\end{aligned}
\end{equation}

Let us first consider the particularly simple case $\beta=1$, for which the background metric \eqref{eq_cone metric} is flat. In this case, there is a single alternative fixed point with $\Delta(\hat{\Psi}^{\alpha-\frac{1}{2}})=\frac{1}{2}-|\frac{1}{2}-\alpha|$. We can perturb the unstable alternative fixed point with the bilinear operator
\begin{equation}
\label{eq_fermion defect bilinear}
    \updelta S_{\text{DCFT}}=\hat{m}\int_{r=0} dt \bar{\hat{\Psi}}^{\frac{1}{2}-\alpha}\hat{\Psi}^{\alpha-\frac{1}{2}}~,~~\text{where}~~\Delta(\bar{\hat{\Psi}}^{\frac{1}{2}-\alpha}\hat{\Psi}^{\alpha-\frac{1}{2}})=2\Delta(\hat{\Psi}^{\alpha-\frac{1}{2}})~.
\end{equation}
This deformation is relevant for $\alpha\neq \frac{1}{2}$, and it triggers a defect RG flow to the standard fixed point, where $\Delta(\hat{\Psi}^{\alpha-\frac{1}{2}})=\frac{1}{2}+|\frac{1}{2}-\alpha|$. Notably, the defect deformation \eqref{eq_fermion defect bilinear} approaches marginality as $\alpha\to \frac{1}{2}$, indicating large finite-size corrections. As we show below, the deformation \eqref{eq_fermion defect bilinear} becomes exactly marginal when $\alpha=\frac{1}{2}$, giving rise to a defect conformal manifold.

\begin{figure}[tb]
\centering
\includegraphics[width=.4\textwidth ]{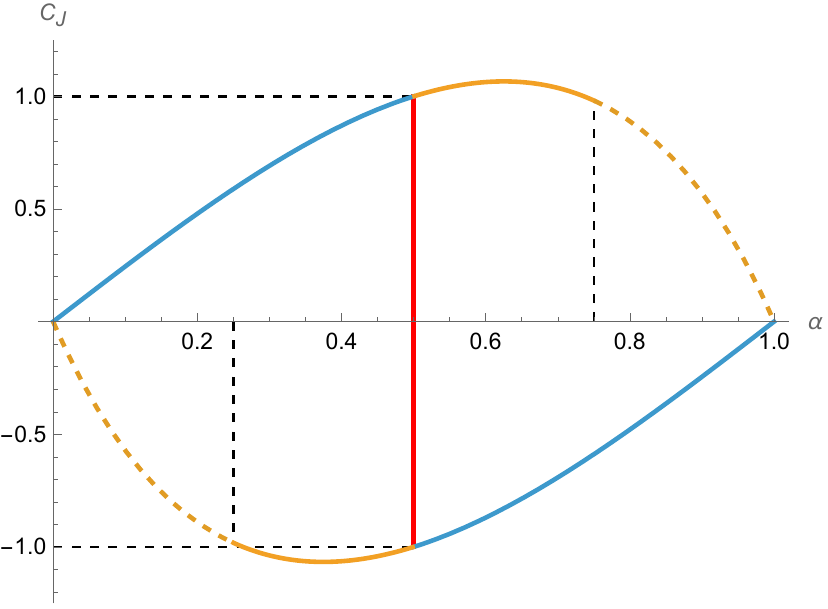}
\hspace{.1 \textwidth }
\includegraphics[width=.4\textwidth ]{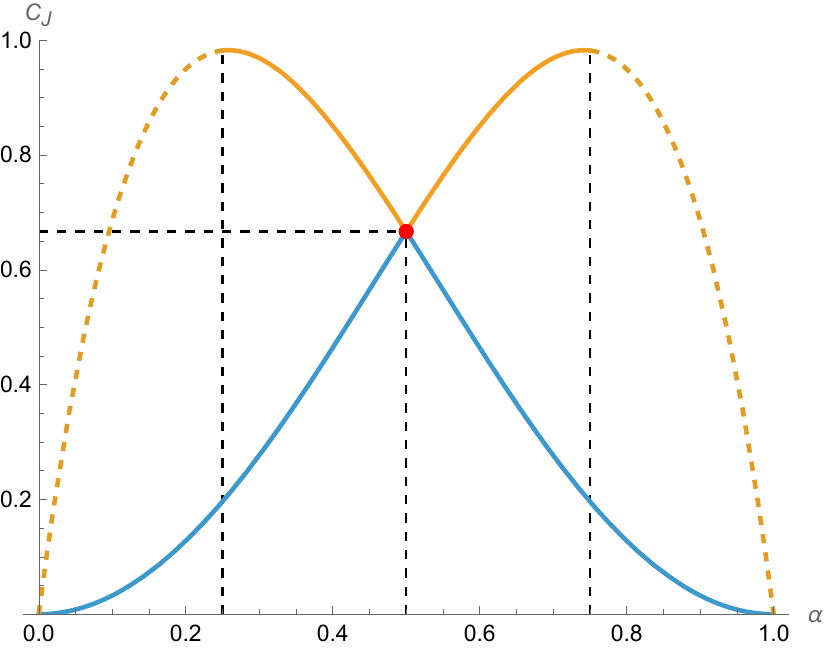}
  \caption[Azimuthal current $C_J$ and energy density $C_T$ of the flat space defect.]{\label{pic_beta=1 one point functions} Azimuthal current $C_J$ and energy density $C_T$ of the defect without conical singularity (i.e., $\beta=1$). We use blue curves to denote values at the standard fixed point, while orange curves denote those at the alternative fixed points. The dashed part of the orange curve represents where the defect quartic interaction becomes relevant. The red line and point mark the values at the defect conformal manifold.}
\end{figure}

The bulk one-point functions of the $U(1)$ current $J_{\mu}$ and the stress-energy tensor $T_{\mu_1\mu_2}$ are particularly useful for understanding the structure of the defect fixed points. They also encode imprints of the crystalline quantum numbers, which we discuss in the next section. 
The functional forms of $\langle J_\mu \rangle$ and $\langle T_{\mu_1\mu_2}\rangle$ are fixed by the conformal algebra 
$\mathfrak{so}(1,2)\times \mathfrak{so}(2)$ up to overall OPE coefficients:
\begin{equation}
    \begin{aligned}
        \langle J_\mu \rangle={}& \langle i \Bar{\Psi} \upgamma_\mu \Psi \rangle=\frac{\delta_{\mu \theta}}{r}C_{J}~;\\
        \langle T_{\mu_1\mu_2}\rangle ={}& \langle \bar{\Psi}  \upgamma_{\mu_1} \partial_{\mu_2} \Psi \rangle=\text{diag}\left(1,-1,2\beta^2r^2\right)\frac{C_T}{r^3}~.
    \end{aligned}
\end{equation}
See also \eqref{eq_bulk polarization 1pt}. Physically, $C_J$ represents the amplitude of the azimuthal current, whereas $C_T$ denotes the local energy density induced by the defect. At a given defect fixed point, $C_J$ and $C_T$ are determined by the emanant flux $\alpha$ and the conical parameter $\beta$. When $\beta=1$, the standard and alternative fixed points in the range $0\leq \alpha<\frac{1}{2}$ yield
\begin{equation}
\label{eq_CJ&CT at beta=1}
    \begin{aligned}
C^{\text{std}}_{J}(\alpha)={}&\frac{\pi}{4}\left(1-4 \alpha ^2\right) \tan (\pi  \alpha )~,~~ {}&&C^\text{alt}_{J}(\alpha)=C_J^{\text{std}}(\alpha)+\pi\big(2\alpha-1\big)\tan{(\pi \alpha)}~;\\
C^{\text{std}}_{T}(\alpha)={}&\frac{\pi}{3}\alpha\left(1-4 \alpha ^2\right) \tan (\pi  \alpha )~,~~{}&&C^\text{alt}_{T}(\alpha)=C_T^{\text{std}}(\alpha)+\pi \big(2\alpha-1\big)^2\tan{(\pi \alpha)}~.
    \end{aligned}
\end{equation}
Time-reversal symmetry implies $C_J(\alpha)=-C_J(1-\alpha)$ and $C_T(\alpha)=C_T(1-\alpha)$, thereby fixing the coefficients for $\frac{1}{2}<\alpha<1$. See Figure \ref{pic_beta=1 one point functions}. At the special point $\alpha=\frac{1}{2}$, we note there is a continuous family of DCFTs for which the azimuthal current takes values in the range $-1\leq C_J\leq 1$. This is precisely the defect conformal manifold associated with the marginal deformation \eqref{eq_fermion defect bilinear}. We also note that the Zamolodchikov distance from a generic point on this manifold to the point $C_J=0$ is given by $\arcsin{|C_J|}$.

\begin{figure*}
\centering
\includegraphics[width=.32\textwidth ]{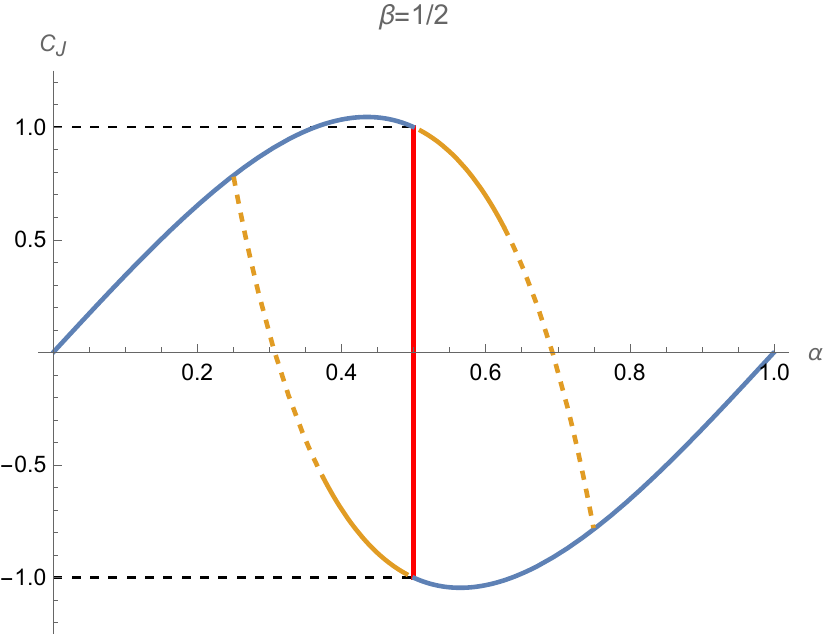}
\hspace{.003 \textwidth }
\includegraphics[width=.32\textwidth ]{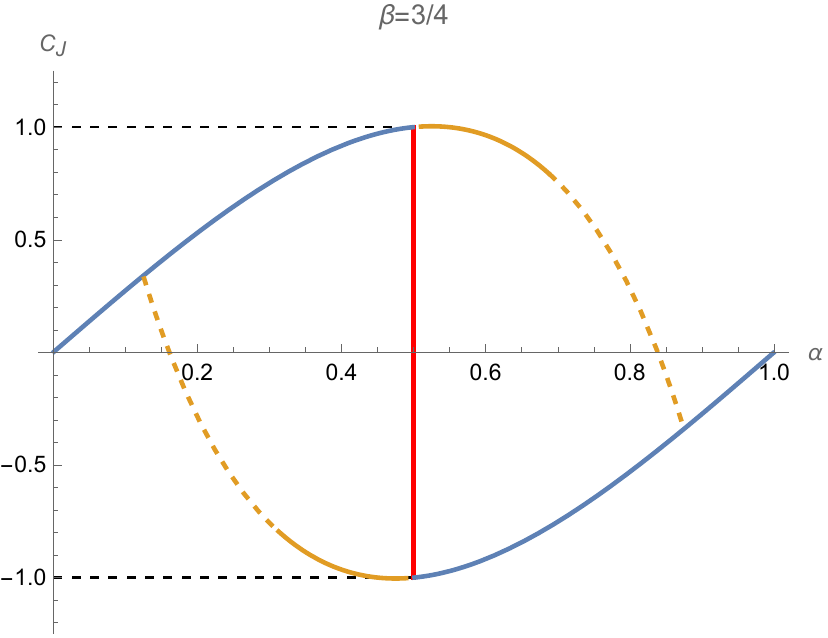}
\hspace{.003 \textwidth }
\includegraphics[width=.32\textwidth ]{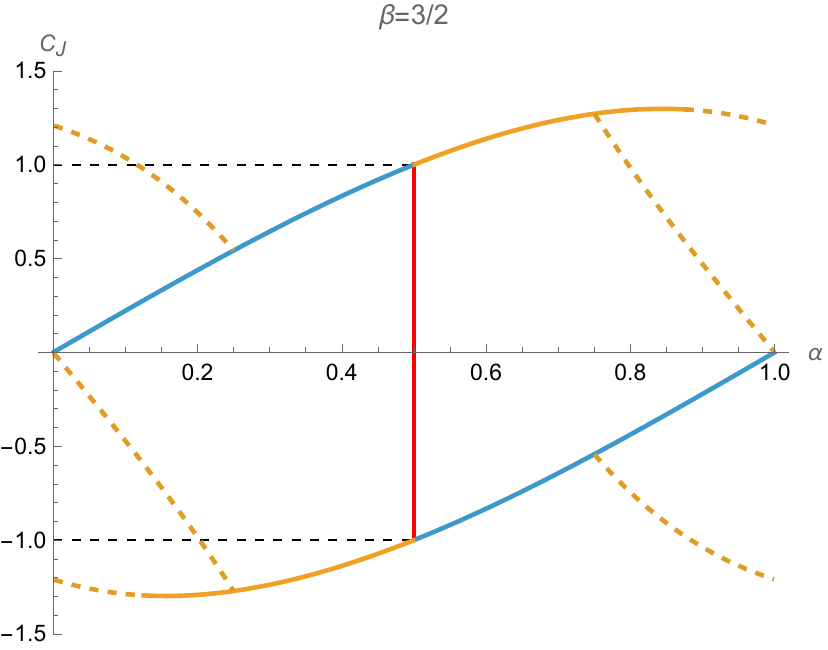}
\includegraphics[width=.32\textwidth ]{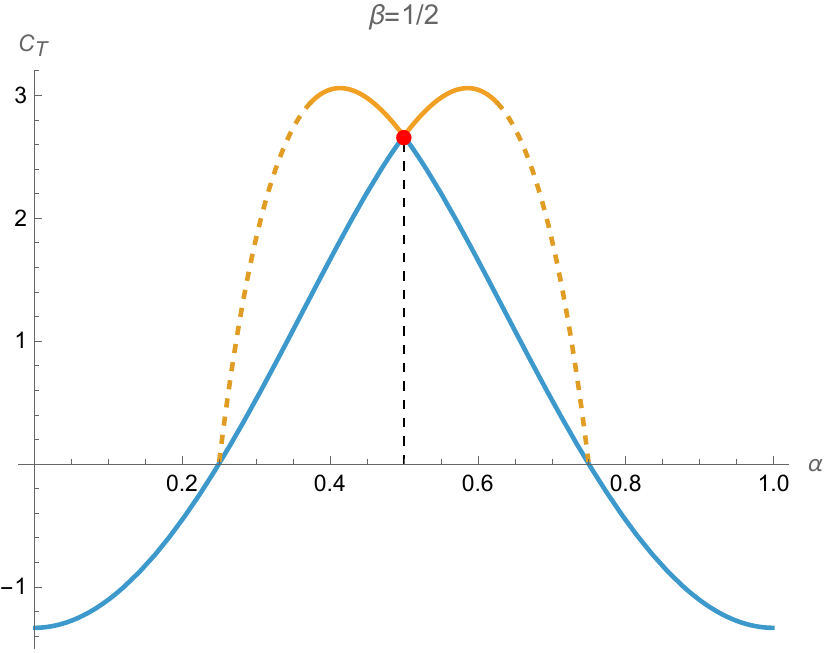}
\hspace{.003 \textwidth }
\includegraphics[width=.32\textwidth ]{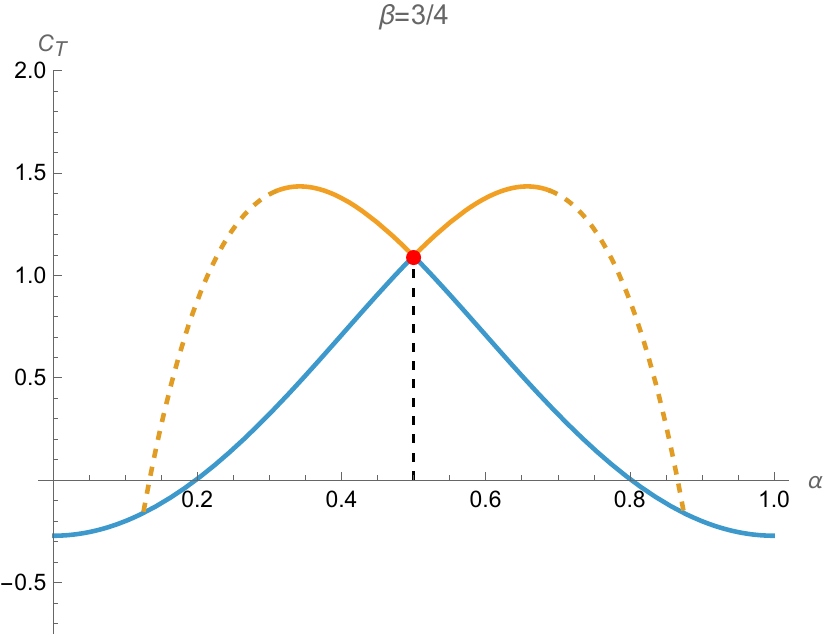}
\hspace{.003 \textwidth }
\includegraphics[width=.32\textwidth ]{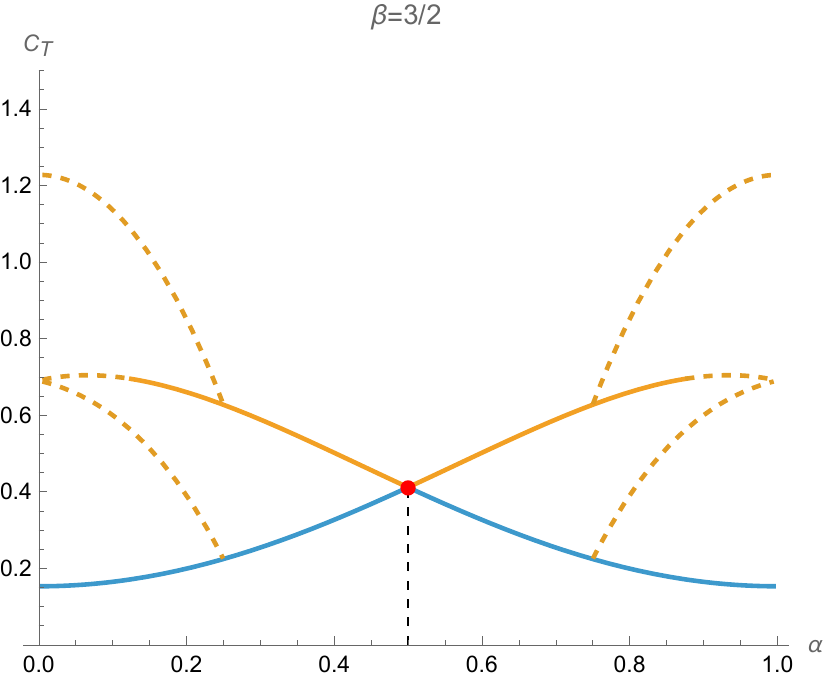}
  \caption[Azimuthal current $C_J$ and conformal weight $C_T$ of conical defects.]{\label{pic_beta=1/2 and beta=3/2 one point functions} Azimuthal current $C_J$ and conformal weight $C_T$ of the defects with conical singularity. From left to right: defects with a conical deficit ($\beta=\frac{1}{2}$ and $\beta=\frac{3}{4}$) and defects with a conical excess ($\beta=\frac{3}{2}$).}
\end{figure*}

Next, we turn to the fixed points of conical defects and their associated current and energy density. At the standard fixed point, $C_J$ as a function of $\alpha$ and $\beta$ admits an integral representation  \begin{equation}
\begin{aligned}
C_J^{\text{std}}={}&-2\lim_{v\to 2}\int_{0}^1\Big(\frac{u^{\frac{\alpha}{\beta}}-u^{-\frac{\alpha}{\beta}}}{u^{\frac{1}{2\beta}}-u^{-\frac{1}{2\beta}}}\Big)\frac{u^{\frac{v}{2}-1}du}{(1-u)^v}~,    
\end{aligned}
\end{equation}
where we take the analytical continuation of the index $v\to 2$. Similarly, $C_T$ at the standard fixed point can be written as
\begin{equation}
\label{eq_conical CT integral}
C^{\text{std}}_T=\frac{4}{\beta^2}\lim_{v\to 2}\int_0^1 \Big(\frac{(\frac{1}{2}-\alpha)(u^{\frac{2\alpha+2}{2\beta}}+u^{-\frac{2\alpha+1}{2\beta}})}{(u^{\frac{1}{2\beta}}-u^{-\frac{1}{2\beta}})^2}+\frac{(\frac{1}{2}+\alpha)(u^{\frac{2\alpha-1}{2\beta}}+u^{\frac{1-2\alpha}{2\beta}})}{(u^{\frac{1}{2\beta}}-u^{-\frac{1}{2\beta}})^2}\Big)\frac{u^{\frac{v}{2}-1}du}{(1-u)^v}~.
\end{equation}
Denoting the spins ${s}'\in \mathbb{Z}+\alpha+\frac{1}{2}$ that are associated with alternative fixed points in \eqref{eq_std and alt fixed points}, we also find
\begin{equation}
\label{eq_alternaive fixed point current and energy}
\begin{aligned}
C^{\text{alt}}_J={}&C^{\text{std}}_J+\frac{2\pi }{\beta}\sum_{{s}'}{s}'\cot\big(\pi|{s}'+\alpha|/\beta\big)~,\\
C^{\text{alt}}_T={}&C^{\text{std}}_T+\frac{4\pi}{\beta^3}\sum_{{s}'}({s}')^2\cot\big(\pi|{s}'+\alpha|/\beta\big)~.
    \end{aligned}
\end{equation}
Interestingly, the zimuthal current takes values in the range $-1\leq C_J\leq 1$ at the special point $\alpha=\frac{1}{2}$, regardless of the parameter $\beta$. It follows that the Zamolodchikov volume of the defect conformal manifold is also independent of the cone angle.

Finally, we comment that the time-reversal symmetry can be realized at $\alpha=0$ and $\alpha=\frac{1}{2}$, imposing $C_J=0$. In this case, the current alone cannot distinguish the half emanant flux from zero flux, but we can still measure the energy density:
\begin{equation}
C^{\text{std}}_T\big(\alpha=\frac{1}{2}\big)-C^{\text{std}}_T\big(\alpha=0\big)=\frac{1}{ \beta^2}\Big[2+\int_{0}^1\Big(\frac{u^{\frac{1}{2\beta}}-1}{u^{\frac{1}{2\beta}}+1}\Big)^2\frac{du}{(u-1)^2}\Big]>0~,
\end{equation}
which, among many other observables, distinguishes the time-reversal symmetric defects.

\subsection{Observables in quantum critical lattice models}

In this section, we compare the azimuthal current and energy density obtained from numerical simulations with the DCFT predictions. We focus on the $m=\pm 1$ quantum critical points of the Qi--Wu--Zhang model \eqref{eq_QWZ Hamiltonian}, both of which are described in the IR by a single Dirac fermion field. See \cite{Barkeshli:2025cjs} for other critical lattice models and further numerical results.

\begin{figure*}[t]
        \subfloat[Lattice with $5766$ sites and a vertex-centered disclination impurity $(\beta=3/4)$.]{
            \includegraphics[width=.46\textwidth]{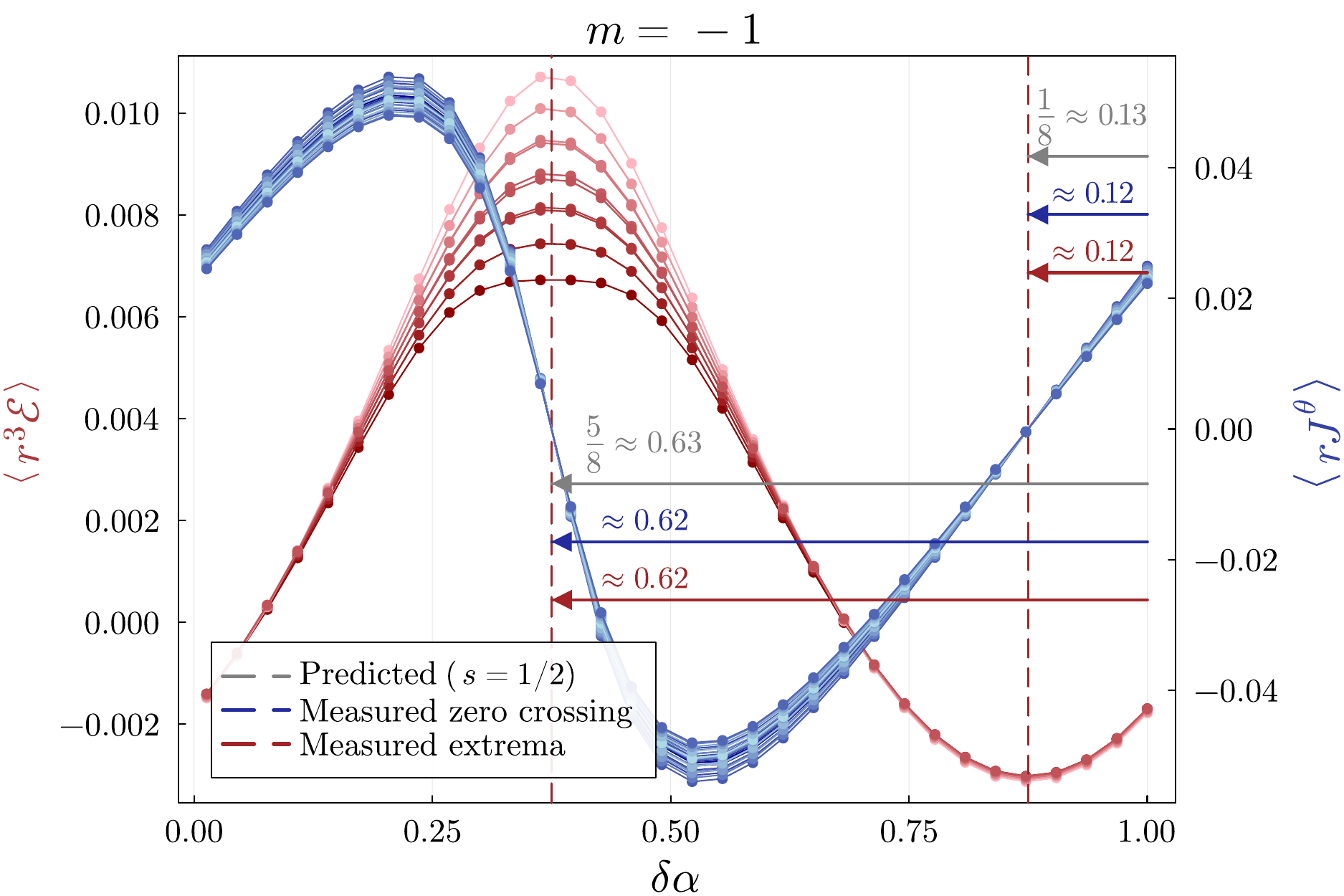}
            \hspace{.05 \textwidth }
            \includegraphics[width=.46\textwidth]{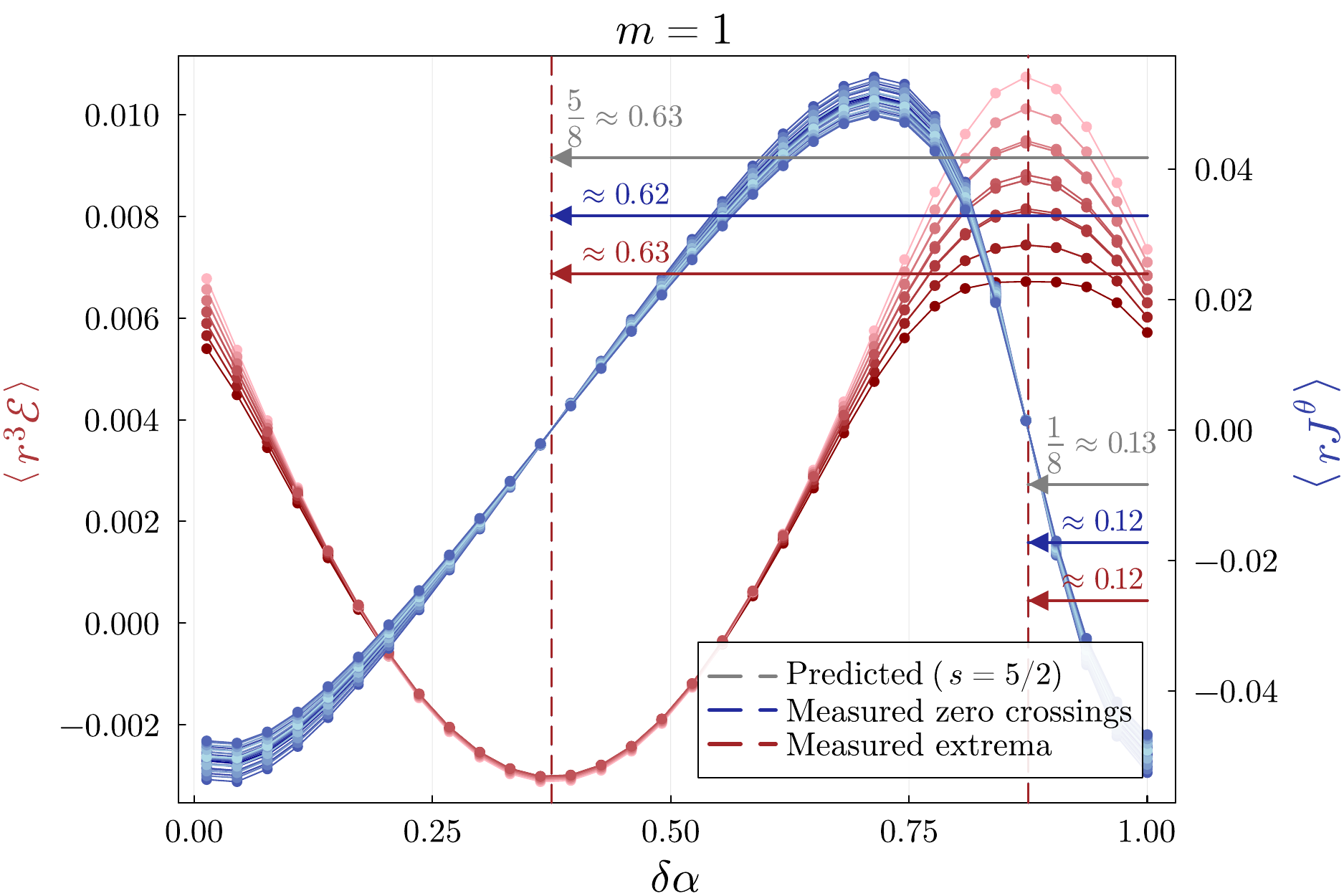}
        }\\
        \subfloat[Lattice with $5582$ sites and a plaquette-centered disclination impurity $(\beta=3/4)$.]{
            \includegraphics[width=.46\linewidth]{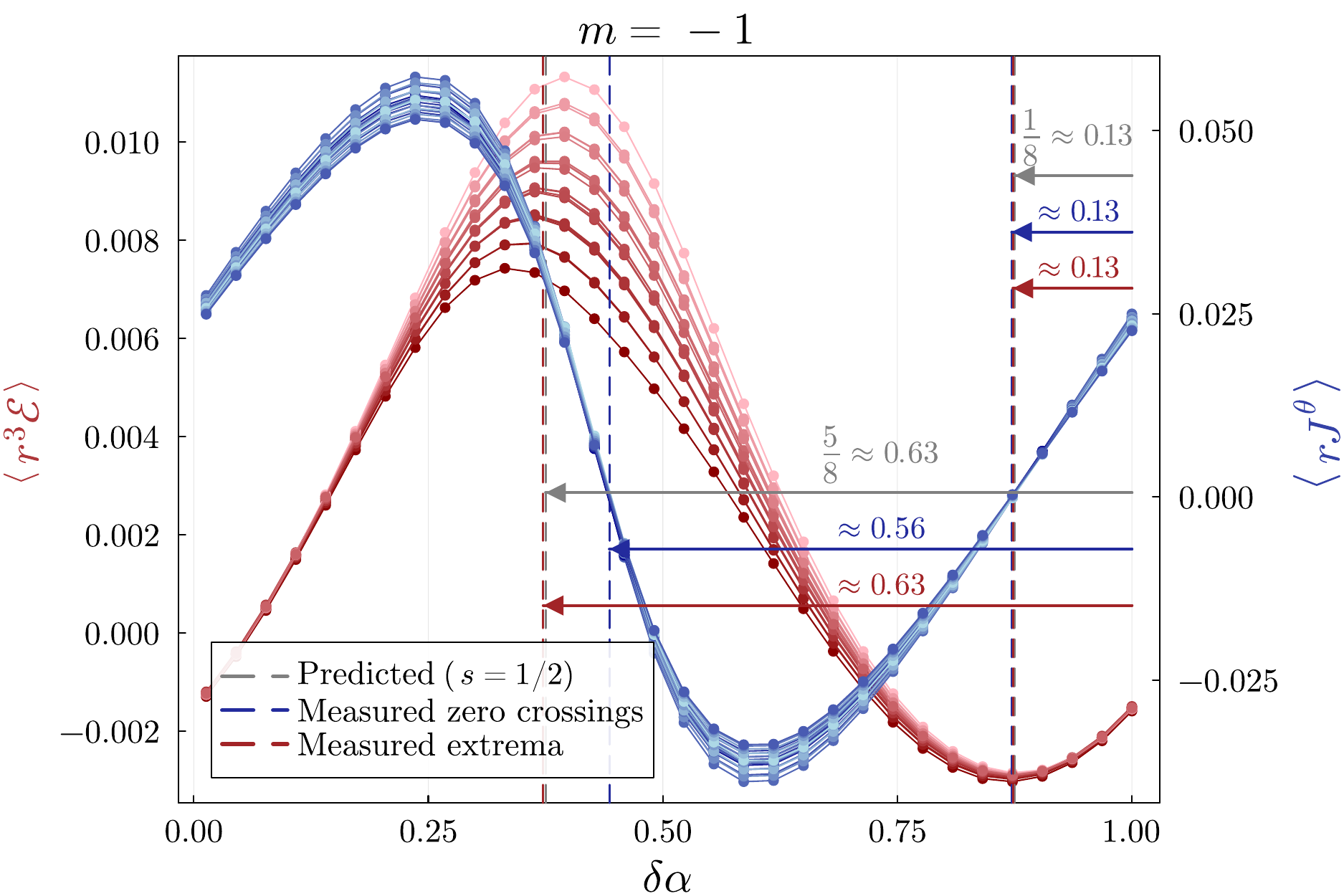}
            \hspace{.05 \textwidth }
            \includegraphics[width=.46\linewidth]{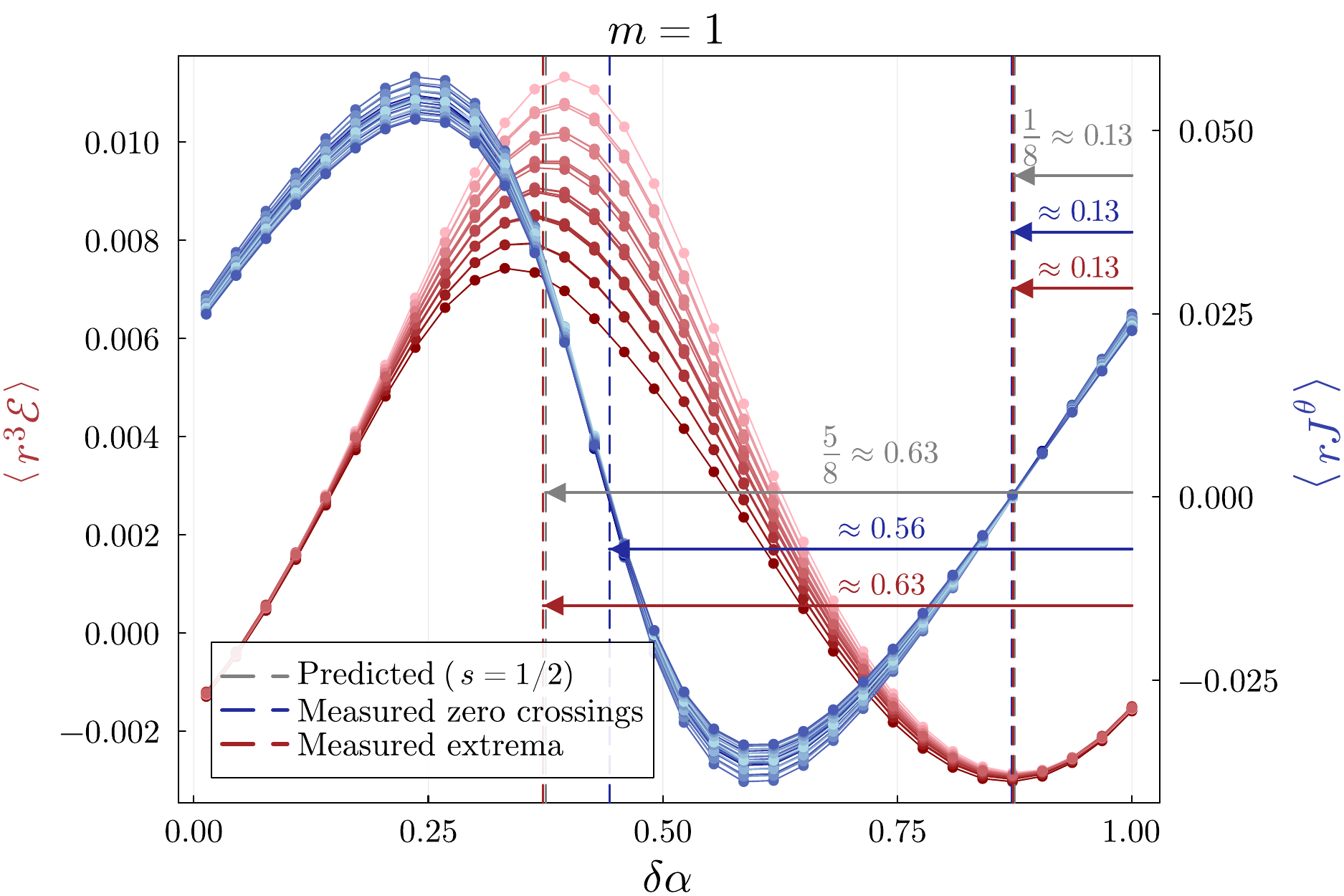}
        }\\
        \subfloat[Lattice with $5460$ sites and a dislocation impurity (unit Burgers vector)]{
            \includegraphics[width=.46\linewidth]{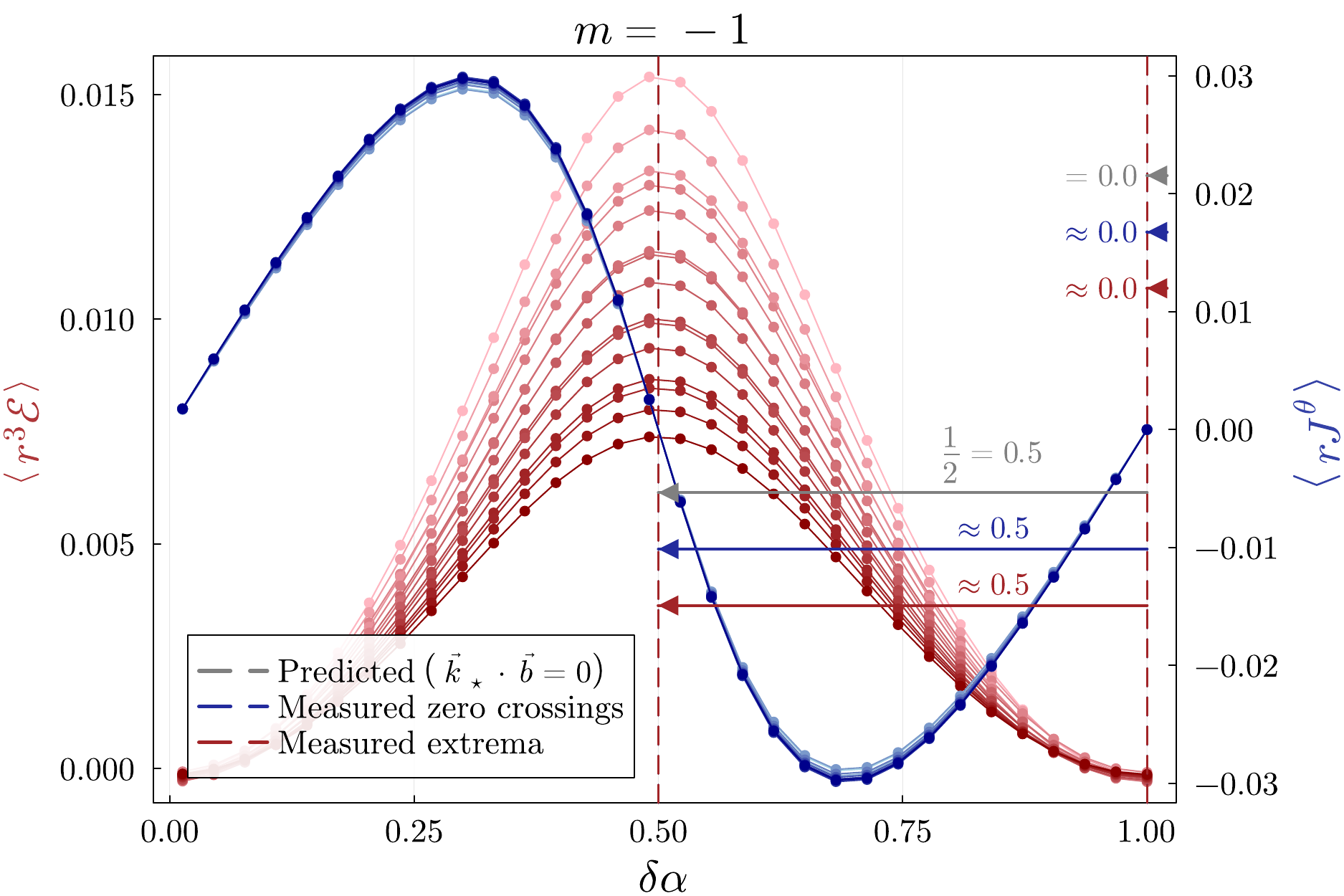}
            \hspace{.05 \textwidth }
            \includegraphics[width=.46\linewidth]{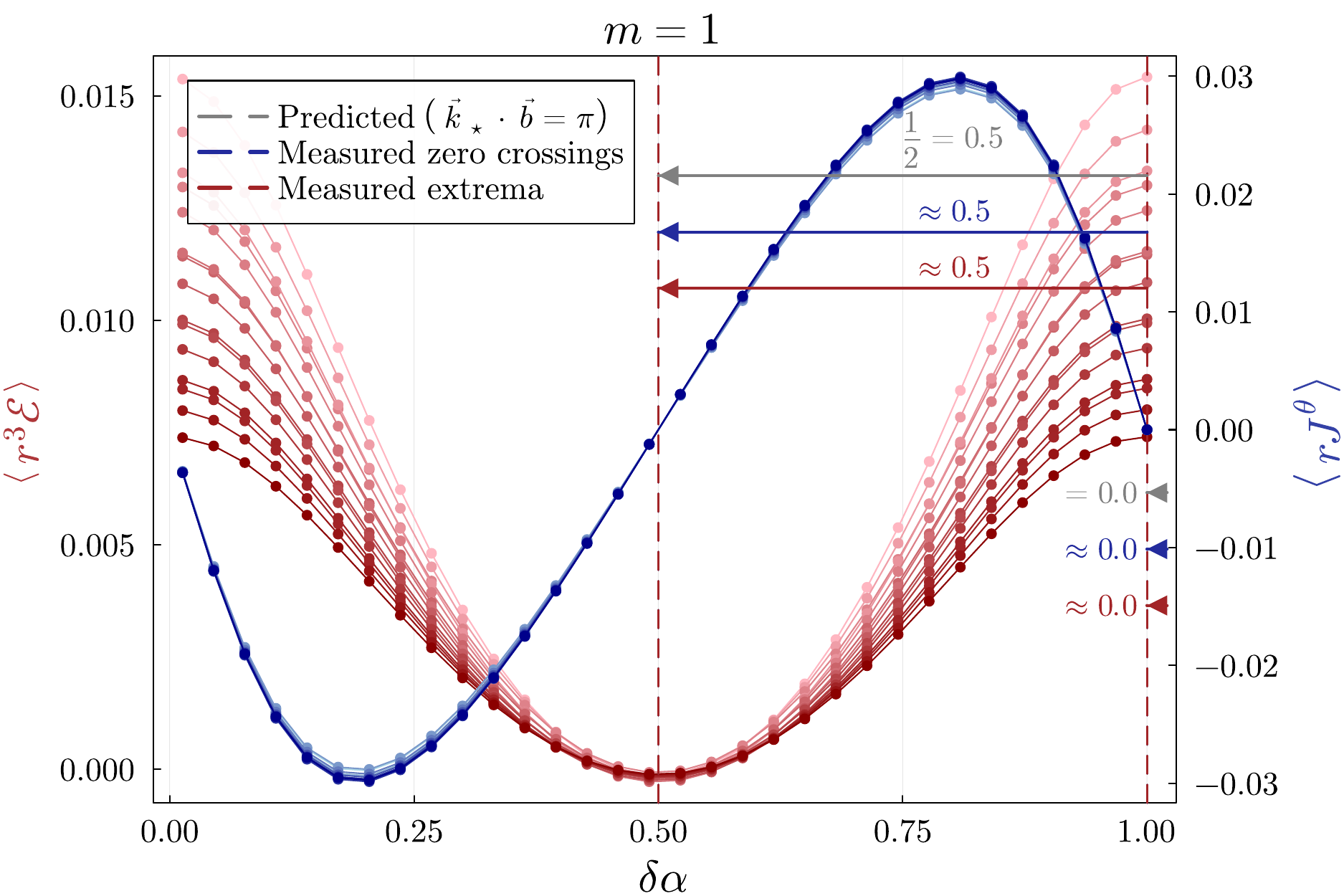}
        }
        \caption[Observables in Qi--Wu--Zhang model with crystalline impurities.]{\label{fig_QWZ_currents}Observables in Qi--Wu--Zhang model with crystalline impurities. In this figure, the shading of each curve indicates the distance from the measurement point to the defect, with lighter curves corresponding to lattice sites closer to the defect.}
    \end{figure*}

We impose an external $U(1)$ magnetic flux, parametrized by $\updelta \alpha$, at the crystalline impurities. By scanning over $0 \leq \updelta\alpha<1$, we obtain the azimuthal current and energy density plots as in Figures \ref{pic_beta=1 one point functions} and \ref{pic_beta=1/2 and beta=3/2 one point functions}. In the presence of emanant fluxes $\alpha_{\mathrm{em}}$ induced by the crystalline impurities, the total flux seen by the IR fermion fields becomes $\alpha=\alpha_{\text{em}}+\updelta\alpha$. For example, consider a disclination impurity with cone angle $2\pi \beta$, for which
\begin{equation}
    \alpha_{\text{em}}=(\beta-1)\textbf{s}_{\vec{o}}~,
\end{equation}
where $\textbf{s}_{\vec{o}}$ is the crystalline quantum number associated with the lattice rotation operator. The curves for $C_J$ and $C_T$ are accordingly shifted horizontally by $\alpha_{\mathrm{em}}$. This is indeed what we observe in Figure \ref{fig_QWZ_currents}. The same method can be applied to other lattice models to extract the crystalline quantum numbers $\textbf{s}_{i;\vec{o}}$ and the Dirac cone momenta by fitting the horizontal shift induced by the emanant flux.

\chapter[Defects and Effective String Theory]{Defects and Effective String Theory}
\label{cha_2}

In the previous chapter, we have seen that defects exhibit rich physics, and themselves probe aspects of the ambient quantum field theory that are often invisible to local operators. We now turn to a special class of defects that are central to understanding the strong forces in the Standard Model, namely confining strings.

Evidence from lattice simulations \cite{Bali:1994de, Bali:1997am} and simplified theoretical models \cite{Nielsen:1973cs, Mandelstam:1974pi} of QCD indicates that the chromoelectric flux between quarks forms string-like excitations with nonzero tension, known as the QCD flux tubes. This provides a basic explanation for the mechanism that confines quarks inside hadrons at long distances. A similar phenomenon also occurs in type-II superconductors, where magnetic monopoles are confined by the Abrikosov vortices \cite{Abrikosov:1956sx}. See \cite{Schwinger:1962tp, Wegner:1971app, Polyakov:1976fu, Fradkin:1978dv, Seiberg:1994rs} for a partial collection of other models exhibiting similar confining strings.

Unlike the examples discussed in Sections \ref{sec_Constraints from bulk dynamics}, \ref{sec_RG flows in the $O(N)$ models}, and \ref{sec_Matching UV and IR defects}, confining strings are typically realized in strongly coupled systems. When local excitations are fully gapped, these string-like excitations are stable against breaking and can only end on external defects or other strings. Studying the dynamics of these defects from a top-down perspective is challenging. Nevertheless, we can classify and study the possible low-energy effective theories of the confining strings. This is the effective string theory \cite{Dietz:1982uc, Polchinski:1991ax,Luscher:2004ib,Drummond:2004yp,Meyer:2006qx,Aharony:2009gg,Aharony:2010db,Aharony:2010cx,Aharony:2011gb,Dubovsky:2012sh,Meineri:2012uav,Aharony:2013ipa,Brambilla:2014eaa,Hellerman:2014cba,Brandt:2016xsp,Hellerman:2017upi,EliasMiro:2019kyf,EliasMiro:2021nul,Caselle:2021eir,Komargodski:2024swh,Lou:2026xqr,Albert:2026fqj}, which we will use extensively throughout this chapter.

The rest of this chapter is organized as follows. Section \ref{sec_Confining strings} provides a brief review of effective string theory, with an emphasis on the open-closed duality \cite{Polchinski:1987tu, Cardy:1989ir}. In Section \ref{sec_Effective theory of baryon junctions}, we introduce baryon junctions in effective string theory. These junctions are important structures that connect multiple confining strings.\footnote{The baryon junctions are directly observed inside baryonic states in lattice simulations of $(3+1)$-dimensional QCD \cite{Takahashi:2000te, Takahashi:2002bw, Bissey:2005sk, Bissey:2006bz}.} We establish the open-closed duality of the baryon junction in Section \ref{sec_Open-closed duality of baryon junctions}. Importantly, we map junctions with nonlinear corrections to the $s$-wave scattering amplitudes between confining string loops.

\section{Confining strings}
\label{sec_Confining strings}

We first review the effective field theory of long confining strings in $d$-dimensional spacetime. From a modern perspective, an infinitely long confining string is a vacuum of the underlying UV theory that spontaneously breaks the Poincar\'e symmetry $ISO(d)$ down to 
\begin{equation}
\label{eq_EST symmetry}
    ISO(2)\times SO(d-2)~.
\end{equation}
Here, the group $ISO(2)$ denotes the residual Poincar\'e symmetry along the string direction, whereas $SO(d-2)$ is the transverse rotation group. Despite the apparent resemblance to \eqref{eq_DCFT symmetry}, we emphasize that the symmetry breaking in \eqref{eq_EST symmetry} is spontaneous, and it gives rise to exactly $(d-2)$ Nambu--Goldstone Bosons (NGBs) \cite{Low:2001bw}, which describe the transverse fluctuations of the string. Throughout this chapter, we assume that there are no other degrees of freedom in the low-energy effective theory of the string other than these NGBs.\footnote{Confining strings in supersymmetric gauge theories often support massless fermionic modes \cite{Auzzi:2003fs, Hanany:2004ea, Edalati:2007vk, Cooper:2014noa}, as dictated by the preserved and spontaneously broken supercharges. These cases are beyond the scope of our discussion.} This assumption applies to a wide class of gapped confining gauge theories, including the pure $SU(N)$ Yang-Mills theory in $(2+1)$ and $(3+1)$ dimensions \cite{Athenodorou:2010cs, Athenodorou:2011rx, Caselle:2011fy, Solberg:2018dtu}.

For later convenience, we adopt the following EFT power-counting convention: an action term is of order $n$ if it scales as $(\text{worldsheet size})^{-n}$. In particular, an order $n$ term in the worldsheet Lagrangian density scales as $O(\partial^{n+2})$, while an order $n$ term localized on a one-dimensional submanifold of the worldsheet scales as $O(\partial^{n+1})$. In the low-energy limit, we retain the effective string action only up to a certain order.

\subsection{Nambu--Goto action}

The dynamics of the confining strings are elusive at first sight, since the underlying field theory is often strongly coupled at long distances. For sufficiently long strings, however, it was shown under mild assumptions that their low-energy dynamics are uniquely determined by the Nambu--Goto action \cite{Luscher:2004ib, Aharony:2009gg, Dubovsky:2012sh, Aharony:2013ipa}. This action describes the shape fluctuations of a string with tension $l_{\text{s}}^{-2}$ and is given by \cite{doi:10.1142/9789812795823_0026, Goto:1971ce}
\begin{equation}
\label{eq_NG action}
S_\text{string}=-l_{\text{s}}^{-2}\int d t d\upsigma  \sqrt{-\det \left(\partial_\upalpha X^\mu \partial_\upbeta X_\mu \right)}+O(\partial^4)~,
\end{equation}
where $X^\mu(t,\upsigma)$ is the embedding of the string worldsheet in $d$-dimensional spacetime. Here, we have also included higher-order corrections starting at order 4. Such corrections include the Polyakov–Kleinert rigidity term \cite{Polyakov:1986cs, Kleinert:1986bk}. 

Taking advantage of the diffeomorphism invariance, we can choose the static gauge of the worldsheet embedding $X_\mu(t,\upsigma)$ as
\begin{equation}
\label{eq_static gauge}
    X_1=t~,~~X_2=\upsigma~,~~\text{and}~~X_i=l_{\text{s}} x_i(t,\upsigma)\text{ for }3\leq i\leq d~.
\end{equation}
In Lorentzian signature, the effective string action \eqref{eq_NG action} under the ghost-free gauge \eqref{eq_static gauge} takes the expansion form
\begin{equation}
\label{eq_NG action 3}
\mathtoolsset{multlined-width=0.9\displaywidth}
\begin{multlined}
S_{\text{string}} =\int dt d\upsigma \left[-\frac{1}{l_{\text{s}}^2}+\frac{1}{2}\left((\partial_t x_i)^2-(\partial_\upsigma x_i)^2\right)\right.\hfill\\
\hfill \left.+\frac{l_{\text{s}}^2}{8}(\partial_t x_i-\partial_\upsigma x_i)^2(\partial_t x_{{i}'}+\partial_\upsigma x_{{i}'})^2+O(\partial^6)\right]~,
 \end{multlined}
\end{equation}
where both the order 0 quadratic term and the order 2 quartic term follow from the Nambu--Goto action.

The action \eqref{eq_NG action 3} can be used to compute the spectrum of a long, closed confining string with high precision. We consider a string that wraps along the periodic $X_2$-direction of circumference $2\pi R$, such that in the static gauge \eqref{eq_static gauge} we have $\upsigma \sim \upsigma+2\pi R$. The energy of a generic closed string state takes the following form \cite{Arvis:1983fp, Luscher:2004ib, Aharony:2010db, Aharony:2013ipa}
\begin{equation}
\label{eq_closed energy level}
E^{\text{closed}}_a=\frac{2\pi R}{l_{\text{s}}^2}+\frac{2}{R}\left(n_a-\frac{d-2}{24}\right)-\frac{l_{\text{s}}^2}{\pi R^3}\left(n_a-\frac{d-2}{24}\right)^2+O(R^{-5})~,
\end{equation}
where the subscript $a$ (and similarly $b$ and $c$ below) labels weakly coupled multi-particle states on the closed string. In equation \eqref{eq_closed energy level}, $n_a$ denotes the average of the left- and right-moving excitation levels, following from the order $0$ free kinetic term in the action \eqref{eq_NG action 3}\footnote{In the free theory approximation of \eqref{eq_closed energy level}, the worldsheet Hamiltonian of the closed string reads \begin{equation*}
    H_{\text{free}}=\sum_{i=3}^d\sum_{n \in \mathbb{N^+}} \frac{n}{R}\left((\alpha^i_{\text{L},n})^{\dagger} \alpha^i_{\text{L},n}+(\alpha^i_{\text{R},n})^{\dagger} \alpha^i_{\text{R},n}\right)-\frac{d-2}{12R}~,
\end{equation*}
where $\alpha^i_{\text{L},n}$ and $\alpha^i_{\text{R},n}$ are the $n$-th annihilation operator left- and right-moving oscillator modes, respectively. Eigenvalues of $H_{\text{free}}$ take the form of $\frac{2}{R}(n_a-\frac{d-2}{24})$, as explained below equation \eqref{eq_closed energy level}.}. For states with vanishing momentum along the $X_2$-direction, we have $n_a \in \mathbb{N}$.

 We note from \eqref{eq_closed energy level} that there exists a large degeneracy between closed string excited states at the precision of $O(R^{-3})$, which is a consequence of the Poincar\'e symmetry and the integrability of the quartic deformation in \eqref{eq_NG action 3}. In the following sections, we will be interested in states that transform in the scalar representation of the transverse rotation group $SO(d-2)$ and carry zero longitudinal momentum. The two lowest-lying states satisfying these conditions are the ground state $\textbf{0}$ (with $n_{\textbf{0}}=0$) and an excited state $\textbf{1}$ (with $n_{\textbf{1}}=1$), while states with excitation levels $n_a\geq 2$ are degenerate.

\subsection{Open string boundary conditions
}
In this subsection, we review the Dirichlet and the Neumann boundary effective theories of open confining strings. See, e.g., \cite{Luscher:2004ib, Aharony:2009gg, Aharony:2010cx} for a systematic and detailed analysis.

Let us consider a semi-infinite string with longitudinal coordinate $X_2=\upsigma\geq 0$, whose worldsheet action is given by \eqref{eq_NG action 3}. The variational principle of the order $0$ kinetic term in \eqref{eq_NG action 3} leads to two canonical choices of boundary conditions at $\upsigma=0$, namely Dirichlet and Neumann. The Dirichlet boundary condition that preserves $SO(d-2)$ rotation symmetry reads
\begin{equation}
\label{eq_Dirichlet bdry condition}
    \text{Dirichlet}~:~~~\partial_tx_i=0~,~~\text{for}~~3\leq i \leq  d~.
\end{equation}
It is also convenient to set the NGBs $x_i=0$ at the Dirichlet boundary $\upsigma=0$. Higher-order corrections of the Dirichlet boundary are constrained by Poincar\'e symmetry, and the leading boundary deformation takes the following form:
\begin{equation}
\label{eq_Dirichlet action}
    S_{\text{Dirichlet}}=\int_{\upsigma=0} dt \left[-m_\text{D}+\kappa_\text{D}(\partial_t\partial_\upsigma x_i)^2+O\left(\partial^6\right)\right]~.
\end{equation}
The $m_\text{D}$ term in \eqref{eq_Dirichlet action} is of order $-1$ according to our EFT counting convention, whereas the $\kappa_\text{D}$ term is of order $3$. The constant $m_\text{D}$ simply shifts the zero-point energy associated with the Dirichlet boundary. In lattice simulations, $m_\text{D}$ represents the scheme-dependent mass of the static quarks. For convenience, we set $m_\text{D}=0$ throughout this paper, i.e., we subtract the quark mass contributions.

On the other hand, the $SO(d-2)$ symmetric Neumann boundary condition reads
\begin{equation}
    \text{Neumann}~:~~~\partial_\upsigma x_i=0~,~~\text{for}~~3\leq i \leq  d~.
\end{equation}
Similarly, we can consider higher-order corrections to the Neumann boundary condition that are compatible with the Poincar\'e symmetry. The leading deformations are as follows 
\begin{equation}
\label{eq_Neumann action}
\begin{aligned}
    S_{\text{Neumann}}=&\int_{\upsigma=0} dt \left[-m_{\text{N}}\sqrt{-(\partial_tX_\mu)^2}+\kappa_\text{N}(\partial_t^2x_i)^2+O\left(\partial^6\right)\right]\\
    =&\int_{\upsigma=0} dt \left[-m_{\text{N}}+\frac{m_{\text{N}}l_{\text{s}}^2}{2}(\partial_t x_i)^2-\frac{m_{\text{N}}l_{\text{s}}^4}{8}(\partial_t x_i)^4+\kappa_\text{N}(\partial_t^2x_i)^2+O\left(\partial^6\right)\right]~.
\end{aligned}
\end{equation}
The first term in \eqref{eq_Neumann action} denotes the worldline length of the dynamical string endpoint, where the parameter $m_{\text{N}}$ is interpreted as the endpoint mass. We note that both the order $-1$ and order $1$ terms in \eqref{eq_Neumann action} are fixed by the mass $m_{\text{N}}$, whereas at order $3$ an additional parameter $\kappa_\text{N}$ can be introduced. Unlike $m_\text{D}$ in \eqref{eq_Dirichlet action}, $m_{\text{N}}$ perturbs the spectrum associated with the Neumann boundary.

\subsection{Open-closed duality}
\label{sec_Open-closed duality}

We now follow \cite{Luscher:2004ib, Meyer:2006qx, Giudice:2009di, Aharony:2010cx} and discuss the duality that associates the open string spectrum with an effective theory of closed string fields. For simplicity, we consider a finite string extending from $X_2=\upsigma=0$ to $X_2=\upsigma=L$, with the Dirichlet boundary condition \eqref{eq_Dirichlet bdry condition} at both ends. This confining string configuration describes a probe meson.

The partition function of the finite open string, denoted by $Z_{\text{meson}}$, can be computed and interpreted in several equivalent ways. In the open channel, the partition function is given by the Boltzmann sum
\begin{equation}
\label{eq_2pt open channel}
   \text{open channel}~:~~~ Z_{\text{meson}}=\sum_{E_{\text{open}}}e^{-\beta E_{\text{open}}}~,
\end{equation}
where $E_{\text{open}}$ denotes the energy eigenvalues of open string states defined on $0\leq \upsigma \leq L$, and $\beta$ is the inverse temperature. The partition function can be obtained by Wick rotating $t\to -i \uptau$ and compactifying $\uptau \sim \uptau+\beta$ in the path-integral of the effective string action \eqref{eq_NG action 3}. In particular, we take the circumference of the Euclidean time circle to be $\beta=2\pi R$. The open string partition function takes the following form
\begin{equation}
\label{eq_meson partition function 1}
Z_{\text{meson}}=\frac{e^{-2\pi R L/l_{\text{s}}^2}}{\left[\eta(\textbf{q})\right]^{(d-2)}}\left[1-\frac{(d-2)\pi l_{\text{s}}^2 }{1152L^2}\ln q\left(2E_4(\textbf{q})+(d-4)(E_{2}(\textbf{q}))^2\right)+O\left(\partial^3\right)\right]~,
\end{equation}
where the modular parameter $\textbf{q}\equiv \exp (-2\pi^2 R/L)$, $\eta(\textbf{q})$ denotes the Dedekind eta function, and $E_2(\textbf{q})$, $E_4(\textbf{q})$ are the Eisenstein series. In the expansion \eqref{eq_meson partition function 1}, terms that scale as $O(L^{-n}R^{-{n}'})$ have been collectively denoted by $O(\partial^{n+{n}'})$. We note that order $n$ terms in the effective action give leading contributions to the partition function that scale as $O(\partial^n)$. For example, corrections to the string endpoints, parametrized by $\kappa_{\text{D}}$ \eqref{eq_Dirichlet action} at $\upsigma=0$ and $\upsigma=L$, are grouped into $O(\partial^3)$ terms in the expansion \eqref{eq_meson partition function 1}.

In the closed channel, the partition function is interpreted as the two-point function of the Polyakov loop operators wrapped along the $X_1$-direction. A wrapped Polyakov loop preserves the Lorentz symmetry in the $(d-1)$-dimensional space spanned by the coordinates $X^{\perp}=(X_2, X_{3\leq i\leq d})$, and it defines a scalar point operator $\Omega (X^{\perp})$. In general, the reducible operator $\Omega$ admits a decomposition into massive particle fields in the gapped effective theory as follows: 
\begin{equation}
\label{eq_Polyakov decomposition}
    \Omega(X^{\perp})= (\pi l_{\text{s}}^2)^{\frac{d-1}{4}}\sum_{a} v_a\sqrt{E^{\text{closed}}_a} \Phi_a(X^{\perp})~,
\end{equation}
where $v_a \in \mathbb{C}$ denote decomposition coefficients, and $\Phi_a$ are $d$-dimensional complex scalar fields with definite masses. In equation \eqref{eq_Polyakov decomposition} (and similarly below), the summation runs over Dirichlet boundary states of a confining closed string with circumference $2\pi R$. These boundary states reduce to Ishibashi states in the free theory limit, and are perturbed by higher-order corrections in the action \eqref{eq_NG action 3} and \eqref{eq_Dirichlet action}. These closed string states satisfy the following properties: (i) they carry zero longitudinal momentum, as required by the Cardy condition \cite{Cardy:1989ir}; and (ii) they transform in the scalar representation of the transverse rotation group $SO(d-2)$, which is preserved by the Dirichlet condition \eqref{eq_Dirichlet bdry condition}. As we noted previously, the two lowest-lying boundary states are the ground state $\textbf{0}$ and the excited state $\textbf{1}$. We have also chosen the normalization in the decomposition \eqref{eq_Polyakov decomposition} such that the coefficients $v_a$ are dimensionless.

We adopt the following effective action ansatz for the dynamics of the closed-string fields $\Phi_a$ in $d$-dimensional space:
\begin{equation}
\label{eq_closed string EFT}
    S_{\text{loops}}=\int d^{d-1}X^\perp\left[\frac{1}{2}\sum_{a}\left(|\partial_\nu\Phi_a|^2+(E^{\text{closed}}_a)^2|\Phi_a|^2\right)+V(\Phi_a,\Phi_b,...)\right]~,
\end{equation}
where $V(\Phi_a,\Phi_b,...)$ denotes interaction terms. The functional $V(\Phi_a,\Phi_b,...)$ is constrained by 1-form symmetries of the confining strings, whether IR emergent or UV exact. We will investigate the structure of $V(\Phi_a,\Phi_b,...)$ in Section \ref{sec_Open-closed duality of baryon junctions}. Notably, the mass of the field $\Phi_a$ corresponding to the closed-string state $a$ is given by \eqref{eq_closed energy level}. From a $d$-dimensional viewpoint, $E^{\text{closed}}_a$ plays the role of the internal mass of a particle at rest.

Following the decomposition \eqref{eq_Polyakov decomposition} and the effective action \eqref{eq_closed string EFT}, we find that the closed channel partition function takes the form
\begin{equation}
\label{eq_2pt closed channel}
\begin{aligned}
   \text{closed channel}~:~~~ Z_{\text{meson}}={}&\langle \Omega^\dagger(X^\perp)\Omega(Y^\perp)\rangle\\
   ={}& \sum_{a}|v_a|^2\frac{(E^{\text{closed}}_a)^{\frac{d-1}{2}}l_{\text{s}}^{d-2}}{\sqrt{\pi}(2L)^{\frac{d-3}{2}}}K_{\frac{d-3}{2}}\left(E^{\text{closed}}_a L\right)~,    
\end{aligned}
\end{equation}
where $|X^\perp-Y^\perp|=L$, and $K_{\alpha}(x)$ denotes the modified Bessel function of the second type. In equation \eqref{eq_2pt closed channel}, we have used the tree-level propagator of massive particles in $d$-dimensional space and omitted possible loop corrections from the interaction terms. Such loop corrections are exponentially suppressed by the mass scale $O(E^{\text{closed}}_a)=O(R/l_{\text{s}}^2)$, and the partition function is therefore heavily dominated by the contribution in \eqref{eq_2pt closed channel}. With the dual modular parameter given by $\tilde{\textbf{q}}\equiv \exp(-2 L/R)$, the partition function \eqref{eq_2pt closed channel} admits the expansion
\begin{equation}
\mathtoolsset{multlined-width=0.9\displaywidth}
\begin{multlined}
\label{eq_meson partition function 2} 
Z_{\text{meson}}=\left(\frac{\pi R}{L}\right)^{\frac{d-2}{2}}\frac{e^{-2\pi R L/l_{\text{s}}^2}}{\tilde{\textbf{q}}^{\frac{d-2}{24}}}\sum_a|v_a|^2\tilde{\textbf{q}}^{n_a}\Big[1-\frac{(d-2)l_{\text{s}}^2}{2\pi R^2}\\
\times \Big(\frac{\ln \tilde{\textbf{q}}}{d-2}\Big(n_a-\frac{d-2}{24}\Big)^2-\Big(n_a-\frac{d-2}{24}\Big)+\frac{d-4}{4\ln \tilde{\textbf{q}}}\Big)+O\left(\partial^3\right)\Big]~.
\end{multlined}
\end{equation}

The open-closed duality identifies the partition functions evaluated in the open channel \eqref{eq_2pt open channel} and in the closed channel \eqref{eq_2pt closed channel}. Consequently, each term in the expansion \eqref{eq_meson partition function 1} is matched to the corresponding term in \eqref{eq_meson partition function 2} by a modular transformation. This leads to strong constraints on the decomposition coefficients $v_a$. By locality, the coefficients $v_a=v_a(l_{\text{s}},R,\kappa_\text{D}, \dots)$ are fixed solely by the local parameters of the Dirichlet boundaries, which are implemented by the Polyakov loops. From \eqref{eq_2pt open channel} and \eqref{eq_meson partition function 2}, we find that
\begin{equation}
\label{eq_v0 and v1}
\begin{aligned}
v_{\textbf{0}}={}&1+\frac{(d-2) l_{\text{s}}^2}{48 \pi  R^2}+O\left(R^{-3}\right)~,\\
v_{\textbf{1}}={}&\sqrt{d-2}\left(1+\frac{(d-26) l_{\text{s}}^2}{48 \pi  R^2}+O\left(R^{-3}\right)\right)~,
    \end{aligned}
\end{equation}
to which we will return in Sections \ref{sec_Open-closed duality of baryon junctions}.

\section{Effective theory of baryon junctions}
\label{sec_Effective theory of baryon junctions}

\begin{figure}[thb]
\centering
\includegraphics[width=.45\textwidth]{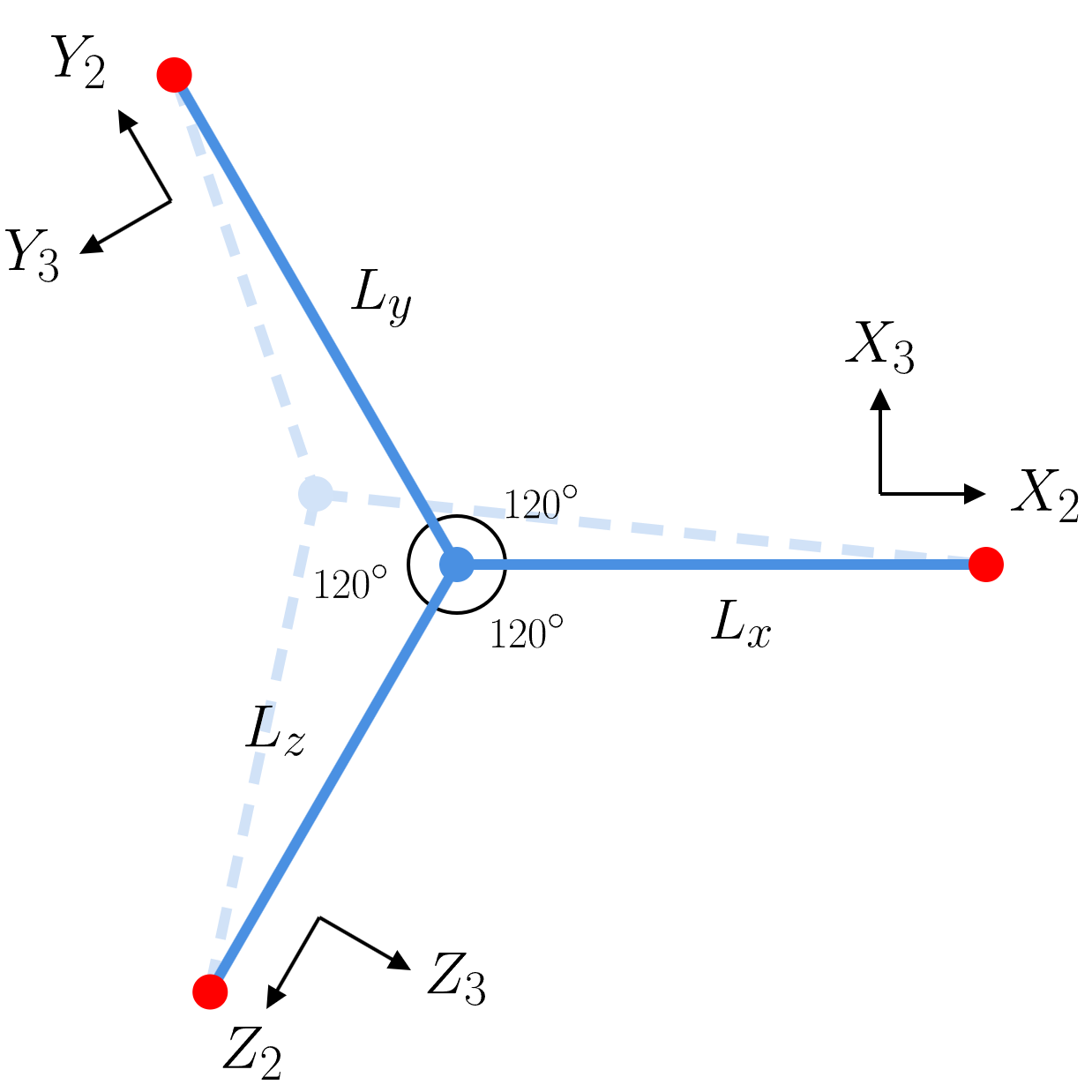}
  \caption[Confining strings in a probe baryon.]{\label{pic_baryon config}Confining strings in a probe baryon. The figure shows the spatial plane determined by the insertions of three static quarks (red points). The confining strings (blue lines) are tied at the dynamical baryon junction (blue point) and terminate on the quarks at their other ends.} 
\end{figure}

In this section, we review the effective field theory of the trivalent baryon junctions introduced in \cite{Komargodski:2024swh}. We consider the confining string configurations as in Figure \ref{pic_baryon config}, which describe probe baryons composed of three identical quarks. Our analysis applies, for example, to baryonic states in $SU(3N)$ pure Yang-Mills theories in $(2+1)$ and $(3+1)$ dimensions.

Classically, the junction position in the space is determined by balancing the tensions $l_{\text{s}}^{-2}$ of the three confining strings. It therefore coincides with the Fermat–Weber point of the triangle whose vertices are set by the quarks. Throughout this chapter, we focus on baryon configurations where the Fermat–Weber point lies within the triangle.

We denote the coordinates of three strings by $X_\mu$, $Y_\mu$, and $Z_\mu$, with their lengths $L_x$, $L_y$, and $L_z$. See also Figure \ref{pic_baryon config}. In the static gauge \eqref{eq_static gauge}, we take $X_2=Y_2=Z_2=\upsigma=0$ to be the joint baryon junction where three strings are connected, while $X_2=\upsigma=L_x$, $Y_2=\upsigma=L_y$, and $Z_2=\upsigma=L_z$ are the string endpoints on the quarks. Furthermore, we use $x_i$, $y_i$, and $z_i$ for the NGBs on the corresponding worldsheet.  In work \cite{Komargodski:2024swh, Lou:2026xqr}, we argued that the effective action of the probe baryon takes the following form:
\begin{equation}
\mathtoolsset{multlined-width=0.9\displaywidth}
\begin{multlined}
\label{eq_baryon action}
S_{\text{baryon}}=\left(S_\text{strings}^{(-2)}+S_\text{strings}^{(0)}+S_\text{strings}^{(2)}\right)+\left(S_\text{junction}^{(-1)}+S_\text{junction}^{(1)}\right)\\
+\left(S_\text{displacement}^{(1)}+S_\text{displacement}^{(2)}\right)+O(\partial^3)~,
\end{multlined}
\end{equation}
which we now unpack term by term.

\subsection{Baryon junction condition}
In the action \eqref{eq_baryon action}, the terms $S_\text{strings}^{(-2)}$, $S_\text{strings}^{(0)}$, and $S_\text{strings}^{(2)}$ collect effective actions of the three strings up to order $2$. As in the single string case \eqref{eq_NG action 3}, the order $-2$ classical term of three strings is as follows:
\begin{equation}
\label{eq_baryon strings order -2}
S_\text{strings}^{(-2)}=-l_{\text{s}}^{-2}\int d t \left (\int_0^{L_x}d \upsigma+\int_0^{L_y}d \upsigma+\int_0^{L_z}d \upsigma \right)~.
\end{equation}
The order $0$ term $S_\text{strings}^{(0)}$ is the quadratic kinetic action that governs the leading quantum fluctuations on the worldsheets:
\begin{equation}
\label{eq_baryon strings order 0}
S_\text{strings}^{(0)}=\frac{1}{2}\int d t \int_0^{L_x}d \upsigma \left[(\partial_t x_i)^2-(\partial_\upsigma x_i)^2\right]+(x\to y)+(x\to z)~.
\end{equation}
At the quark endpoints, the NGBs $x_i$, $y_i$, and $z_i$ are subject to the Dirichlet boundary condition \eqref{eq_Dirichlet bdry condition}. On the other hand, the rigid geometry of confining strings at the baryon junction leads to the constraints:
\begin{equation}
\label{eq_junction conditions 1}
x_3+y_3+z_3=0~,~~\text{and}~~x_j=y_j=z_j~~\text{for}~~4\leq j \leq d~.
\end{equation}
It is convenient to introduce the linear combinations $\xi_i^{[1]}=x_i+y_i+z_i$, $\xi_i^{[2]}=x_i-y_i$, and $\xi_i^{[3]}=x_i+y_i-2z_i$ for $3\leq i \leq d$. In these new variables, the geometry constraints \eqref{eq_junction conditions 1} and the variational principle of \eqref{eq_baryon strings order 0} determine the following boundary condition at $\upsigma=0$:
\begin{equation}
\label{eq_junction conditions 2}
\partial_t\xi_{2}^{[1]}=\partial_t\xi_{j}^{[2]}=\partial_t\xi_{j}^{[3]}=0~, ~~\text{and}~~\partial_\upsigma\xi_2^{[2]}= \partial_\upsigma\xi_2^{[3]}=\partial_\upsigma\xi^{[1]}_j=0~~\text{for}~~4\leq j \leq d~.
\end{equation}
Higher-order corrections to the baryon junction boundary condition \eqref{eq_junction conditions 2} arise from interaction terms localized at $\upsigma=0$. As we will show in Section \ref{sec_Open-closed duality of baryon junctions}, these interactions are strongly constrained by Poincar\'e symmetry and open-closed duality. Finally, the order $2$ term $S_\text{strings}^{(2)}$ contains the quartic interactions on the worldsheets:
\begin{equation}
\label{eq_baryon strings order 2}
S_\text{strings}^{(2)}=\frac{l_{\text{s}}^2}{8}\int d t \int_0^{L_x}d \upsigma (\partial_t x_i-\partial_\upsigma x_i)^2(\partial_t x_{{i}'}+\partial_\upsigma x_{{i}'})^2+(x\to y)+(x\to z)~,
\end{equation}
which follows from the Nambu--Goto action, as in equation \eqref{eq_NG action 3}.

The terms $S_\text{junction}^{(-1)}$ and $S_\text{junction}^{(1)}$ in equation \eqref{eq_baryon action} describe the worldline dynamics of the baryon junction. Let $W_\mu$ denote the spacetime coordinates of the baryon junction point. The leading worldline action compatible with Poincar\'e symmetry then takes the form:
\begin{equation}
\label{eq_baryon junction mass}
    S_\text{junction}=-m_{\text{j}}\int_{\upsigma=0} dt\sqrt{-(\partial_t W_\mu)^2}= S_\text{junction}^{(-1)}+ S_\text{junction}^{(1)}+O(\partial^3)~.
\end{equation}
We have introduced a new EFT parameter $m_{\text{j}}$ in the action \eqref{eq_baryon junction mass}, which is interpreted as the classical mass of the baryon junction. Expanding the worldline action \eqref{eq_baryon junction mass} in the NGBs $x_i$, $y_i$, and $z_i$, we find the order $-1$ term
\begin{equation}
\label{eq_baryon junction order -1}
    S_\text{junction}^{(-1)}=-m_{\text{j}}\int_{\upsigma=0} dt~,
\end{equation}
together with the order $1$ term
\begin{equation}
\label{eq_baryon junction order 1}
\mathtoolsset{multlined-width=0.9\displaywidth}
\begin{multlined}
S_\text{junction}^{(1)}=\frac{m_{\text{j}}l_{\text{s}}^2}{18}\int_{\upsigma=0} dt\left[(\partial_tx_j+\partial_ty_j+\partial_tz_j)^2\right.\hfill\\
\hfill\left.+(\partial_tx_3+\partial_ty_3-2\partial_tz_3)^2+3(\partial_tx_3-\partial_ty_3)^2\right]~.
\end{multlined}
\end{equation}
In equation \eqref{eq_baryon junction order 1}, we have chosen the linear combinations of $x_i$, $y_i$, and $z_i$ that survive the geometry constraint \eqref{eq_junction conditions 1}.

Unlike the open string boundaries reviewed in Section \ref{sec_Confining strings}, the baryon junction is allowed to fluctuate in the longitudinal directions of the confining strings. The longitudinal displacements of the three strings at $\upsigma=0$ are determined by the geometry in Figure \ref{pic_baryon config}, and are given by
\begin{equation}
\label{eq_longitudinal displacement}
    \updelta X_2=\frac{l_\text{s}}{\sqrt{3}}(z_3-y_3)~,~~\updelta Y_2=\frac{l_\text{s}}{\sqrt{3}}(x_3-z_3)~,~~\text{and}~~\updelta Z_2=\frac{l_\text{s}}{\sqrt{3}}(y_3-x_3)~.
\end{equation}
These displacements couple to EFT operators that generate infinitesimal deformations of the worldsheet boundary, analogous to the DCFT displacement operators \eqref{eq_displacement def}. For example, let us consider a single confining string whose endpoint fluctuates infinitesimally in the longitudinal direction with $\upsigma \geq \updelta X_2(t)$. Expanding the worldsheet action \eqref{eq_NG action 3} in powers of $\updelta X_2$, we identify the following effective coupling at the boundary:
\begin{equation}
\label{eq_ws displacement def}
    S_{\text{displacement}}=\sum_{n\in \mathbb{N}}\int_{\upsigma=0} dt(-\updelta X_2)^{n+1}\partial_\upsigma^{n}\left(-\frac{1}{l_{\text{s}}^2}+\frac{1}{2}\left((\partial_t x_i)^2-(\partial_\upsigma x_i)^2\right)+O(\partial^4)\right)~.
\end{equation}
In the baryon junction case, the longitudinal displacements are promoted to dynamical fields as in equation \eqref{eq_longitudinal displacement}, while the worldsheet action up to order $2$ is dictated by \eqref{eq_baryon strings order -2}, \eqref{eq_baryon strings order 0}, and \eqref{eq_baryon strings order 2}. The geometric constraint $\updelta X_2+\updelta Y_2+\updelta Z_2=0$ ensures that the order $-1$ coupling of the longitudinal fluctuations is trivial. The leading order $1$ coupling is therefore
\begin{equation}
\label{eq_baryon displacement order 1}
\mathtoolsset{multlined-width=0.9\displaywidth}
\begin{multlined}
S_\text{displacement}^{(1)}=\frac{l_{\text{s}}}{2\sqrt{3}}\int_{\upsigma=0} dt \left[(y_3-z_3)\left((\partial_t x_i)^2-(\partial_\upsigma x_i)^2\right)\right.\hfill\\
\hfill\left. {}+(z_3-x_3)\left((\partial_t y_i)^2-(\partial_\upsigma y_i)^2\right)+(x_3-y_3)\left((\partial_t z_i)^2-(\partial_\upsigma z_i)^2\right)\right]~,
\end{multlined}
\end{equation}
followed by the subleading order $2$ term,
\begin{equation}
\label{eq_baryon displacement order 2}
\mathtoolsset{multlined-width=0.9\displaywidth}
\begin{multlined}
S_\text{displacement}^{(2)}=\frac{l_{\text{s}}^2}{12}\int_{\upsigma=0} dt \left[(y_3-z_3)^2\partial_\upsigma\left((\partial_t x_i)^2-(\partial_\upsigma x_i)^2\right)\right.\hfill\\
\hfill\left. {}+(z_3-x_3)^2\partial_\upsigma\left((\partial_t y_i)^2-(\partial_\upsigma y_i)^2\right)+(x_3-y_3)^2\partial_\upsigma\left((\partial_t z_i)^2-(\partial_\upsigma z_i)^2\right)\right]~,
\end{multlined}
\end{equation}
For simplicity, we have combined the longitudinal contributions from the three worldsheets in equations \eqref{eq_baryon displacement order 1} and \eqref{eq_baryon displacement order 2}.

To summarize, we have classified terms in the effective action of the probe baryon up to order $2$. We find that the effective action \eqref{eq_baryon action} is determined by two EFT parameters, namely the classical string tension $l_{\text{s}}^{-2}$ and the classical baryon junction mass $m_{\text{j}}$. New EFT parameters may appear at higher orders; for example, the $\kappa_\text{D}$ term in \eqref{eq_Dirichlet action} at the quark endpoints is of order $3$. The quantum fluctuations of the probe baryon are fully controlled by $l_{\text{s}}^{-2}$ and $m_{\text{j}}$ up to order $2$, whose physical implications we will discuss in the upcoming sections.

\subsection{Free-field limit and junction stability}
\label{sec_Free-field limit}

We now analyze the spectrum of the probe baryon system in Figure \ref{pic_baryon config} by considering the Euclidean path-integral of the effective action \eqref{eq_baryon action}. As in Section \ref{sec_Open-closed duality}, we compute this partition function by Wick rotating $t\to -i \tau$ and compactifying $\tau\sim \tau +2\pi R$. The result admits the following perturbative expansion:
\begin{equation}
\label{eq_perturbative expansion}
\mathtoolsset{multlined-width=0.9\displaywidth}
\begin{multlined}
Z_{\text{baryon}}=e^{-S_\text{strings}^{(-2)}-S_\text{junction}^{(-1)}}Z_{\text{baryon}}^{(0)}\biggl[1-\langle S_\text{junction}^{(1)}\rangle-\langle S_\text{strings}^{(2)}\rangle-\langle S_\text{displacement}^{(2)}\rangle\hfill\\
\hfill +\frac{1}{2}\langle(S_\text{junction}^{(1)})^2\rangle+\frac{1}{2}\langle(S_\text{displacement}^{(1)})^2\rangle+O\left(\partial^3\right)\biggr]~,
\end{multlined}
\end{equation}
where we have defined
\begin{equation}
\label{eq_Z^0 definition}
    Z_{\text{baryon}}^{(0)}\equiv \int dx_idy_idz_i e^{-S_\text{strings}^{(0)}}~,~~\text{and}~~\langle {\cdots}\rangle\equiv \frac{1}{Z_{\text{baryon}}^{(0)}}\int dx_idy_idz_i ({\cdots})e^{-S_\text{strings}^{(0)}}~.
\end{equation}
The classical contributions to the path-integral are given by $S_\text{strings}^{(-2)}$ in equation \eqref{eq_baryon strings order -2} and $S_\text{junction}^{(-1)}$ in equation \eqref{eq_baryon junction order -1}. We readily find that
\begin{equation}
\label{eq_baryon classical contribution}
    S_\text{strings}^{(-2)}=\frac{2\pi R}{l_{\text{s}}^2}(L_x+L_y+L_z)~,~~\text{and}~~S_\text{junction}^{(-1)}=2\pi R m_{\text{j}}~.
\end{equation}
We have organized the quantum fluctuations in the path-integral \eqref{eq_perturbative expansion} according to their EFT order, where $Z_{\text{baryon}}^{(0)}$ is the leading order $0$ contribution. Among the higher-order corrections, we note that the expectation values of parity-odd terms vanish. For example, $\langle S_\text{displacement}^{(1)}\rangle=0$ and $\langle S_\text{junction}^{(1)}S_\text{displacement}^{(1)}\rangle=0$. 

In the rest of this section, we compute $Z_{\text{baryon}}^{(0)}$ and $\langle S_\text{junction}^{(1)}\rangle$, thereby fixing the partition function up to order $1$. This amounts to treating the NGBs $x_i$, $y_i$, and $z_i$ as free fields.

The order $0$ effective action \eqref{eq_baryon strings order 0} is simply the free kinetic action for the NGBs $x_i$, $y_i$, and $z_i$. The partition function $\mathcal{Z}^{(0)}$ with the baryon junction condition \eqref{eq_junction conditions 1} and \eqref{eq_junction conditions 2} can be evaluated as follows: we first fix the spacetime trajectory of the point-like junction, then integrate over the NGB fluctuations around the saddle point determined by this trajectory, and finally perform the quantum-mechanical path-integral over the junction worldline. See \cite{Jahn:2003uz, Pfeuffer:2008mz, Bakry:2014gea, Lou:2026xqr} for a detailed treatment. For convenience, we define the modular parameters
\begin{equation}
    \textbf{q}_x\equiv e^{-\frac{2\pi^2R}{L_x}}~,~~\textbf{q}_y\equiv e^{-\frac{2\pi^2R}{L_y}}~,~~\textbf{q}_z\equiv e^{-\frac{2\pi^2R}{L_z}}~,
\end{equation}
and the Gaussian path-integral in \eqref{eq_Z^0 definition} yields 
\begin{equation}
\label{eq_partition function order 0 general}
\mathtoolsset{multlined-width=0.9\displaywidth}
\begin{multlined}
Z_{\text{baryon}}^{(0)}=\frac{({L_x^{-1}+L_y^{-1}+L_z^{-1})^{\frac{3-d}{2}}}}{(2\pi R)^{\frac{d-1}{2}}\big(\eta(\textbf{q}_x)\eta(\textbf{q}_y)\eta(\textbf{q}_z)\big)^{d-2}}\sqrt{\frac{L_x L_y L_z}{L_x+L_y+L_z}}\prod_{n\in\mathbb{N}^+}\Bigg[\tanh{(\frac{nL_x}{R})}\\
  \times \tanh{(\frac{nL_y}{R})}\tanh{(\frac{nL_z}{R})}\frac{\left(\coth{(\frac{nL_x}{R})}+\coth{(\frac{nL_y}{R})}+\coth{(\frac{nL_z}{R})}\right)^{3-d}}{\tanh{(\frac{nL_x}{R})}+\tanh{(\frac{nL_y}{R})}+\tanh{(\frac{nL_z}{R})}}\Bigg]~.
\end{multlined}
\end{equation}

To the best of our knowledge, the partition function \eqref{eq_partition function order 0 general} with generic $L_x$, $L_y$, and $L_z$ does not admit a closed form. There are, however, special cases where the ratios of the three confining string lengths are rational:
\begin{equation}
\begin{aligned}
    \text{rational baryon}~:&~~L_x=N_xL~,~~L_y=N_yL~,~~L_z=N_zL,\\
    &~~\text{s.t.}~~N_x,N_y,N_z\in\mathbb{N}^+~~\text{and}~~\text{gcd}(N_x,N_y,N_z)=1~.
\end{aligned}
\end{equation}
Let us first note several useful mathematical facts concerning $N_x$, $N_y$, and $N_z$. In particular, consider the following two polynomials:
\begin{equation}
\label{eq_rational points}
\begin{aligned}
    P_1(z)\equiv{}& 1+\frac{1}{3}(z^{N_x}+z^{N_y}+z^{N_z})-\frac{1}{3}(z^{N_x+N_y}+z^{N_y+N_z}+z^{N_z+N_x})-z^{N_x+N_y+N_z}~,\\
    P_2(z)\equiv{}& 1-\frac{1}{3}(z^{N_x}+z^{N_y}+z^{N_z})-\frac{1}{3}(z^{N_x+N_y}+z^{N_y+N_z}+z^{N_z+N_x})+z^{N_x+N_y+N_z}~.
\end{aligned}
\end{equation}
Both $P_1(r)$ and $P_2(r)$ have all of their roots on the unit circle $|z|=1$ \cite{lakatos2004self}. We can therefore write the factorization
\begin{equation}
    P_1(z)=(1-z)\prod_{u}\big(e^{2\pi i u}-z\big)~,~~
    P_2(z)=(1-z)^2\prod_{v}\big(e^{2\pi i v}-z\big)~,
\end{equation}
where $0<u<1$ denotes the phases of the $(N_x+N_y+N_z-1)$ non-identity roots of $P_1$, and $0<v<1$ denotes the phases of the $(N_x+N_y+N_z-2)$ non-identity roots of $P_2$. In terms of these roots, we find that the baryon partition function \eqref{eq_partition function order 0 general} takes the form
\begin{equation}
\label{eq_rational baryon partition function}
    Z^{(0)}_{\text{baryon}}=\frac{1}{\big(\eta(\textbf{q})\big)^{2d-5}}\Bigg(\prod_u \frac{\textbf{q}^{\frac{1}{48}-\frac{(u-\frac{1}{2})^2}{4}}}{(\textbf{q}^u;\textbf{q})_{\infty}}\Bigg)\Bigg(\prod_v \frac{\textbf{q}^{\frac{1}{48}-\frac{(v-\frac{1}{2})^2}{4}}}{(\textbf{q}^v;\textbf{q})_{\infty}}\Bigg)^{d-3}~,
\end{equation}
where $\textbf{q}\equiv \exp(-2\pi^2R/L)$. Equation \eqref{eq_rational baryon partition function} indicates that the probe baryon system \eqref{eq_baryon action} has a unique ground state. Moreover, we find the ground state energy
\begin{equation}
\mathtoolsset{multlined-width=0.9\displaywidth}
\begin{multlined}
    E_{\text{GS}}=(N_x+N_y+N_z)\frac{L}{l_{\text{s}}^2}+m_{\text{j}}-\frac{(2d-5)\pi}{24L}\hfill\\
    \hfill-\frac{\pi}{4L}\sum_{u}\Big((u-\frac{1}{2})^2-\frac{1}{12}\Big)-\frac{(d-3)\pi}{4L}\sum_{v}\Big((v-\frac{1}{2})^2-\frac{1}{12}\Big)+O(L^{-2})~.
\end{multlined}
\end{equation}

When the three confining strings have equal length, their fluctuation modes within the string plane decouple from those perpendicular to the plane in the free-field limit. We refer to this particularly simple configuration as
\begin{equation}
\label{eq_equilateral def}
    \text{equilateral baryon}~:~~L_x=L_y=L_z=L~.
\end{equation}
The partition function \eqref{eq_partition function order 0 general} in such cases can be written in terms of (quasi-)modular forms of the parameters $\textbf{q}$ and $\sqrt{\textbf{q}}$ as follows:
\begin{equation} 
\label{eq_linear 1}
Z_{\text{baryon}}^{(0)}=\frac{1}{(\eta(\sqrt{\textbf{q}}))^{d-1}(\eta(\textbf{q}))^{d-4}}~.
\end{equation}

We now examine the stability of the baryon junction with the worldline kinetic term \eqref{eq_baryon junction order 1}. For equilateral baryons, the equations of motion can be solved separately for modes within the string plane and for modes perpendicular to it. We denote the corresponding mode frequencies by $w_{\para}$ and $w_{\perp}$, respectively. The dispersion relation then follows from the quadratic actions \eqref{eq_baryon junction order 1} and \eqref{eq_baryon strings order 0} :
\begin{equation}
\label{eq_massive vertex quantization equation}
    \cos (\omega_{\para} L)=\frac{2}{3}m_\text{j}l_\text{s}^2\omega_{\para} \sin (\omega_{\para} L)~,~~
    \cos (\omega_{\perp} L)=\frac{1}{3}m_\text{j}l_\text{s}^2\omega_{\perp}\sin (\omega_{\perp} L) ~.
\end{equation}
Depending on the scale and sign of $m_{\text{j}}$, there are essential differences in solutions to \eqref{eq_massive vertex quantization equation}:
\begin{itemize}
\item{$m_{\text{j}}<0$ :} The dispersion relation \eqref{eq_massive vertex quantization equation} admits imaginary solutions with $\omega_{\para}, \omega_{\perp}\sim il_\text{s}^2/|m_{\text{j}}|$ when $|m_{\text{j}}|\ll L/l_\text{s}^{2}$. These tachyonic modes do not immediately imply that the junction is unstable. In particular, when $|m_{\text{j}}|\lesssim 1/l_\text{s}$, the wavelengths of these tachyonic modes exceed the EFT cutoff, and junction stability must be examined with higher-order corrections taken into account. Interestingly, a junction mass satisfying $-1/l_\text{s}\lesssim m_\text{j}<0$ is reported in certain large-$N$ gauge theories \cite{Imamura:2004tf}.

When the junction mass is negative and parametrically large $|m_{\text{j}}|\gg 1/l_\text{s}$, we find perturbative tachyonic modes in the EFT. In this regime, the junction is clearly unstable, while the endpoints of these tachyonic directions are beyond the scope of our discussion.

\item{$0\leq m_\text{j} \lesssim 1/l_\text{s}$ :} The solutions to the dispersion relation \eqref{eq_massive vertex quantization equation} satisfy $\omega_{\para}, \omega_{\perp}\sim 1/L$. These solutions are positive half-integers in units of $\pi/L$ with small corrections. This is the most relevant regime to Yang-Mills theory, where $m_\text{j}\sim l_\text{s}\sim \Lambda_{\text{YM}}$ \cite{Takahashi:2000te, Takahashi:2002bw, Jahn:2003uz, Koma:2017hcm, Caselle:2025elf}.

\item{$1/l_\text{s}\ll m_\text{j} \ll L/l_\text{s}^2$ :} The solutions to the dispersion relation \eqref{eq_massive vertex quantization equation} separate into two regimes: low energy modes with $\omega_{\para}, \omega_{\perp}\ll 1/m_\text{j}l_\text{s}^2$ remain positive half-integers in units of $\pi/L$, up to small corrections. By contrast, high energy modes with $ 1/m_\text{j}l_\text{s}^2\ll \omega_{\para}, \omega_{\perp}\ll 1/l_\text{s}$ become positive integers in units of $\pi/L$, again up to small corrections.

\item{$m_\text{j} \gg L/l_\text{s}^2$ :} The solutions to the dispersion relation \eqref{eq_massive vertex quantization equation} are positive integers in units of $\pi/L$, up to small corrections. In addition, there are semiclassical modes with $\omega_{\para}, \omega_{\perp} \sim 1/l_\text{s}\sqrt{m_\text{j}L}$. These modes correspond to a heavy vertex oscillating in the classical potential without creating waves on the string. 
\end{itemize}
In the low-energy limit, we assume that the worldsheet scales $L$, $R$ are larger than any power combination of the intrinsic effective string scales $l_\text{s}$, $m_\text{j}$. We thereby find that our perturbative analysis applies as long as the baryon junction is stable with $m_\text{j} \gtrsim -1/l_\text{s}$.

Analogous to \eqref{eq_partition function order 0 general}, the order 1 correction to the partition function also has an infinite-sum representation \cite{Lou:2026xqr}
\begin{equation}
\label{eq_partition function order 1 general}
\mathtoolsset{multlined-width=0.9\displaywidth}
\begin{multlined}
    \langle S^{(1)}_\text{junction}\rangle=\frac{m_\text{j} l_{\text{s}}^2}{R}\sum_{n\in\mathbb{N}^+}\left[\frac{(d-3)n}{\coth{(\frac{nL_x}{R})}+\coth{(\frac{nL_y}{R})}+\coth{(\frac{nL_z}{R})}
    }+\frac{4n}{3}\tanh{(\frac{nL_x}{R})}\right.\hfill\\
 \hfill \left.\times \tanh{(\frac{nL_y}{R})}\tanh{(\frac{nL_z}{R})}\frac{\coth{(\frac{nL_x}{R})}+\coth{(\frac{nL_y}{R})}+\coth{(\frac{nL_z}{R})}}{\tanh{(\frac{nL_x}{R})}+\tanh{(\frac{nL_y}{R})}+\tanh{(\frac{nL_z}{R})}}\right]~.
\end{multlined}
\end{equation}
For equilateral baryons, this result can likewise be written in terms of (quasi-)modular forms of $\textbf{q}$ and $\sqrt{\textbf{q}}$: 
\begin{equation}
\label{eq_linear 2}
    \langle S_\text{junction}^{(1)}\rangle=\frac{(d+1)m_\text{j}l_{\text{s}}^2}{144L}\ln{\textbf{q}} \left(2E_2(\textbf{q})-E_2(\sqrt{\textbf{q}}) \right)~.
\end{equation}

\subsection{An accidental $\mathbb{Z}_2$ symmetry}

\begin{figure}[thb]
\centering
\includegraphics[width=.9\textwidth]{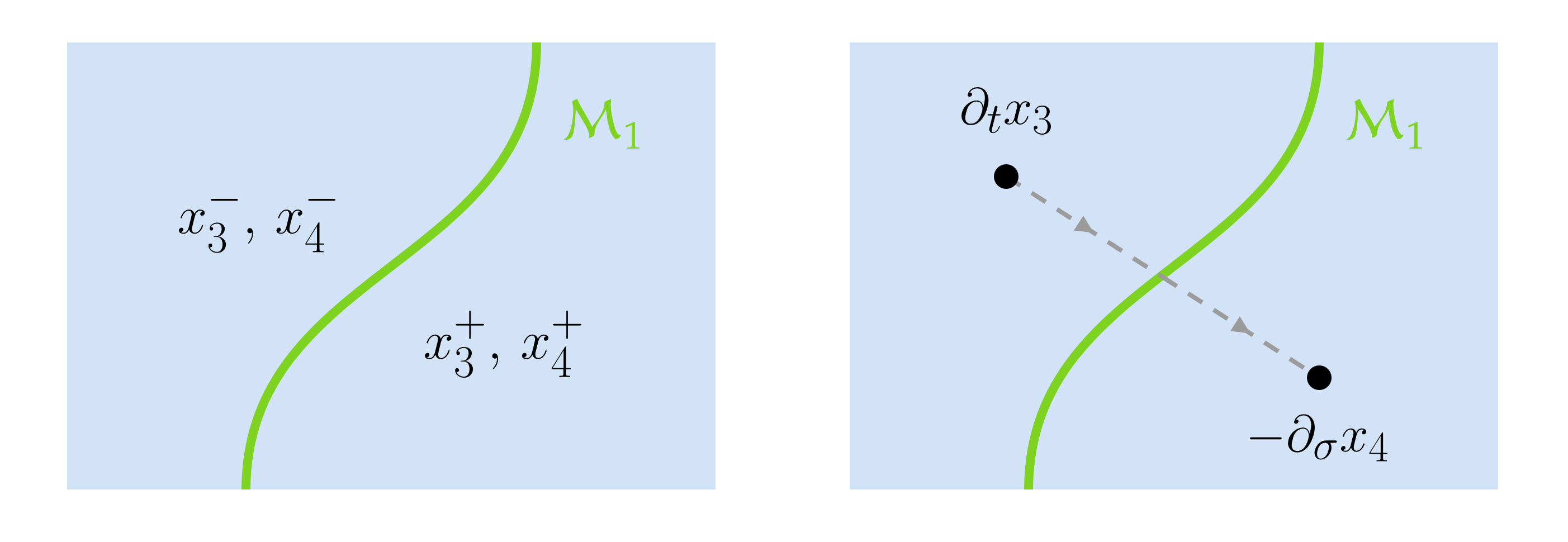}
  \caption[Topological defect line $\mathcal{D}$ on the confining string worldsheet.]{Topological defect line $\mathcal{D}$ on the confining string worldsheet. Left: the NGB field profiles are taken to be discontinuous across the line $\mathcal{M}_1$, and the defect action is given by the bilinear coupling between $x_{i}^-$ and $x_{i}^+$. Right: local operators transform according to equation \eqref{eq_symmetry rule} upon crossing the defect line. The $\mathbb{Z}_2$ symmetry transformation rule can be schematically written as $(\partial_tx_3,\partial_\upsigma x_3)\xrightarrow{\mathcal{D}} (-\partial_\upsigma x_4,-\partial_t x_4)\xrightarrow{\mathcal{D}} (\partial_tx_3,\partial_\upsigma x_3)$.} \label{pic_symmetry defect}
\end{figure}

In this section, we discuss an accidental $\mathbb{Z}_2$ symmetry of baryon junctions in 4-dimensional spacetime (i.e., $d=4$) that is broken by the junction mass $m_\text{j}$. To see this, we consider a class of line defects living on the confining string worldsheet with the effective action \eqref{eq_NG action 3}. The defect $\mathcal{D}$ is defined by coupling the NGB fields across the line manifold $\mathcal{M}_1$ as follows:
\begin{equation}
\label{eq_duality defect}
    \mathcal{D}(\mathcal{M}_1 )\equiv \exp{\left(i\int_{\mathcal{M}_1} (x_3^-dx_4^+-x_4^-dx_3^+)\right)}=\exp\left(i S^{(0)}_{\text{defect}}\right)~,
\end{equation}
where $x_{i}^-$ denotes the field profile on the left-hand side
of $\mathcal{M}_1$, while $x_{i}^+$ denotes the profile on the right-hand side. See also Figure \ref{pic_symmetry defect}. This construction is analogous to the S-duality defects \eqref{eq_maxwell duality defect action} in the Maxwell theory. The insertion of the defect $\mathcal{D}$ preserves the transverse rotation group $SO(2)$ and modifies the equations of motion for the NGB fields. It follows from the variational principle that the fields on the two sides of the defect are matched according to
\begin{equation}
\label{eq_symmetry rule}
    \big(\partial_t x_3^-,\partial_\upsigma x_3^-,\partial_t x_4^-,\partial_\upsigma x_4^-\big)=\big(\partial_\upsigma x_4^+,\partial_t x_4^+,-\partial_\upsigma x_3^+,-\partial_t x_3^+\big)~.
\end{equation}
For the probe baryon configurations shown in Figure \ref{pic_baryon config}, we can similarly define defects $\mathcal{D}$ on the confining string worldsheets using variables $y_i^\pm$ and $z_i^\pm$, respectively.

Up to order $0$, the baryon effective action \eqref{eq_baryon action} is governed by the quadratic kinetic term \eqref{eq_baryon strings order 0} and the junction condition \eqref{eq_junction conditions 2}.  One readily verifies that the worldsheet stress-energy tensor is continuous across $\mathcal{D}$ at this order. Equation \eqref{eq_duality defect} therefore defines a topological defect line in the free theory, which generates a $\mathbb{Z}_2$ global symmetry on the worldsheet. In particular, the symmetry transformation rule for local operators is given by \eqref{eq_symmetry rule}.\footnote{Local vertex operators (e.g., $\exp{(ik_ix_i)}$) are mapped to semi-local twist operators under this $\mathbb{Z}_2$ symmetry. These twist operators are characterized by the field monodromy $x_i\to x_i+\text{const}$ around the operator insertion.}

\begin{figure}[thb]
\centering
\includegraphics[width=\textwidth]{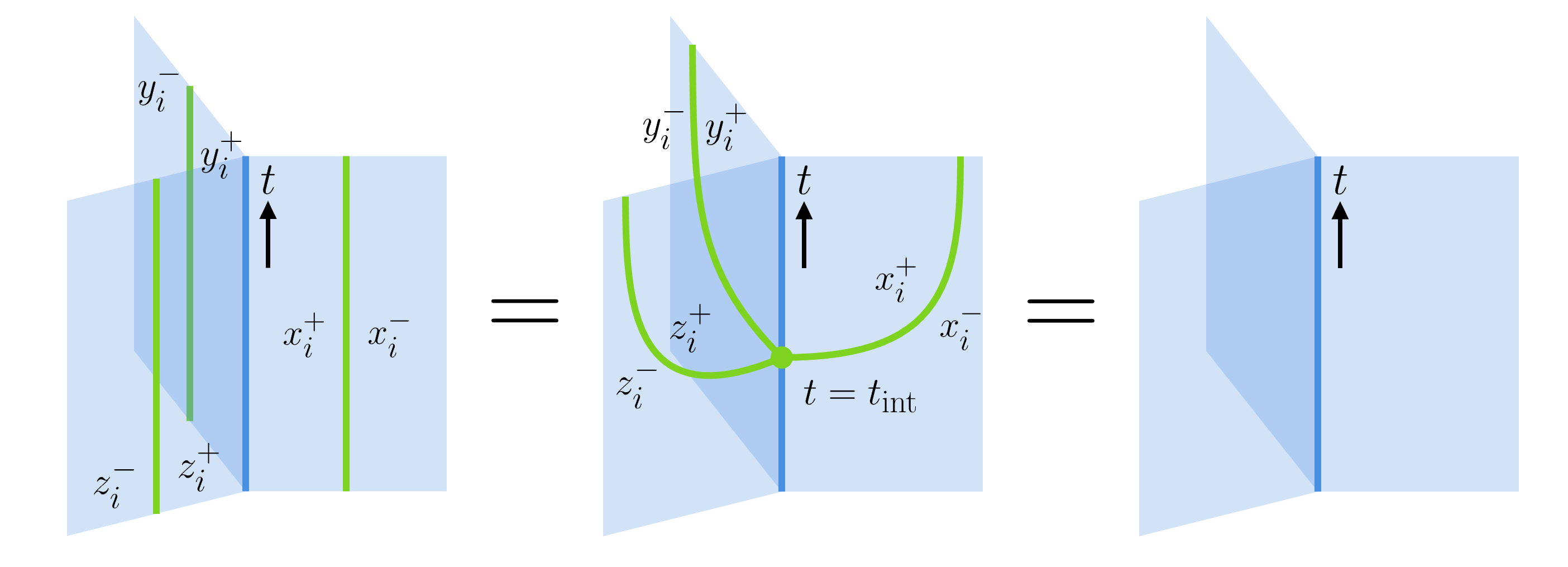}
  \caption{Merging the topological defect lines with the baryon junction worldline.} \label{pic_junction fusion}
\end{figure}

Notably, the baryon junction condition \eqref{eq_junction conditions 2} preserves the $\mathbb{Z}_2$ symmetry generated by the topological defect line $\mathcal{D}$.\footnote{For effective string theories in $d$-dimensional spacetime, we can analogously define a line defect that preserves the transverse rotation group $SO(d-2)$ as follows:
\begin{equation*}
    {\mathcal{D}}'(\mathcal{M}_1 )\equiv \exp{\left(i\int_{\mathcal{M}_1} x_i^-dx_i^+\right)}~,~~\text{where}~~3\leq i \leq d~.
\end{equation*}
The baryon junction condition \eqref{eq_junction conditions 2}, however, is not invariant under the symmetry generated by ${\mathcal{D}}'$.} The key observation is as follows: the linear combinations of NGB fields $\xi_3^{[1]}$, $\xi_4^{[2]}$, and $\xi_4^{[3]}$ satisfy Dirichlet boundary conditions at the junction, while the other combinations $\xi_3^{[2]}$, $\xi_3^{[3]}$, and $\xi_4^{[1]}$ satisfy Neumann boundary conditions. The $\mathbb{Z}_2$ symmetry \eqref{eq_duality defect} exchanges Dirichlet and Neumann conditions and acts by a $\pi/2$ rotation on the spatial indices transverse to the confining string, therefore leaving the junction condition invariant. In the free theory limit, the baryon junction worldline can be merged with the defect $\mathcal{D}$ across the three connected worldsheets as in Figure \ref{pic_junction fusion}. The $\mathbb{Z}_2$ symmetry acts on operators localized at the baryon junction through the topological intersection of the junction worldline with the line defects. 

We now examine whether the $\mathbb{Z}_2$ symmetry generated by the defect $\mathcal{D}$ persists in the full effective theory \eqref{eq_baryon action}. On the confining string worldsheets, we note that the quartic interaction in equation \eqref{eq_baryon strings order 2} is invariant under the $\mathbb{Z}_2$ transformation \eqref{eq_symmetry rule}. The leading symmetry-breaking interaction on the worldsheets arises at order $4$ and is controlled by the Polyakov–Kleinert rigidity term.

On the other hand, the baryon junction worldline preserves the $\mathbb{Z}_2$ symmetry if its intersection with defect $\mathcal{D}$ remains topological. We first note that coupling the worldsheets on the two sides of the defect using \eqref{eq_duality defect} requires the longitudinal displacements $\updelta X_2$, $\updelta Y_2$, and $\updelta Z_2$ in \eqref{eq_longitudinal displacement} to be continuous across the defect line. This leads to the following consistency condition at the intersection point $t=t_{\text{int}}$ (see Figure \ref{pic_junction fusion}):
\begin{equation}
\label{eq_gluing condition}
    \text{gluing condition}~:~x_3^+=x_3^-~,~y_3^+=y_3^-~,\text{ and }z_3^+=z_3^-\text{ at }(t,\upsigma)=(t_\text{int},0)~,
\end{equation}
which can be implemented using Lagrange multipliers at the intersection point.

We consider the worldline stress-energy tensor $\hat{T}$ as in \eqref{eq_displacement def}. More explicitly, it is defined by the local divergence of the worldsheet stress-energy tensors $T_{\upalpha\upbeta}[x_i]$, $T_{\upalpha\upbeta}[y_i]$, and $T_{\upalpha\upbeta}[z_i]$ as follows:
\begin{equation}
    \partial^\upalpha T_{\upalpha t}[x_i]+\partial^\upalpha T_{\upalpha t}[y_i]+\partial^\upalpha T_{\upalpha t}[z_i]=-\delta(\upsigma)\partial_t\hat{T}~.
\end{equation}
The operator $\hat{T}$ represents the energy stored at the baryon junction, which vanishes at order $0$. In the full effective theory, $\hat{T}$ receives corrections from the actions \eqref{eq_baryon junction order 1}, \eqref{eq_baryon displacement order 1}, \eqref{eq_baryon displacement order 2}, and other higher-order terms. For example, the contribution from longitudinal displacements on the $X$-string follows from \eqref{eq_ws displacement def} and takes the form
\begin{equation}
\label{eq_local tensor part}
    \hat{T}\supset \sum_{n\in \mathbb{N}}(-\updelta X_2)^{n+1}\partial_\upsigma^{n}\left(\frac{1}{l_{\text{s}}^2}+\frac{1}{2}\left((\partial_t x_i)^2+(\partial_\upsigma x_i)^2\right)+O(\partial^4)\right)~.
\end{equation}
Using the transformation rule \eqref{eq_symmetry rule} and the condition \eqref{eq_gluing condition}, we find that the corrections to the stress-energy tensor $\hat{T}$ in \eqref{eq_local tensor part} are continuous across the intersection point $t=t_\text{int}$ up to order $2$. The discontinuity at $t=t_{\text{int}}$ arises from the worldline kinetic term \eqref{eq_baryon junction order 1} and is given by
\begin{equation}
\label{eq_discontinuity}
\mathtoolsset{multlined-width=0.9\displaywidth}
\begin{multlined}
\hat{T}^+-\hat{T}^- =\frac{m_\text{j}l_{\text{s}}^2}{18}\Big[(\partial_tx_4^++\partial_ty_4^++\partial_tz_4^+)^2-(\partial_\upsigma x_3^++\partial_\upsigma y_3^++\partial_\upsigma z_3^+)^2\hfill\\
\phantom{\tilde{T}\big|_{t>t_\text{int}}-\tilde{T}\big|_{t<t_\text{int}}=} +(\partial_tx^+_3+\partial_ty^+_3-2\partial_tz^+_3)^2-(\partial_\upsigma x^+_4+\partial_\upsigma y^+_4-2\partial_\upsigma z^+_4)^2\hfill
\\
\hfill+3(\partial_tx^+_3-\partial_ty^+_3)^2-3(\partial_\upsigma x^+_4-\partial_\upsigma y^+_4)^2\Big]+O\left(\partial^4\right)~,
\end{multlined}
\end{equation}
where $\hat{T}^+$ and $\hat{T}^-$ denote the stress-energy tensor evaluated on the two sides of the intersection point. The discontinuity \eqref{eq_discontinuity} quantifies the obstruction to moving the intersection point topologically along the baryon junction worldline, thereby signaling the $\mathbb{Z}_2$ symmetry breaking.

We conclude that the baryon junction mass $m_{\text{j}}$ is the leading parameter that breaks the $\mathbb{Z}_2$ symmetry \eqref{eq_duality defect} in the $4$-dimensional effective string theory. Unlike the boundary parameters in \eqref{eq_Dirichlet action} and \eqref{eq_Neumann action}, the baryon junction mass can be defined without reference to external probes (e.g., static quarks) and is therefore intrinsic to the underlying UV theory. Conversely, if the $\mathbb{Z}_2$ symmetry is exactly realized in the UV theory, it imposes strong constraints on the baryon junction mass, forcing $m_\text{j}=0$.

\section{Open-closed duality of baryon junctions}
\label{sec_Open-closed duality of baryon junctions}

In this section, we establish the open-closed duality for the junctions of confining strings. To set the stage, it is useful to recall the dual channels of a finite open string reviewed in Section \ref{sec_Open-closed duality}. The open-closed duality identifies the thermal partition function of a probe meson \eqref{eq_2pt open channel} with the two-point correlation function of Polyakov loop operators \eqref{eq_2pt closed channel}. We assume that the Polyakov operators (i.e., the quark endpoints of the confining strings) can be decomposed into massive particles in $(d-1)$-dimensional space. By matching the partition function \eqref{eq_meson partition function 1} with \eqref{eq_meson partition function 2} via modular transformations, we fix the decomposition coefficients up to leading orders as in equation \eqref{eq_v0 and v1}.

We now turn to the probe baryon configuration in Figure \ref{pic_baryon config}. In the open channel, the baryon partition function $Z_{\text{baryon}}$ takes the form
\begin{equation}
\label{eq_baryon open channel}
    \text{open channel}~:~~~Z_{\text{baryon}}=\sum_{E_{\text{baryon}}} e^{-\beta E_{\text{baryon}}}~,
\end{equation}
where $E_{\text{baryon}}$ denotes the energy eigenvalues of the baryon states. The open-channel partition function was partially computed in Section \ref{sec_Free-field limit}, with $\beta=2\pi R$. In this section, we continue that calculation and evaluate the order 2 nonlinear corrections.

With three external quarks, we argue that the closed channel of the baryon partition function is given by the three-point function of Polyakov operators:
\begin{equation}
\label{eq_baryon closed channel}
    \text{closed channel}~:~~~Z_{\text{baryon}}=\langle\Omega(X^\perp)\Omega(Y^\perp)\Omega(Z^\perp)\rangle~,
\end{equation}
where $X^\perp$, $Y^\perp$, and $Z^\perp$ denote positions in $(d-1)$-dimensional space. We adopt the convention
\begin{equation}
    X^\perp=(L_x,0,\dots)~,~~Y^\perp=(-\frac{1}{2}L_y,\frac{\sqrt{3}}{2}L_y,\dots)~,~~\text{and}~~Z^\perp=(-\frac{1}{2}L_z,-\frac{\sqrt{3}}{2}L_z,\dots)~,
\end{equation}
where we have omitted the $(d-3)$ directions perpendicular to the string plane, and the Fermat-Weber point of the $X^\perp Y^\perp Z^\perp$ triangle is at the origin. In this section, we show that the three-point function \eqref{eq_baryon closed channel} is dominated by the tree-level $s$-wave scattering of the massive particles. Using the open-closed
duality, we further determine the leading contributions to the coupling constants governing these scattering processes.

\subsection{$S$-wave scattering and selection rules}

Higher-point functions of the Polyakov loop operators are determined by the potential $V(\Phi_a,..)$ in the effective action \eqref{eq_closed string EFT}. For trivalent baryon junctions in Figure \ref{pic_baryon config}, we have assumed that the underlying $d$-dimensional gauge theory is endowed with a $\mathbb{Z}_3$ $1$-form symmetry, either exact in the UV or emergent in the IR.  Upon dimensional reduction, this $1$-form symmetry descends to a $\mathbb{Z}_3$ $0$-form acting on the complex scalar fields $\Phi_a$ in $(d-1)$-dimensional space. We note that the $\mathbb{Z}_3$-symmetric potential $V(\Phi_a,\Phi_b,\dots)$ admits cubic interaction vertices of the form:
\begin{equation}
\label{eq_cubic coupling def}
\mathtoolsset{multlined-width=0.9\displaywidth}
\begin{multlined}
    V(\Phi_a,\Phi_b,\dots)\supset \frac{1}{2}\left(\frac{3}{2\pi}\right)^{\frac{d-1}{2}} (\sqrt{\pi}l_\text{s})^{\frac{d-4}{2}}\sum_{a,b,c} \sqrt{E^{\text{closed}}_aE^{\text{closed}}_bE^{\text{closed}}_c}\hfill\\
   \hfill  \times \Big[C_{abc}^{(0)}\Phi_a\Phi_b\Phi_c+C_{abc}^{(2)}\partial_{\nu} \Phi_a\partial_{\nu}\Phi_b\Phi_c+(\text{c.c.})+O\left(\partial^4\right)\Big]~,
\end{multlined}
\end{equation}
where $C^{(n)}_{abc}$ denote the order $n$ cubic coupling constants, fully symmetric in the massive mode indices $a$, $b$, $c$. The various prefactors in equation \eqref{eq_cubic coupling def} are chosen so that $C^{(0)}_{abc}$ are dimensionless and geometric factors are canceled, as we will see below. 

\begin{figure}[thb]
\centering
\includegraphics[width=\textwidth]{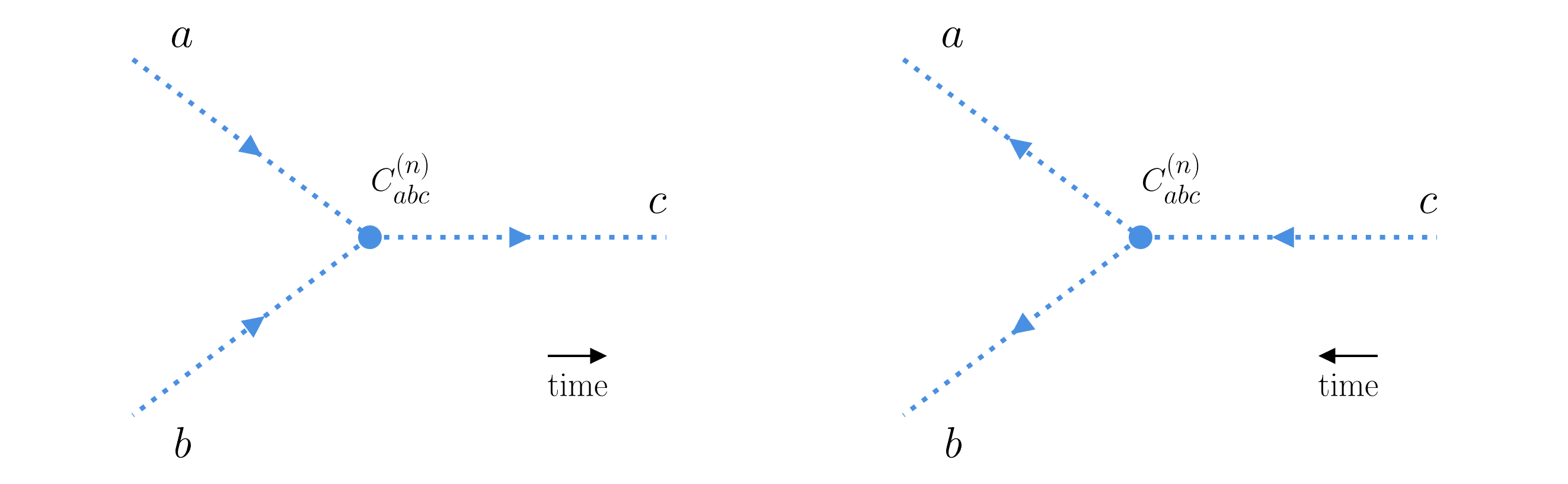}
  \caption[Tree-level scattering amplitudes of massive modes in the closed channel.]{Tree-level scattering amplitudes of massive modes in the closed channel. Interpreting one direction of the $d$-dimensional space as time, the cubic interaction vertices in equation \eqref{eq_cubic coupling def} describe the $2\to 1$ fusion process (Left)
and the $1\to 2$ decay process (Right). The external legs in this figure are generally off-shell, since all modes in the closed channel have approximately the same mass \eqref{eq_closed energy level}. } \label{pic_scattering amp}
\end{figure}

At tree-level in the perturbation theory, these interaction vertices determine scattering amplitudes between massive particles in $(d-1)$-dimensional space. See also Figure \ref{pic_scattering amp}. In particular, the coupling constants $C_{abc}^{(0)}$ are associated with the $s$-wave scattering. Likewise, $C_{abc}^{(2)}$ correspond to the $p$-wave scattering, while higher-spin scatterings (e.g.\ $d$-wave) arise at order $O(\partial^4)$ in the potential \eqref{eq_cubic coupling def}. We find that the following interaction vertices become total derivatives upon using the equations of motion:
\begin{equation}
2\sum_{a,b,c}C_{abc}^{(2)}\partial_{\nu} \Phi_a\partial_{\nu}\Phi_b\Phi_c=\sum_{a,b,c}C_{abc}^{(2)}\Big(\partial_{\nu}(\partial_{\nu} \Phi_a\Phi_b\Phi_c)-(\partial_{\nu}^2\Phi_a)\Phi_b\Phi_c\Big)~.
\end{equation}
As far as local kinetic terms are concerned, the coupling constants $C_{abc}^{(2)}$ can be removed by redefining $C^{(0)}_{abc}$ and the higher-spin terms. We therefore find that the tree-level amplitudes in the three-point function $\langle\Phi_a\Phi_b\Phi_c\rangle$ are determined entirely by $s$-wave scatterings up to order $3$.

As in Section \ref{sec_Open-closed duality}, we argue that the loop contributions in the three-point function \eqref{eq_baryon closed channel} are exponentially suppressed. The partition function up to the order $3$ is thereby given by the tree-level Feynman integral 
\begin{equation}
\label{eq_baryon closed channel Feynman integral}
\mathtoolsset{multlined-width=0.9\displaywidth}
\begin{multlined}
Z_\text{baryon}=\sum_{a,b,c}\int d^{d-1}W^\perp l_\text{s}^{1-d}C_{abc}^{(0)}v_av_bv_c\Bigg(\frac{(E^{\text{closed}}_a)^{\frac{d-1}{2}}l_\text{s}^{d-2}}{\sqrt{\pi}(2L_{wx})^{\frac{d-3}{2}}}K_{\frac{d-3}{2}}\left(E^{\text{closed}}_a L_{wx}\right)\Bigg)\\
\times \Bigg(\frac{(E^{\text{closed}}_b)^{\frac{d-1}{2}}l_\text{s}^{d-2}}{\sqrt{\pi}(2L_{wy})^{\frac{d-3}{2}}}K_{\frac{d-3}{2}}\left(E^{\text{closed}}_b L_{wy}\right)\Bigg) \Bigg(\frac{(E^{\text{closed}}_c)^{\frac{d-1}{2}}l_\text{s}^{d-2}}{\sqrt{\pi}(2L_{wz})^{\frac{d-3}{2}}}K_{\frac{d-3}{2}}\left(E^{\text{closed}}_c L_{wz}\right)\Bigg)~,
\end{multlined}
\end{equation}
where $L_{wx}=|X^\perp-W^\perp|$, $L_{wy}=|Y^\perp-W^\perp|$, and $L_{wz}=|Z^\perp-W^\perp|$. In equation \eqref{eq_baryon closed channel Feynman integral}, we have employed the decomposition \eqref{eq_Polyakov decomposition} of the Polyakov operators and used the massive propagator \eqref{eq_2pt closed channel} derived from the effective action \eqref{eq_closed string EFT}. 

The Feynman integral \eqref{eq_baryon closed channel Feynman integral} is heavily dominated by its saddle point contribution. For each set of closed string states, the saddle point is given by the $(d-1)$-dimensional vector $W^\perp=(W_1,W_2,\dots)$ that maximizes the integrand of \eqref{eq_baryon closed channel Feynman integral}. Finding such a vector reduces to the Fermat-Weber problem, which we solve perturbatively in powers of $R^{-1}\sim \beta^{-1}$. Denoting the energy levels of the closed string states in \eqref{eq_closed energy level} by $n_a$, $n_b$, and $n_c$, we find that 
\begin{equation}
\label{eq_Fermat point}
    W^\perp_\text{saddle}=\frac{Ll_\text{s}^2}{3 \pi R^2}\bigl(2 n_{a}-n_{b}-n_{c},\sqrt{3}(n_{b}-n_{c}),0,\dots \bigr)+O\left(R^{-3}\right)~.
\end{equation}
Near this saddle point, equation \eqref{eq_baryon closed channel Feynman integral} reduces to a Gaussian integral at the leading order. It is convenient to introduce the following dual modular parameters: 
\begin{equation}
    \tilde{\textbf{q}}_x\equiv e^{-\frac{2L_x}{R}}~,~~\tilde{\textbf{q}}_y\equiv e^{-\frac{2L_y}{R}}~,~~\text{and}~~\tilde{\textbf{q}}_z\equiv e^{-\frac{2L_z}{R}}~,
\end{equation}
so that the partition function \eqref{eq_baryon closed channel Feynman integral} takes the form \cite{Komargodski:2024swh}
\begin{equation}
\mathtoolsset{multlined-width=0.9\displaywidth}
\begin{multlined}
\label{eq_closed_channel_final}
Z_\text{baryon}=\left(\frac{3\pi R}{L_x+L_y+L_z}\right)^{d-\frac{5}{2}}
\left(\frac{(L_x+L_y+L_z)^2}{3(L_xL_y+L_yL_z+L_zL_x)}\right)^{\frac{d}{2}-2}
\\
\times\frac{e^{-2\pi R(L_x+L_y+L_z)/l_\text{s}^2}}{2^{\frac{d-1}{2}}(\tilde{\textbf{q}}_x\tilde{\textbf{q}}_y\tilde{\textbf{q}}_z)^{\frac{d-2}{24}}}
\sum_{a,b,c}\left(C^{(0)}_{abc}+O(\partial^2)\right)v_{a}v_{b}v_{c}
\tilde{\textbf{q}}_x^{n_{a}}\tilde{\textbf{q}}_y^{n_{b}}\tilde{\textbf{q}}_z^{n_{c}}~.
\end{multlined}
\end{equation}

The open-closed duality of the baryon junctions identifies the two representations \eqref{eq_perturbative expansion} and \eqref{eq_closed_channel_final} of the partition function. This allows us to fix the coupling constants $C^{(0)}_{abc}$ by consistency conditions, even though the underlying confining gauge theory is strongly coupled. Indeed, the open channel terms in equations \eqref{eq_partition function order 0 general} and \eqref{eq_partition function order 1 general} admit expansions in $\tilde{\textbf{q}}_x$, $\tilde{\textbf{q}}_y$, and $\tilde{\textbf{q}}_z$ using the modular transformation. Putting together $Z_{\text{baryon}}^{(0)}$, $\langle S^{(1)}_{\text{junction}}\rangle$, and the classical contribution \eqref{eq_baryon classical contribution} to the partition function, we find the four coupling constants involving the closed string states $\textbf{0}$ and $\textbf{1}$: 
\begin{equation}
\label{eq_coupling const 1}
\mathtoolsset{multlined-width=0.9\displaywidth}
\begin{multlined}
C^{(0)}_{\textbf{0}\textbf{0}\textbf{0}}=e^{-2 \pi  m_\text{j} R}\left[1+\frac{(d+1) m_\text{j} l_\text{s}^2}{36 R}+O\left(R^{-2}\right)\right]~,\hfill\\
C^{(0)}_{\textbf{0}\textbf{0}\textbf{1}}=\frac{e^{-2 \pi m_\text{j} R}}{3\sqrt{d-2}}\left[(d-4)+\frac{(d+1)(d+20)m_\text{j} l_\text{s}^2}{36 R}
+O\left(R^{-2}\right)\right]~,\hfill\\
C^{(0)}_{\textbf{0}\textbf{1}\textbf{1}}=\frac{e^{-2 \pi m_\text{j} R}}{9(d-2)}\left[(d^2-4 d+8)
+\frac{(d^3+45 d^2-236 d-88) m_\text{j} l_\text{s}^2}{36 R}
+O\left(R^{-2}\right)\right]~,\hfill\\
C^{(0)}_{\textbf{1}\textbf{1}\textbf{1}}=\frac{e^{-2 \pi m_\text{j} R}}{27(d-2)^{\frac{3}{2}}}\biggl[(d-4)(d^2+6d-19)\hfill\\
\hfill +\frac{(d^4+73d^3-544d^2+2360d-3360) m_\text{j} l_\text{s}^2}{36 R}
+O\left(R^{-2}\right)\biggr]~.
\end{multlined}
\end{equation}
These coupling constants in \eqref{eq_coupling const 1} depend on the effective string parameters $l_\text{s}$ and $m_{j}$, as well as on the radius $R$ of Polyakov loops. They are independent of the operator positions (i.e., $L_x$, $L_y$, and $L_z$), as expected from the locality of the interaction vertices in \eqref{eq_cubic coupling def}.

Notably, the $\mathbb{Z}_2$ symmetry \eqref{eq_duality defect} constrains the interaction vertices in the effective potential $V(\Phi_a, \Phi_b,\dots )$ and implies selection rules for scattering processes. For example, the closed string ground state $\textbf{0}$ is even under the $\mathbb{Z}_2$ transformation, while the excited state $\textbf{1}$ is odd. By the $\mathbb{Z}_2$ symmetry, the coupling constants in \eqref{eq_cubic coupling def} must satisfy
\begin{equation}
\label{eq_selection rule}C^{(0)}_{\textbf{0}\textbf{0}\textbf{1}}=C^{(0)}_{\textbf{1}\textbf{1}\textbf{1}}= 0~,
\end{equation}
in agreement with the explicit results \eqref{eq_coupling const 1} with $d=4$ and $m_\text{j}=0$. In general, we expect the $\mathbb{Z}_2$ symmetry to be accidental, with the baryon junction acquiring a nonzero mass $m_\text{j}\sim l_\text{s}^{-1}$. The $s$-wave scattering processes in 4 dimensions are associated with different scales, corresponding to the $\mathbb{Z}_2$ symmetry-preserving (SP) and symmetry-breaking (SB) channels:
\begin{equation}
    \Lambda^{\text{SP}}_{\text{$s$-wave}}\sim Rl_\text{s}^{-2}e^{-\frac{4\pi }{3} m_\text{j} R}~,~~\text{and}~~\Lambda^{\text{SB}}_{\text{$s$-wave}}\sim m_\text{j}^{\frac{2}{3}}R^{\frac{1}{3}}l_\text{s}^{-\frac{2}{3}}e^{-\frac{4\pi }{3} m_\text{j}R}~.
\end{equation}

Around the saddle point \eqref{eq_Fermat point}, we can incorporate non-Gaussian corrections to the Feynman integral \eqref{eq_baryon closed channel Feynman integral} through a perturbative expansion. This is carried out in \cite{Lou:2026xqr}. In particular, we find that the $s$-wave scattering amplitude corresponding to the equilateral baryon reads
\begin{equation}
\label{eq_closed channel partition function 2}
\mathtoolsset{multlined-width=0.9\displaywidth}
\begin{multlined}
   Z_\text{baryon}=\left(\frac{\pi R}{L}\right)^{d-\frac{5}{2}} \frac{e^{-6 \pi R L / l_\text{s}^2}}{2^{\frac{d-1}{2}} \tilde{\textbf{q}}^{\frac{d-2}{8}}}\sum_{a,b,c}C_{abc}^{(0)}v_{a}v_{b}v_{c}\tilde{\textbf{q}}^{n_a+n_b+n_c}\bigg\{1-\frac{l_\text{s}^2}{6\pi R^2}\bigg[ \frac{4d^2-32d+59}{4\ln\tilde{\textbf{q}}}\hfill \\
    +\frac{(2d-5)}{8}\big(d-2-8(n_a+n_b+n_c)\big)+\bigg(\frac{(d-2)^2}{64}-\frac{d-2}{4}(n_a+n_b+n_c)\\
   +6\left(n_a^2+n_b^2+n_c^2\right)-\left(n_a+n_b+n_c\right)^2\bigg) \ln \tilde{\textbf{q}}\bigg]+O\left(\partial^4\right)\bigg\}~,
\end{multlined}
\end{equation}
where we have $\tilde{\textbf{q}}_x=\tilde{\textbf{q}}_y=\tilde{\textbf{q}}_z=\tilde{\textbf{q}}$. The locality of the interaction vertices in \eqref{eq_cubic coupling def} requires the coupling constants $C_{abc}^{(0)}$ to be independent of the confining string length $L\sim \ln{\tilde{\textbf{q}}}$. For the open-closed duality to be consistent, the terms scaling as $(\ln{\tilde{\textbf{q}}})^1$ and $(\ln{\tilde{\textbf{q}}})^{-1}$ in equation \eqref{eq_closed channel partition function 2} must match exactly with those obtained from the open-channel calculation at order $2$. In the next section, we will show that this is indeed the case.

\subsection{Nonlinear corrections to the partition function}

\begin{figure}[thb]
\centering
\includegraphics[width=\textwidth]{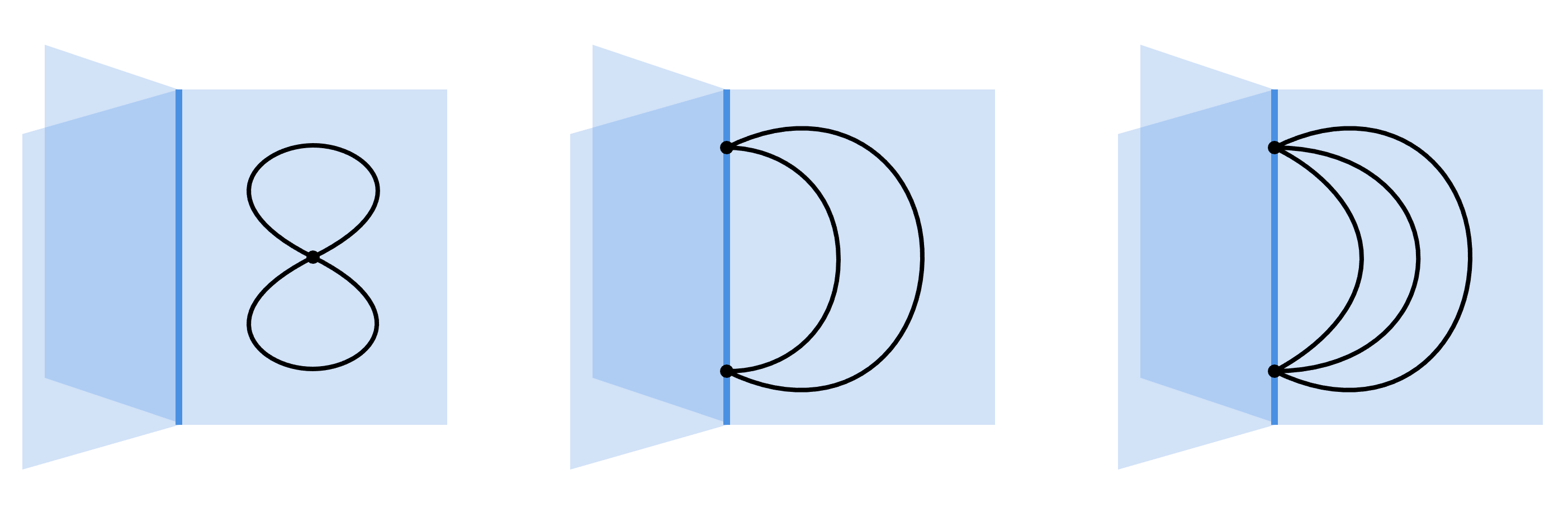}
  \caption[Worldsheet loop diagrams.]{Worldsheet loop diagrams. The black lines denote the propagators of the NGB fields $x_i$, $y_i$, and $z_i$. See appendix in \cite{Lou:2026xqr} for the explicit forms of these propagators. Left: $\langle S^{(2)}_{\rm strings}\rangle$ is given by the 2-loop diagrams on the confining string worldsheets; Middle: $\langle (S^{(1)}_{\rm junction})^2\rangle$ is given by the 1-loop diagrams on the baryon junction worldline; Right: $\langle (S^{(1)}_{\rm displacement})^2\rangle$ is given by the 2-loop diagrams on the baryon junction worldline.} \label{pic_feynman diagram}
\end{figure}

As shown in Figure \ref{pic_feynman diagram}, the order 2 nonlinear corrections to the partition function \eqref{eq_perturbative expansion} are evaluated by loop diagrams either on confining string worldsheets or localized on the baryon junction worldline. The loop integrals in $\langle S^{(2)}_{\rm displacement}\rangle$, $\langle S^{(2)}_{\rm strings}\rangle$, $\langle (S^{(1)}_{\rm junction})^2\rangle$, and $\langle (S^{(1)}_{\rm displacement})^2\rangle$ are generally UV divergent. In \cite{Lou:2026xqr}, we adopt the zeta-function regularization and extract the finite contribution from these loop integrals. Below, we present the results for equilateral baryons and comment on their implications for open-closed duality.

We start with the quartic interaction \eqref{eq_baryon displacement order 2} at the baryon junction arising from longitudinal displacements. Under zeta-function regularization, we find
\begin{equation}
\label{eq_nonlinear 1}
    \langle S_\text{displacement}^{(2)}\rangle=0~.
\end{equation}
The quartic interaction \eqref{eq_baryon strings order 2} on the confining string worldsheets follows from the Nambu--Goto action. Using zeta-function regularization once again, we obtain $\langle S^{(2)}_{\rm strings}\rangle$ in terms of the Eisenstein series
\begin{equation}
\label{eq_nonlinear 2}
\mathtoolsset{multlined-width=0.9\displaywidth}
\begin{multlined}
\langle S_\text{strings}^{(2)}\rangle=\frac{\pi l_{\text{s}}^2}{13824 L^2}(\ln \textbf{q})\Big[2(d+1) E_4(\sqrt{\textbf{q}})+4(d-4)(d-10)\big((E_2(\textbf{q})\big)^2\hfill\\
\phantom{\langle S_\text{strings}^{(2)}\rangle=\frac{\pi l_\text{s}^2}{13824 L^2}(\ln \textbf{q})\Big[}+24(d-4)E_4(\textbf{q})+(d^2-4d-1)(E_2(\sqrt{\textbf{q}}))^2\hfill\\
\hfill+4(d-1)(d-4) E_2(\sqrt{\textbf{q}}) E_2(\textbf{q})\Big]~.
\end{multlined}
\end{equation}
Here, the result is expressed in terms of $\textbf{q}=\exp{(-2\pi^2 R/L)}$. In the open channel, equation \eqref{eq_nonlinear 2} represents subleading corrections to the probe baryon energy levels $E_{\text{baryon}}$. After being weighted by the Boltzmann factor $\exp(-\beta E_{\text{baryon}})$, these corrections scale with positive powers of $\ln{\textbf{q}}\sim \beta$. Alternatively, we can recast \eqref{eq_nonlinear 2} in terms of the dual variable $\tilde{\textbf{q}}=\exp{(-2L/R)}$ using the modular transformation. We find that
\begin{equation}
\label{eq_nonlinear 2 dual}
\mathtoolsset{multlined-width=0.9\displaywidth}
\begin{multlined}
\langle S_\text{strings}^{(2)}\rangle=\frac{(4d^2-28d+47)l_{\text{s}}^2}{24\pi R^2(\ln\tilde{\textbf{q}})}\hfill\\
\phantom{\langle S_\text{strings}^{(2)}\rangle=}+\frac{l_\text{s}^2}{144\pi R^2}\Big[(d-4)(2d-11)E_2(\tilde{\textbf{q}})+2(2d^2-9d+3)E_2(\tilde{\textbf{q}}^2)\Big]\hfill\\
\phantom{\langle S_\text{strings}^{(2)}\rangle=}+\frac{l_\text{s}^2}{3456\pi R^2}(\ln\tilde{\textbf{q}})\Big[8(d+1)E_4(\tilde{\textbf{q}}^2)+4(d^2-4d-1)\big(E_2(\tilde{\textbf{q}}^2)\big)^2\hfill\\
\phantom{\langle S_\text{strings}^{(2)}\rangle=+\frac{l_\text{s}^2}{3456\pi R^2}(\ln\tilde{\textbf{q}})\Big[}+6(d-4)E_4(\tilde{\textbf{q}})+(d-4)(d-10)\big(E_2(\tilde{\textbf{q}})\big)^2\hfill\\
\hfill+4(d-1)(d-4)E_2(\tilde{\textbf{q}})E_2(\tilde{\textbf{q}}^2)\Big]~,
\end{multlined}
\end{equation}
where it splits into terms that scale as $(\ln \tilde{\textbf{q}})^{-1}$, $(\ln \tilde{\textbf{q}})^{0}$ and $(\ln \tilde{\textbf{q}})^{1}$. As we will show shortly,  the $(\ln \tilde{q})^{0}$ term in \eqref{eq_nonlinear 2} encodes corrections to the local interaction vertices $C_{abc}^{(0)}$ in the closed channel, whereas the $(\ln \tilde{\textbf{q}})^{-1}$ and $(\ln \tilde{\textbf{q}})^{1}$ terms follow from the tree-level Feynman integral \eqref{eq_baryon closed channel Feynman integral}.

The kinetic action of the baryon junction \eqref{eq_baryon junction order 1} generates the following nonlinear corrections at order $2$:
\begin{equation}
\label{eq_nonlinear 3}
\mathtoolsset{multlined-width=0.9\displaywidth}
\begin{multlined}
\langle(S_\text{junction}^{(1)})^2\rangle=\frac{(d+5) m_\text{j}^2l_{\text{s}}^4}{216 L^2} (\ln \textbf{q})\Big[2 E_2(\textbf{q})-E_2(\sqrt{\textbf{q}})\Big]\hfill\\
\phantom{\langle(S_\text{junction}^{(1)})^2\rangle=}+\frac{m_\text{j}^2l_{\text{s}}^4}{20736L^2}(\ln \textbf{q})^2\Big[2 (d+5) E_4(\sqrt{\textbf{q}})+(d^2-9)(E_2(\sqrt{\textbf{q}}))^2\hfill\\
\phantom{\langle(S_\text{junction}^{(1)})^2\rangle=+\frac{m_\text{j}^2l_{\text{s}}^4}{20736L^2}(\ln \textbf{q})^2\Big[}-8(d+5)E_4(\textbf{q})-4 (d+1)^2 E_2(\sqrt{\textbf{q}})E_2(\textbf{q})\hfill\\
\hfill +4 (d^2+4d+11) (E_2(\textbf{q}))^2\Big]~.
\end{multlined}
\end{equation}
As in \eqref{eq_nonlinear 2}, \eqref{eq_nonlinear 3} represents subleading corrections to the probe baryon energy levels, with the corresponding terms scaling as $\ln{\textbf{q}}\sim \beta$ and $(\ln{\textbf{q}})^2\sim \beta^2$. In the dual variable $\tilde{\textbf{q}}$, we find 
\begin{equation}
\label{eq_nonlinear 3 dual}
\mathtoolsset{multlined-width=0.9\displaywidth}
\begin{multlined}
\langle(S_\text{junction}^{(1)})^2\rangle=\frac{m_\text{j}^2l_{\text{s}}^4}{1296R^2}\Big[8(d+5)E_4(\tilde{\textbf{q}}^2)-2(d+5)E_4(\tilde{\textbf{q}})-4(d+1)^2 E_2(\tilde{\textbf{q}}) E_2(\tilde{\textbf{q}}^2)\hfill\\
 \hfill   +(d^2+4d+11)\bigl(E_2(\tilde{\textbf{q}})\bigr)^2+4(d^2-9)\bigl(E_2(\tilde{\textbf{q}}^2)\bigr)^2\Big]~,
\end{multlined}
\end{equation}
so that it scales as $(\ln {\tilde{\textbf{q}}})^0$, indicating that the baryon junction mass $m_\text{j}$ affects only the local interaction vertices in the effective action \eqref{eq_closed string EFT}.

Finally, the longitudinal displacements \eqref{eq_baryon displacement order 1} gives rise to the nonlinear correction
\begin{equation}
\label{eq_nonlinear 4}
\mathtoolsset{multlined-width=0.9\displaywidth}
\begin{multlined}
\langle(S_\text{displacement}^{(1)})^2\rangle=\frac{\pi l_{\text{s}}^2}{1728L^2}(\ln \textbf{q})\Big[(E_2(\sqrt{\textbf{q}}))^2+4(d-4)(E_2(\textbf{q}))^2\hfill\\
\hfill-E_4(\sqrt{\textbf{q}})-4(d-4)E_4(\textbf{q})\Big]~.
\end{multlined}
\end{equation}
Unlike the corrections from the world line kinetics (i.e., $\langle S_\text{junction}^{(1)}\rangle$ and $\langle(S_\text{junction}^{(1)})^2\rangle$), \eqref{eq_nonlinear 4} contains contributions scaling as $(\ln \tilde{\textbf{q}})^{-1}$, $(\ln \tilde{\textbf{q}})^{0}$ and $(\ln \tilde{\textbf{q}})^{1}$ when written in terms of the variable $\tilde{\textbf{q}}$:
\begin{equation}
\label{eq_nonlinear 4 dual}
\mathtoolsset{multlined-width=0.9\displaywidth}
\begin{multlined}
\langle(S_\text{displacement}^{(1)})^2\rangle=\frac{(d-3)l_\text{s}^2}{3\pi R^2(\ln\tilde{\textbf{q}})}+\frac{l_\text{s}^2}{18\pi R^2}\Big[(d-4)E_2(\tilde{\textbf{q}})+2E_2(\tilde{\textbf{q}}^2)\Big]+\frac{l_\text{s}^2}{432\pi R^2}(\ln\tilde{\textbf{q}})\hfill\\
\hfill\times \Big[4\big(E_2(\tilde{\textbf{q}}^2)\big)^2+(d-4)\big(E_2(\tilde{\textbf{q}})\big)^2-(d-4)E_4(\tilde{\textbf{q}})-4E_4(\tilde{\textbf{q}}^2)\Big]~.
\end{multlined}
\end{equation}

As a side remark, our analysis yields predictions for the baryon spectrum that can be tested in lattice simulations. In particular, we consider the ground state energy of an equilateral baryon \eqref{eq_equilateral def}. From the open-channel interpretation and results in \eqref{eq_linear 1}, \eqref{eq_linear 2}, \eqref{eq_nonlinear 1}, \eqref{eq_nonlinear 2}, \eqref{eq_nonlinear 3}, and \eqref{eq_nonlinear 4}, we find that
\begin{equation}
\label{eq_gs energy to order 2}
\mathtoolsset{multlined-width=0.9\displaywidth}
\begin{multlined}
E_{\text{GS}}=\frac{3L}{l_\text{s}^2}+m_\text{j}-\frac{(d-3)\pi}{16L}-\frac{ (d+1)\pi  m_\text{j} l_\text{s}^2}{144 L^2}\hfill\\
\hfill+\frac{(d+5)\pi m_\text{j}^2l_\text{s}^4}{432L^3}-\frac{(d-3)^2\pi^2l_\text{s}^2}{1536L^3}+O\left(L^{-4}\right)~.
\end{multlined}
\end{equation}
Notably, the quantum corrections in \eqref{eq_gs energy to order 2} are fixed by the two classical parameters (i.e., the string tension $l_\text{s}^{-2}$ and the junction mass $m_\text{j}$) up to the next-to-next-to-leading order. This is a consequence of the Poincar\'e symmetry and diffeomorphism invariance on the confining string worldsheet.

On the other hand, by matching the closed-channel interpretation \eqref{eq_closed channel partition function 2} with the results  \eqref{eq_nonlinear 1}, \eqref{eq_nonlinear 2 dual}, \eqref{eq_nonlinear 3 dual}, and \eqref{eq_nonlinear 4 dual}, we determine the two cubic coupling constants $C^{(0)}_{\textbf{0}\textbf{0}\textbf{0}}$ and $C^{(0)}_{\textbf{0}\textbf{0}\textbf{1}}$. We find that
\begin{equation}
\label{eq_coupling const 2}
\mathtoolsset{multlined-width=0.9\displaywidth}
\begin{multlined}
C^{(0)}_{\textbf{0}\textbf{0}\textbf{0}}=e^{-2 \pi m_\text{j}R }\left[1+\frac{(d+1) m_\text{j} l_\text{s}^2}{36 R}+\frac{5(d-2)l_\text{s}^2}{144\pi R^2}+\frac{(d+1)^2 m_\text{j}^2l_\text{s}^4}{2592R^2}+O\left(R^{-3}\right)\right]~,\hfill\\
C^{(0)}_{\textbf{0}\textbf{0}\textbf{1}}=\frac{e^{-2 \pi m_\text{j} R }}{3\sqrt{d-2}}\left[(d-4)+\frac{(d+1)(d+20)m_\text{j} l_\text{s}^2}{36 R}
+\frac{(d-4)(5d-178)l_\text{s}^2}{144\pi R^2}\right.\hfill\\
\left.{}+\frac{(d^3+46d^2-487d-2836)m_\text{j}^2l_\text{s}^4}{2592R^2}+O\left(R^{-3}\right)\right]~.
\end{multlined}
\end{equation}
which are independent of $L\sim \ln{\tilde{\textbf{q}}}$. In particular, we note that $C^{(0)}_{\textbf{0}\textbf{0}\textbf{1}}=O(R^{-3})$ when $d=4$ and $m_\text{j}=0$. This agrees with the selection rule \eqref{eq_selection rule} that follows from the $\mathbb{Z}_2$ symmetry \eqref{eq_duality defect} of the low-energy effective theory. 

\chapter[Defects in Atomic Quantum Gases]{Defects in Atomic Quantum Gases}

\label{cha_3}

While defects in relativistic quantum field theories exhibit rich and intriguing physics, as we have discussed extensively in Chapters \ref{cha_1} and \ref{cha_2}, experimental setups often involve nonrelativistic systems that are Galilean invariant. In particular, atomic quantum gases provide versatile platforms for realizing and probing many-body phenomena in controllable settings. Prominent examples of these nonrelativistic systems include Bose--Einstein condensates confined in traps and ultracold Fermi gases with tunable interactions. For reviews of this broad and active field of research, see, e.g., \cite{Lewenstein:2007yee, pethick2008bose, Bloch:2008zzb, Giorgini:2008zz}.

The framework of RG flow extends naturally to nonrelativistic quantum gases. At long distances, the dynamics of atomic degrees of freedom are described by quantum fields with couplings that generally run under scale transformations. Fixed points of such RG flows are commonly referred to as nonrelativistic CFTs. The Jackiw--Pi models \cite{Jackiw:1990mb, Jackiw:1990tz, Doroud:2015fsz, Doroud:2016mfv} furnish an important class of examples. These models describe the dynamics of abelian and non-abelian anyons in 2-dimensional space and exhibit emergent nonrelativistic conformal symmetry. We refer the reader to \cite{Henkel:1992xs, Nishida:2007pj, Nikolic:2007zz, Nishida:2010tm, Goldberger:2014hca, Pal:2018idc, Boisvert:2025hex} for representative works on formal aspects of nonrelativistic CFTs, and to \cite{Baiguera:2023fus} for a modern review.

In this chapter, we discuss universal aspects of defects in atomic quantum gases using the formalism developed in \cite{Raviv-Moshe:2024yzt}. We also study a concrete example with direct experimental relevance, drawn from \cite{Cuomo:2023vvd}. The sections are organized as follows. Section \ref{sec_Galilean field theories} provides a pedagogical review of Galilean symmetry and the extended Schrödinger symmetry. In Section \ref{sec_Conformal defects in sch field theories}, we study the long-distance dynamics of point-like impurities immersed in dilute quantum gases. In particular, we apply DCFT methods to conformal defects in free and unitary Fermi gases. Finally, in Section \ref{sec_Giant superfluid vortices}, we study the effective theory of superfluids that emerge from quantum gases at large particle density. We analyze the giant vortex, a macroscopic defect recently observed in superfluid experiments \cite{guo2020supersonic}.

Before concluding this preface, we highlight a change in convention: throughout this chapter, we use $d$ to denote the dimension of the space. This is clearly more convenient for nonrelativistic systems, where time and space play distinct roles. For reasons that will become clear shortly, we will also refer to nonrelativistic CFTs as Schrödinger field theories in what follows.

\section{Galilean field theories}
\label{sec_Galilean field theories}

Just as the dynamics of relativistic field theories are governed by Poincaré symmetry, the nonrelativistic systems are subject to the Galilean symmetry. In $d$-dimensional space, the simply connected part of the Galilean group is generated by spatial translations $P_\mu$, $\mathfrak{so}(d)$ spatial rotations $J_{\mu_1\mu_2}$, Galilean boosts $B_\mu $, and time translation $H$. Both the translations and boosts transform as vectors under the rotation, with the corresponding commutation relations 
\begin{equation}
\label{eq_Galilean algebra 1}
    \begin{aligned}
        [J_{\mu_1\mu_2},J_{\mu_1\mu_2}]={}&i\big(\delta_{\mu_1\mu_3}J_{\mu_2\mu_4}-\delta_{\mu_2\mu_3}J_{\mu_1\mu_4}+\delta_{\mu_1\mu_4}J_{\mu_3\mu_2}-\delta_{\mu_2\mu_4}J_{\mu_3\mu_1}\big)~;\\
        [J_{\mu_1\mu_2},P_{\mu_3}]={}&i\big(\delta_{\mu_1\mu_3}P_{\mu_2}-\delta_{\mu_2\mu_3}P_{\mu_1}\big)~;\\
        [J_{\mu_1\mu_2},B_{\mu_3}]={}&i\big(\delta_{\mu_1\mu_3}B_{\mu_2}-\delta_{\mu_2\mu_3}B_{\mu_1}\big)~.
    \end{aligned}
\end{equation}
Time translations are generated by the Hamiltonian $H$, which commutes with spatial translations and rotations. Galilean boosts, however, act as $x_\mu\to x_\mu+v_\mu t$ and therefore involve the time coordinate. This leads to the commutation relation
\begin{equation}
\label{eq_Galilean algebra 2}
    [H,B_\mu]=-iP_\mu~.
\end{equation}
With all other commutators vanishing, \eqref{eq_Galilean algebra 1} and \eqref{eq_Galilean algebra 2} define the Galilean algebra acting on spacetime coordinates.

The Galilean algebra can be extended to the Bargmann algebra by introducing a central element in the commutation relation between spatial translations and Galilean boosts:
\begin{equation}
\label{eq_Galilean algebra 3}
        [B_{\mu_1},P_{\mu_2}]= i\delta_{\mu_1\mu_2}M~.
\end{equation}
The central charge $M$ has the physical interpretation of nonrelativistic mass and generates the $U(1)$ particle number symmetry. The fact that Galilean invariant systems are associated with a $U(1)$ symmetry might be odd at first sight. Denoting the speed of light by $c$, one finds that anti-particles decouple from the spectrum in the $c\to \infty$ limit of relativistic systems. The Hilbert space consequently decomposes into superselection sectors of fixed particle number, giving rise to the $U(1)$ symmetry.

As a simple example of a Galilean field theory, we consider the action of a free complex scalar field
\begin{equation}
\label{eq_free Sch scalar}
    S_{\text{free}}=\int dtd^dx\Big(i\phi^\dagger\partial_t\phi-\frac{|\partial_\mu \phi|^2}{2m}\Big)~,
\end{equation}
where the parameter $m$ is defined by $[M,\phi^\dagger]=m \phi^\dagger$. This theory also has an enhanced scaling symmetry, which we discuss below.

\subsection{Schrödinger symmetry}

We now consider the analogue of conformal symmetry in nonrelativistic systems. The Schrödinger algebra $\mathfrak{sch}(d)$ is defined by extending the Bargmann algebra in $d$-dimensional space with dilation $D$ and special conformal transformation $C$. Notably, the dispersion relation of the free model \eqref{eq_free Sch scalar} has dynamical critical exponent $z=2$.\footnote{In Lifshitz theories \cite{Hornreich:1975zz, Ardonne:2003wa, Kachru:2008yh, Chen:2009ka,Horava:2009uw}, the dispersion relation generally takes the form $\omega\propto|k|^{z}$, where $\omega$ is the frequency and $|k|$ is the spatial wavelength.} We thus consider scaling transformations that act differently on the spatial and temporal directions. In particular, generators of the Galilean transformations are of the weight
\begin{equation}
    \label{eq_Sch algebra 1}
    [D,P_\mu]=iP_\mu~,~~[D,H]=2iH~,~~[D,B_\mu]=-iB_\mu~.
\end{equation}
When acting on local operators, $H$ and $P_\mu$ raise their scaling dimensions by $2$ and $1$, while $B_\mu$ reduces scaling dimensions by $1$. The special conformal transformation $C$ completes the Schrödinger algebra through the additional nontrivial commutation relations
\begin{equation}
\label{eq_Sch algebra 2}
    [D,C]=-2iC~,~~[H,C]=-iD~,~~[C,P_\mu]=iB_\mu~,
\end{equation}
so that it reduces scaling dimensions by $2$ when acting on local operators.

Dynamics of Schrödinger field theories are strongly constrained by the $\mathfrak{sch}(d)$ symmetry \cite{Henkel:1992xs, Nishida:2007pj, Goldberger:2014hca}. In particular, we consider the two-point functions of an operator $\mathcal{O}$ that transform as a scalar under $\mathfrak{so}(d)$. The nonrelativistic mass $m(\mathcal{O})$ and the scaling dimension $\Delta(\mathcal{O})$ of the operator is given by the commutation relations 
\begin{equation}
\begin{aligned}
[M,\mathcal{O}]={}&m(\mathcal{O}) \mathcal{O}~,\\
[D, \mathcal{O}] = {}&i(2t\partial_t+x^\mu\partial_\mu+\Delta(\mathcal{O}))\mathcal{O}~.    
\end{aligned}
\end{equation}
Parallel to the conformal descendants in relativistic theories, we can define descendants of $\mathcal{O}$ by recursively acting on it with $P_\mu$ and $H$. On the other hand, a primary operator with nonzero mass\footnote{The notion of a primary operator is ambiguous when $m(\mathcal{O})=0$, since $P_\mu$ and $B_\mu$ commute in these cases. To avoid confusion, we reserve the term bulk/defect primaries only for operators with nonzero charge under the particle-number symmetry.} satisfies $[ B_\mu, \mathcal{O} ] = [ C, \mathcal{O} ] =0$, where
\begin{equation}
\label{eq_sch primary def}
\begin{aligned}
[B_\mu, \mathcal{O} ]={}& \left(-it\partial_\mu+x_\mu m(\mathcal{O})\right)\mathcal{O}~,\\
[C, \mathcal{O}] ={}&\Big(-itx^\mu \partial_{\mu}-it^2\partial_t-it\Delta (\mathcal{O})+\frac{|x|^2}{2}m(\mathcal{O}) \Big)\mathcal{O}~.
\end{aligned}
\end{equation}

It is convenient to introduce the Wick rotation of the time coordinate $t\to-i \tau$, so that the convergence of the correlation functions becomes rapidly apparent. Let us now consider the two-point function $\langle \mathcal{O} \mathcal{O}^\dagger  \rangle$. Due to the absence of antiparticles in the spectrum, operators in Schrödinger field theories can be normally ordered without encountering singularities \cite{Pal:2018idc, Boisvert:2025hex}. It follows that, for a non-identity operator, either $\mathcal{O}|0\rangle =0$ or $\mathcal{O}^\dagger|0\rangle =0$, where $|0\rangle$ denotes the vacuum. We will assume $\mathcal{O}|0\rangle =0$ with $m(\mathcal{O}^\dagger)=-m(\mathcal{O})>0$, so that the two-point function $\langle \mathcal{O} (\tau,x)\mathcal{O}^\dagger(\tilde{\tau},\tilde{x})  \rangle \propto \Theta(\tau-\tilde{\tau})$ yields a Euclidean time ordering.

The explicit functional form of $\langle \mathcal{O} \mathcal{O}^\dagger  \rangle$ is also constrained by the translation and rotation symmetries of $\mathfrak{sch}(d)$ in a straightforward way. For primary operators, it is further fixed up to an overall constant by the conditions from \eqref{eq_sch primary def}, such that 
\begin{equation}
\label{eq_bulk primary basis}
    \langle \mathcal{O} (\tau,x)\mathcal{O}^\dagger(\tilde{\tau},\tilde{x})  \rangle=C_{\mathcal{O}\mathcal{O}^\dagger}\frac{\Theta(\tau-\tilde{\tau})}{|\tau-\tilde{\tau}|^{\Delta(\mathcal{O})}}\exp\Big(-m(\mathcal{O}^\dagger)\frac{|x-\tilde{x}|^2}{2|\tau-\tilde{\tau}|}\Big)~.
\end{equation}
By contrast, three-point and higher-point functions in Schr\"{o}dinger field theories depend on crossratios, and their functional forms can not be uniquely determined from the symmetry algebra $\mathfrak{sch}(d)$.

Another key aspect of Schr\"{o}dinger field theories is the state/operator correspondence. We introduce the oscillator frame \cite{Goldberger:2014hca}:
\begin{equation}
\label{eq_oscillator frame}
    t_{\text{osc}}=\frac{1}{\omega}\arctan{(\omega t)}~, ~~~(x_\text{osc})_\mu=\frac{x_\mu}{\sqrt{1+\omega^2 t^2}},
\end{equation}
where $\omega$ denotes the harmonic frequency. Time translations in this frame are generated by
\begin{equation}
    H_{\text{osc}}=H+\omega^2 C~.
\end{equation}
Importantly, the operator $H_{\text{osc}}$ takes the form of the quantum mechanical Hamiltonian for particles in a harmonic trap. We consider the state
\begin{equation}
\label{eq_sch state/operator correspondence 1}
|\mathcal{O}^\dagger\rangle_\text{osc} \equiv e^{-H/\omega}\mathcal{O}^\dagger(t=0,x=0)\left|0\right\rangle~,~~\text{s.t.}~~H_\text{osc} |\mathcal{O}^\dagger\rangle_\text{osc} =\omega \Delta(\mathcal{O})|\mathcal{O}^\dagger\rangle_\text{osc}~.
\end{equation}
Here, the scaling dimension of the operator $\mathcal{O}$ is interpreted as the energy of the quantum state $|\mathcal{O}^\dagger\rangle_\text{osc}$ in units of the trap frequency $\omega$. Moreover, we consider the two-point function
\begin{equation}
\label{eq_sch state/operator correspondence 2}
    \Psi_\mathcal{O}(\tau_{\text{osc}}, x_{\text{osc}})\equiv \langle 0|\mathcal{O}(\tau_{\text{osc}}, x_{\text{osc}}) |\mathcal{O}^\dagger\rangle_{\text{osc}} \propto \exp\Big(-\omega \Delta(\mathcal{O})\tau_{\text{osc}}-\omega m(\mathcal{O}^\dagger)\frac{|x_{\text{osc}}|^2}{2}\Big)~,
\end{equation}
where the Wick-rotation is $t_\text{osc}\to -i \tau_{\text{osc}}$. This is precisely the Hartree-Fock wave function of a many-body bound state in the harmonic trap. We therefore see that \eqref{eq_sch state/operator correspondence 1} and \eqref{eq_sch state/operator correspondence 2} establish the nonrelativistic state/operator correspondence. 

\subsection{Unitary Fermi gases}
As noted previously, the free model \eqref{eq_free Sch scalar} provides a simple example of a Schrödinger field theory. In this section, we introduce an interacting model with broad experimental applications; see \cite{o2002observation, Thomas:2005zz, Giorgini:2008zz, Nascimbene:2009spz} and many references therein.

Unitary Fermi gases are systems of fermionic atoms with short-range attractive interactions tuned near the an $s$-wave Feshbach resonance \cite{Tiesinga:1993zza, timmermans1999feshbach, Chin:2010crf}. For a dilute ultracold Fermi gas, we denote by $k_\text{F}$ the Fermi momentum, which sets the typical interatomic distance, and by $a_s$ the $s$-wave scattering length, which characterizes the interaction strength. In $3$-dimensional space, this system has two perturbative regimes: the Bardeen–Cooper–Schrieffer limit, where $a_s<0$ and $|a_sk_\text{F}|\ll 1$, and the Bose--Einstein condensate limit, where $a_s>0$ and $|a_sk_\text{F}|\ll 1$. In the cross-over regime, where $|a_s|\gg k_\text{F}^{-1}$, the $s$-wave scattering length exceeds the interatomic distance. In this limit, the system approaches the unitarity bound, and the $\mathfrak{sch}(3)$ Schrödinger symmetry emerges \cite{Nikolic:2007zz, Nishida:2007pj}.

Unitary Fermi gases are strongly coupled in $3$-dimensional space, which makes their dynamics particularly challenging to analyze. Nevertheless, they become perturbatively accessible when the spatial dimension $d\gtrsim 2$ or $d \lesssim 4$.

When $d\gtrsim 2$, unitary Fermi gases admit a perturbative Gross-Neveu type description. Let us now consider the case of a single-flavor Fermi gas. We represent the two spin components of the fermionic atoms by Grassmann fields $\psi_\uparrow$ and $\psi_\downarrow$. At zero particle density, the interacting Fermi gas is described by the action
\begin{equation}
\label{eq_d=2 bulk Largragian}
S_{\text{Fermi}}=\int dtd^dx\Big(\sum_{ \varsigma =\uparrow,\downarrow}\psi_{\varsigma}{}^\dagger\big(i \partial_t+\frac{\partial_\mu^2}{2m}\big)\psi_\varsigma+\frac{m}{\lambda}\Lambda^{d-2}\Phi^\dagger\Phi-\big(\Phi^\dagger \psi_\downarrow \psi_\uparrow+\text{h.c.}\big)\Big)~,
\end{equation}
where $\Lambda$ sets a UV scale, $\lambda$ is the coupling constanat, and $\Phi$ denotes an auxiliary scalar field. We also note that the equation of motion identifies $\Phi^\dagger=\lambda_\text{b}\psi_\uparrow{}^\dagger\psi_\downarrow{}^\dagger$ as a composite two-body operator. Crucially, the RG flow associated with the coupling $\lambda$ can be determined by considering loop Feynman diagrams in the two-body sector and is therefore 1-loop exact \cite{Nikolic:2007zz}. The corresponding beta function reads
\begin{equation}
\label{eq_d=2_bulk_RG_flow}
   - \beta(\lambda)= -\bar{\epsilon}\lambda-\frac{\lambda^2}{2\pi}+O({\bar\epsilon}^3)~,~~\text{where}~~\bar{\epsilon}\equiv d -2\ll 1~. 
\end{equation}
Here we have omitted geometric factors that affect only the $O({\bar\epsilon}^3)$ terms, while keeping all orders in $\lambda$. This beta function admits a stable trivial fixed point and a multicritical interacting fixed point with $\lambda_\text{fixed}=-2\pi \bar{\epsilon}+O(\bar{\epsilon}^2)$. We summarize that the scaling dimensions of the one-body and two-body operators at each fixed points as follows
\begin{equation}
\begin{aligned}
    \text{Free Fermi gas}~:{}&~~\lambda=0~,~~\Delta(\psi_{\varsigma})=\frac{d}{2}~,~~\Delta(\Phi)=d~;\\
    \text{Unitary Fermi gas}~:{}&~~\lambda=\lambda_\text{fixed}<0~,~~\Delta(\psi_{\varsigma})=\frac{d}{2}~,~~\Delta(\Phi)=2~.    
\end{aligned}
\end{equation}
For the one-body operator $\psi_{\varsigma}$, a non-renormalization theorem following from particle-number symmetry states that its scaling dimension is the same at the trivial and interacting fixed points \cite{Boisvert:2025hex}. Moreover, at the interacting fixed point, the 1-loop exactness of the RG flow gives the exact result $\Delta(\Phi)=2$.

When $d\lesssim 4$, the two-body operator $\Phi$ in the unitary Fermi gas approaches the scalar unitarity bound $\Delta\geq d/2$ \cite{Pal:2018idc}. The field $\Phi$ may therefore be regarded as a free scalar, up to perturbative corrections. Indeed, the interacting Fermi gas in this case is described by the action \footnote{To avoid double counting loop diagrams in the two-body sector, we introduce the counterterm
\begin{equation*}
   \updelta  S_{\text{Fermi}}=-\int dtd^dx\Big(\Phi^\dagger\big(i \partial_t+\frac{\partial_\mu^2}{4m}\big)\Phi\Big)~.
\end{equation*}
The resulting Feynman rules from \eqref{eq_d=4 bulk Largragian} then agree with those from \eqref{eq_d=2 bulk Largragian}.}
\begin{equation}
\label{eq_d=4 bulk Largragian}
S_{\text{Fermi}}=\int dtd^dx\Big(\sum_{ \varsigma =\uparrow,\downarrow}\psi_{\varsigma}{}^\dagger\big(i \partial_t+\frac{\partial_\mu^2}{2m}\big)\psi_\varsigma+\Phi^\dagger\big(i \partial_t+\frac{\partial_\mu^2}{4m}\big)\Phi-\frac{\Lambda^{\frac{4-d}{2}}}{m}\left(g\Phi^\dagger \psi_\downarrow \psi_\uparrow+\text{h.c.}\right)\Big)~,
\end{equation}
where $g$ is a Yukawa-type coupling constant. As in \eqref{eq_d=2_bulk_RG_flow}, the beta function associated with the 1-loop exact RG flow of $g$ reads
\begin{equation}
\label{eq_d=4 bulk coupling RG}
-\beta(|g|^2)=\epsilon |g|^2-\frac{|g|^4}{8\pi^2}+O\left(\epsilon^3\right)~,~~\text{where}~~\epsilon=4-d\ll 1~.
\end{equation}
This gives a stable interacting fixed point with $|g|_\text{fixed}=8\pi^2 \epsilon+O\left(\epsilon^2\right)$, corresponding to the unitary Fermi gas. At first sight, it may seem confusing that the unitary Fermi gas appears multicritical in \eqref{eq_d=2_bulk_RG_flow}, whereas it appears stable in \eqref{eq_d=4 bulk coupling RG}. This is resolved by noting that the model \eqref{eq_d=4 bulk Largragian} flows to the free Fermi gas under the relevant deformation
\begin{equation}
    \updelta S_{\text{Fermi}}=-\upmu \int dtd^dx\, \Phi^\dagger \Phi~,
\end{equation}
where $\upmu$ denotes the chemical potential for the particle represented by the field $\Phi$.

\section{Conformal defects in Schrödinger field theories}
\label{sec_Conformal defects in sch field theories}

To understand the dynamics of a heavy defect immersed in atomic quantum gases, we now develop the DCFT formalism for Schrödinger field theories. In particular, we focus on defects that are point-like in space and extended along the time direction. Physically, these defects can model impurity atoms pinned by optical lattices and coupled locally to the ambient bath. See \cite{knap2012time, spethmann2012dynamics, bauer2013realizing} for a limited selection of references on this broad research area.

The microscopic details of the defect and the ambient quantum gas vary from one experimental setup to another. Our goal is to determine what universalities, if any, emerge at long distances. We first note that the maximal conformal subalgebra preserved by a point-like impurity is
\begin{equation}
\label{eq_defect sch symmetry}
    \mathfrak{so}(2,1)\times \mathfrak{so}(d) \subset \mathfrak{sch}(d)~.
\end{equation}
This is precisely the same conformal algebra \eqref{eq_DCFT symmetry} of line defects in relativistic CFTs. We therefore adopt the same conventions as in Section \ref{sec_conformal defects} and denote a defect primary of transverse spin $s$ by $\hat{\mathcal{O}}^s$. Additionally, we assume that the defect preserves the particle number symmetry $U(1)$, thereby excluding local sources or sinks of atoms.

\subsection{Bulk-to-defect OPE}

We begin by setting our conventions. Without loss of generality, we take the defect to be located at $x_\mu=0$. We also perform the Wick rotation to the time coordinate, so that $t\to -i\tau$. We normalize the defect operator $\hat{\mathcal{O}}^{s}$ such that its two-point function takes the form
\begin{equation}
\label{eq_sch defect primary basis}
    \langle \hat{\mathcal{O}}^s(\tau)\hat{\mathcal{O}}^{s}{}^{\dagger}(\tilde{\tau})  \rangle=\frac{\Theta(\tau-\tilde{\tau})}{|\tau-\tilde{\tau}|^{\Delta(\hat{\mathcal{O}}^s)}}~,
\end{equation}
where the Euclidean time ordering follows from the $U(1)$ particle number symmetry.

The Schrödinger symmetry imposes strong constraints on the bulk-to-defect two-point function $\langle \mathcal{O} \hat{\mathcal{O}}^{s}{}^\dagger\rangle $. The argument closely parallels the derivation of the relativistic result \eqref{eq_bulk scalar to defect scalar}. By the particle number symmetry, $\langle \mathcal{O} \hat{\mathcal{O}}^{s}{}^\dagger\rangle=0$ unless $m(\mathcal{O} )=m(\hat{\mathcal{O}}^{s})$, which we assume from now on. Using Ward identities following from \eqref{eq_sch primary def}, we further fix the two-point function of primary operators up to an overall coefficient:
\begin{equation}
\label{eq_sch bulk-defect two point function}
    \langle \mathcal{O} (\tau,x ) \hat{\mathcal{O}}^{s}{}^\dagger (\tilde{\tau})  \rangle=C_{\mathcal{O}}{}^{\hat{\mathcal{O}}^s}\frac{\Theta(\tau-\tilde{\tau})Y_{s}(x_\mu/|x|)}{|\tau-\tilde{\tau}|^{\Delta(\hat{\mathcal{O}}^s)}|x|^{\Delta(\mathcal{O})-\Delta(\hat{\mathcal{O}}^s)}}\exp\Big(-\frac{m(\mathcal{O}^\dagger)|x|^2}{2|\tau-\tilde{\tau}|}\Big)~.
\end{equation}
Here, $C_{\mathcal{O}}{}^{\hat{\mathcal{O}}^s}$ is the corresponding bulk-to-defect OPE coefficient. The two-point function \eqref{eq_sch bulk-defect two point function} implies that, as the bulk scalar operator $\mathcal{O}$ approaches the defect at $x_\mu=0$, its admits the following decomposition
\begin{equation}
\label{eq_Sch bulk-to-defect OPE}
    \mathcal{O} (\tau,x )=\sum_{\hat{\mathcal{O}}^s}\frac{Y_{s}(x_\mu/|x|)}{|x|^{\Delta(\mathcal{O})-\Delta(\hat{\mathcal{O}}^s)}}C_{\mathcal{O}}{}^{\hat{\mathcal{O}}^s}\sum_{n\geq 0}\frac{(-\frac{m(\mathcal{O}^\dagger)}{2}|x|^2\partial_\tau)^n}{n!(\Delta(\hat{\mathcal{O}}^s))_n} \hat{\mathcal{O}}^s(\tau)~.
\end{equation}
A careful treatment of the convergence of the operator equation \eqref{eq_Sch bulk-to-defect OPE}, as well as the associated crossing equation, lies beyond the scope of our discussion. We refer the reader to \cite{Goldberger:2014hca, Boisvert:2025hex} for relevant studies.

Under the state/operator correspondence, the defect operator $\hat{\mathcal{O}}^s$ is mapped to an energy eigenstate in a harmonic trap deformed by delta-function potentials at the center. This can be seen as follows. Since the defect preserves the $\mathfrak{so}(2,1)$ conformal algebra, we can construct defect states in the oscillator frame \eqref{eq_oscillator frame}, in direct analogy with \eqref{eq_sch state/operator correspondence 1}, as follows:
\begin{equation}
|\hat{\mathcal{O}}^s{}^\dagger\rangle_\text{osc} \equiv e^{-H/\omega}\hat{\mathcal{O}}^s{}^\dagger(t=0)\left|0\right\rangle~,~~\text{s.t.}~~H_\text{osc} |\hat{\mathcal{O}}^s{}^\dagger\rangle_\text{osc} =\omega \Delta(\hat{\mathcal{O}}^s)|\hat{\mathcal{O}}^s{}^\dagger\rangle_\text{osc}~.
\end{equation}
The scaling dimensions of defect operators therefore correspond to energy levels measured in units of the harmonic frequency $\omega$, as in the case of bulk operators.

In the oscillator frame, the two-point function of primary operators \eqref{eq_sch bulk-defect two point function}  takes the form
\begin{equation}
\label{eq_defect wave function}
\begin{aligned}
\Psi_{\hat{\mathcal{O}}^s}(\tau_\text{osc},x_{\text{osc}}) {}&\equiv \langle0| \mathcal{O}(\tau_\text{osc},x_\text{osc}) |\hat{\mathcal{O}}^s\rangle_\text{osc}\\
{}&\propto \frac{Y_s(x_\mu/|x|)}{|x_\text{osc}|^{\Delta(\mathcal{O})-\Delta(\hat{\mathcal{O}}^s)}}\exp\Big(-\omega\Delta(\hat{\mathcal{O}}^s)\tau_\text{osc}-\omega m(\mathcal{O}^\dagger)\frac{|x_\text{osc}|^2}{2}\Big)~.  
\end{aligned}
\end{equation}
This is precisely the Hartree-Fock wave function of a bound state with angular momentum $s$ and falloff $|x_\text{osc}|^{\Delta(\hat{\mathcal{O}}^s)-\Delta(\mathcal{O})}$ near the trap center. What about the defect descendant operators $\partial_\tau^{n}\hat{\mathcal{O}}^s$? Under the state/operator correspondence, these operators are mapped to excitations of the breathing modes \cite{Werner:2006zz, Werner:2006zzb}. The creation $(+)$ and annihilation $(-)$ operators for these modes in the oscillator frame are given by
\begin{equation}
    L^\pm_{\text{osc}}\equiv\frac{1}{2\omega}H-\frac{\omega}{2}C\pm iD~.
\end{equation}

\subsection{One-body operators and $s$-wave resonance}

We now discuss general properties of defect operators with unit charge under the $U(1)$ particle number symmetry. For concreteness, we present the argument in the Fermi gas model \eqref{eq_d=2 bulk Largragian}: 
\begin{equation}
\label{eq_example largragian}
S_{\text{Fermi}}=\int d\tau d^dx\Big(\sum_{ \varsigma =\uparrow,\downarrow}\psi_{\varsigma}{}^\dagger\big( \partial_\tau-\frac{\partial_\mu^2}{2m}\big)\psi_\varsigma+\frac{\lambda}{m}\Lambda^{2-d}\psi_\uparrow{}^\dagger\psi_\downarrow{}^\dagger\psi_\downarrow\psi_\uparrow\Big)~.
\end{equation}
It follows from the absence of anti-particles that correlation functions of one-body operators, such as
$\langle \psi_{\varsigma}\psi_{\tilde{\varsigma}}{}^\dagger\rangle$,
are independent of the bulk coupling $\lambda$. We can thus work either in the free Fermi gas with $\lambda=0$, in the unitary Fermi gas with $\lambda=\lambda_\text{fixed}<0$, or in more complicated models with additional matter fields. In the absence of defects, we find
\begin{equation}
\label{eq_free field propagator}
\begin{aligned}
    &\langle \psi_{\varsigma}(\tau,x )\psi_{\tilde{\varsigma}}{}^\dagger(\tilde{\tau},\tilde{x})\rangle=\delta_{\varsigma\tilde{\varsigma}}G(\tau-\tilde{\tau},x-\tilde{x};m)~,\\
    &\text{where}~~G(\tau,x;m)\equiv \Theta(\tau)\left(\frac{m}{2\pi \tau}\right)^{d/2}\exp\Big(-\frac{m |x|^2}{2\tau}\Big)~.
\end{aligned}
\end{equation}

A simple example of the defect preserving the $SU(2)$ spin rotation symmetry is given by
\begin{equation}
\label{eq_sch defect action example}
    S_{\text{impurity}}=\frac{\hat{\lambda}}{m}\Lambda^{2-d}\int_{x=0} d\tau (\psi_{\uparrow}{}^\dagger\psi_{\uparrow}+\psi_{\downarrow}{}^\dagger\psi_{\downarrow})~,
\end{equation}
which can be viewed as a local spike in the chemical potential, or equivalently as the two-body interaction between the fermionic atoms and the heavy impurity. More generally, one may include other interaction vertices or additional degrees of freedom localized on the defect. For the one-body sector, however, it suffices to consider the bilinear deformation in \eqref{eq_sch defect action example}.

We denote the defect primaries with nonrelativistic mass $m$ by $\psi_{\varsigma}^{s}{}^{\dagger}$ and consider the bulk-to-defect two-point function $\langle \psi_\varsigma \psi_{\tilde{\varsigma}}^{s}{}^{\dagger}\rangle $. As for correlation functions of bulk one-body operators, applying the equation of motion to the $\psi_\varsigma$ field yields
\begin{equation}
\label{eq_sch 1bd constraint}
    \big( \partial_\tau-\frac{\partial_\mu^2}{2m}\big)\langle\psi_\varsigma (\tau,x)\psi_{\tilde{\varsigma}}^{s}{}^{\dagger} (\tilde{\tau})\rangle\propto C_{\psi_\varsigma}{}^{\psi_{\tilde{\varsigma}}^s}\big(\Delta (\psi_{\varsigma}^{s})+\frac{d}{2}+s-2\big)\big(\Delta (\psi_{\varsigma}^{s})-\frac{d}{2}-s\big)=0~,
\end{equation}
where we have used \eqref{eq_sch bulk-defect two point function}. We therefore conclude that defect primaries with nonzero OPE coefficients must have either
$\Delta(\hat{\psi}_{\varsigma}^{s})=\frac{d}{2}+s$ or $\Delta(\hat{\psi}_{\varsigma}^{s})=2-\frac{d}{2}-s$. This argument runs in close parallel to the analysis of DCFTs in relativistic free theories \cite{Lauria:2020emq, Herzog:2022jqv, Shao:2025qvf}. In nonrelativistic theories, however, the $U(1)$ particle number symmetry allows us to extend the argument to interacting cases.

We refer to defect RG fixed points with
$\Delta(\hat{\psi}_{\varsigma}^{0})=2-\frac{d}{2}$ and $C_{\psi_\varsigma}{}^{\psi_{\tilde{\varsigma}}^0}\neq 0$
as $s$-wave resonances. As in Section \ref{sec_Matching UV and IR defects}, the unitarity of the $\mathfrak{so}(2,1)$ conformal algebra requires $\Delta(\hat{\psi}_{\varsigma}^{s})\geq 0$. We thus find that $s$-wave resonances only exist when the spatial dimension $d\leq 4$. In the model \eqref{eq_sch defect action example}, the $s$-wave resonance fixed point is obtained by fine-tuning the coupling $\hat{\lambda}$ associated with the defect bilinear deformation to criticality. In this case, the bulk two-point function $\langle \psi_{\varsigma}\psi_{\tilde{\varsigma}}{}^\dagger\rangle$ in the presence of a resonant conformal defect at $x_\mu=0$ is given by
\begin{equation}
\label{eq_exact defect propagator}
\begin{aligned}
    &\langle \psi_{\varsigma}(\tau,x )\psi_{\tilde{\varsigma}}{}^\dagger(\tilde{\tau},\tilde{x})\rangle=\delta_{\varsigma\tilde{\varsigma}}\Big(G(\tau-\tilde{\tau},x-\tilde{x};m)+\updelta G(\tau-\tilde{\tau},x,\tilde{x};m)\Big)~,~~\text{where}\\
    &\updelta G(\tau,x,\tilde{x};m)=-\Theta(\tau)  \frac{m \sin \big(\frac{d\pi}{2}\big) \Gamma \big(\frac{d}{2}\big) }{\pi^{\frac{d}{2}+1} \tau (|x||\tilde{x}|)^{\frac{d}{2}-1} }K_{1-\frac{d}{2}}\big(m |x||\tilde{x}|/\tau\big)\exp\Big(-\frac{m
   }{2\tau}(|x|^2+|\tilde{x}|^2)\Big) ~.
\end{aligned}
\end{equation}

Our analysis of one-body operators applies to a broad class of conformal defects in Schrödinger field theories with local interactions. An important class of exceptions is provided by Schrödinger field theories with inverse-square nonlocal potentials \cite{deAlfaro:1976vlx, Kaplan:2009kr, Moroz:2009nm}. Such potentials modify the equation of motion \eqref{eq_sch 1bd constraint}; we leave these exceptional cases for future work.

\subsection{Defects in free and unitary Fermi gases}

While the bulk interaction can be neglected for the dynamics of one-body operators, this is no longer true for many-body operators. In this section, we analyze the two-body operators in the defect model \eqref{eq_sch defect action example}.

\begin{figure}[thb]
\centering
  \includegraphics[width=\textwidth ]{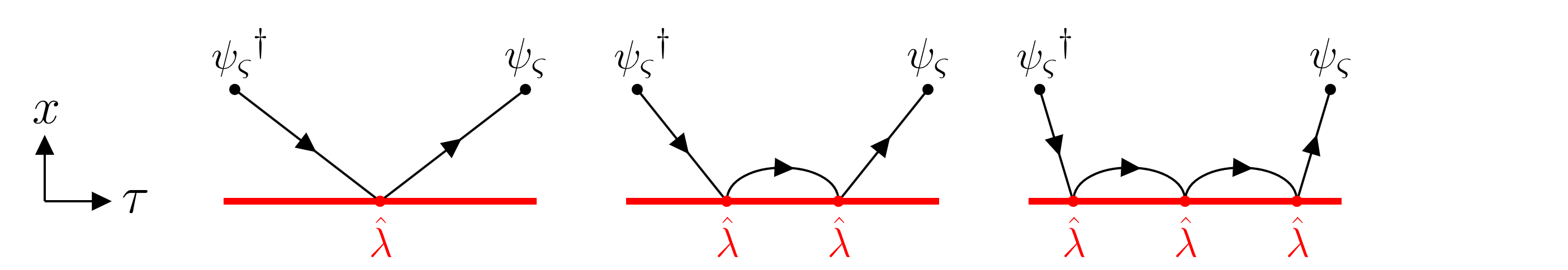}
  \caption[Ladder diagrams induced by the defect bilinear deformation.]{\label{pic_defect chemical potential} Ladder diagrams induced by the defect bilinear deformation. The red line denotes the defect extending in the time direction, while the directed black lines represent the free-field propagators in \eqref{eq_free field propagator}.}
\end{figure}

We begin by considering the defect immersed in a free Fermi gas. The RG flow of the defect coupling $\hat{\lambda}$ associated with the bilinear deformation \eqref{eq_sch defect action example} is determined by the loop diagrams in Figure \ref{pic_defect chemical potential}. Importantly, these loop contributions form a recursive set of ladder diagrams as a consequence of the $U(1)$ particle-number symmetry. This indicates the one-loop exactness of the defect RG flow, and we find the beta function
\begin{equation}
\label{eq_d=2 defect coupling RG}
-\beta(\hat{\lambda})=-\bar{\epsilon} \hat{\lambda}-\frac{\hat{\lambda}^2}{\pi }+O\left(\bar{\epsilon}^3\right)~,~~\text{where}~~\bar{\epsilon}\equiv d -2\ll 1~. 
\end{equation}
As in \eqref{eq_d=2_bulk_RG_flow}, we have omitted geometric factors at $O\left(\bar{\epsilon}^3\right)$, while keeping all orders in $\hat{\lambda}$. The beta function \eqref{eq_d=2 defect coupling RG} admits a stable trivial fixed point and a multicritical fixed point with $\hat{\lambda}_\text{fixed}=-\pi \bar{\epsilon}+O(\bar{\epsilon}^2)$. The latter fixed point, with $\hat{\lambda}=\hat{\lambda}_\text{fixed}<0$, is precisely the $s$-wave resonance, for which the bulk two-point function is given by \eqref{eq_exact defect propagator}.

Let us consider the defect primaries appearing in the OPE \eqref{eq_Sch bulk-to-defect OPE} of the two-body operator $\Phi\propto \psi_{\downarrow}\psi_{\uparrow}$. For the free Fermi gas, the theory remains Gaussian with the defect bilinear deformation. Denoting these two-body defect primaries by $\hat{\Phi}^s$, we find the scaling dimensions
\begin{equation}
\label{eq_sch 2bd op dim}
    \begin{aligned}
    \text{Free Fermi gas}~:~~{}&\Delta(\hat{\Phi}^{0})=2\Delta(\hat{\psi}_{\varsigma}^0)=4-d~;\\
    {}&\Delta(\hat{\Phi}^{s\geq 1})=\Delta(\hat{\psi}_{\varsigma}^s)+\Delta(\hat{\psi}_{\varsigma}^0)=2+s~.
    \end{aligned}
\end{equation}
Under the state/operator correspondence, this implies that the two-body wave function is given by the Slater determinant of the corresponding one-body wave functions. The energy of the two-body state is therefore simply the sum of the one-body energies. We also note that $\Delta(\hat{\Phi}^{s\geq 1})$ gives the energy of the spinning two-body ground state, where one particle is bound to the impurity and the other rotates around it.

We next consider the defect immersed in a unitary Fermi gas. Since the defect RG flow \eqref{eq_d=2 defect coupling RG} can be determined from one-body operators, it is independent of bulk interactions and consequently
remains one-loop exact in the unitary Fermi gas. In what follows, we focus on the $s$-wave resonance with the critical defect coupling $\hat{\lambda}=\hat{\lambda}_\text{fixed}<0$.

\begin{figure}[thb]
\centering
  \includegraphics[width=\textwidth ]{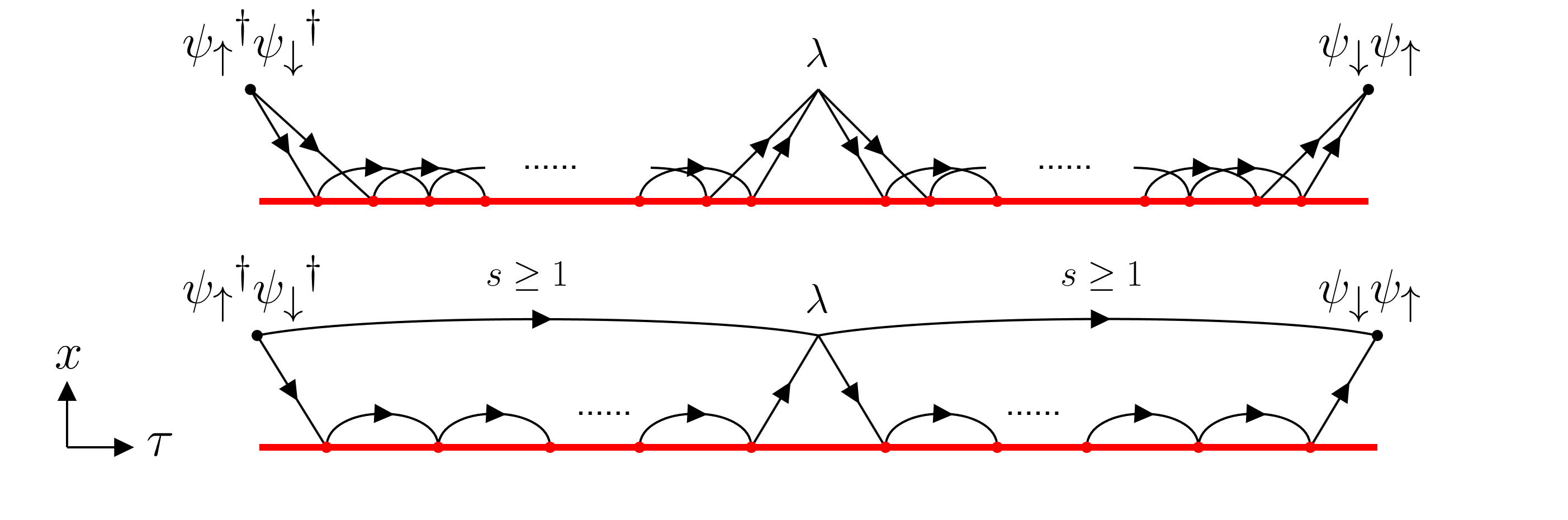}
  \caption[Feynman diagrams of the defect two-body operator for $d \gtrsim 2$.]{\label{pic_d=2 defect 1-loop} Leading Feynman diagrams contributing to the anomalous dimension of the defect two-body operator for $d \gtrsim 2$. Here and in Figure \ref{pic_d=4 defect 1-loop}, we include the infinite ladder diagrams induced by the defect bilinear deformation.}
\end{figure}

As we reviewed in Section \ref{sec_Galilean field theories}, the interaction between fermionic atoms becomes perturbative when the spatial dimension $d\gtrsim 2$ or $d\lesssim 4$. We first use the description \eqref{eq_d=2 bulk Largragian}, where $\bar{\epsilon}=d-2\ll 1$ and the bulk coupling $\lambda=O(\bar{\epsilon})$ is small. In the presence of the defect, the leading $O(\lambda)$ loop correction to the two-point function $\langle \Phi\Phi^\dagger\rangle$ is given by the Feynman diagrams in Figure \ref{pic_d=2 defect 1-loop}. The corresponding Feynman integrals are collected in the appendices of
\cite{Raviv-Moshe:2024yzt}. In the limit where the operator $\Phi$ approaches the defect, we expand the two-point function using the bulk-to-defect OPE \eqref{eq_Sch bulk-to-defect OPE}. We identify the perturbed scaling dimensions of the two-body defect primaries as follows
\begin{equation}
\label{eq_sch 2bd defect dimension 1}
    \begin{aligned}
    \text{Unitary Fermi gas at }d=2+\bar{\epsilon}~:~~{}&\Delta(\hat{\Phi}^{0})=2-2\bar{\epsilon}+O\left(\bar{\epsilon}^2\right)~,\\
    {}&\Delta(\hat{\Phi}^{s\geq 1})=2+s-\frac{\bar{\epsilon}}{2^{s-1}}+O\left(\bar{\epsilon}^2\right)~.
    \end{aligned}
\end{equation}
Interestingly, we find that corrections to the energy levels of the spinning two-body state are exponentially suppressed at large spin $s$. This is because these corrections are controlled by the wave-function overlap between the particle bound to the impurity and the particle rotating around it, which becomes exponentially
small at large $s$.

\begin{figure}[thb]
\centering
\includegraphics[width=\textwidth ]{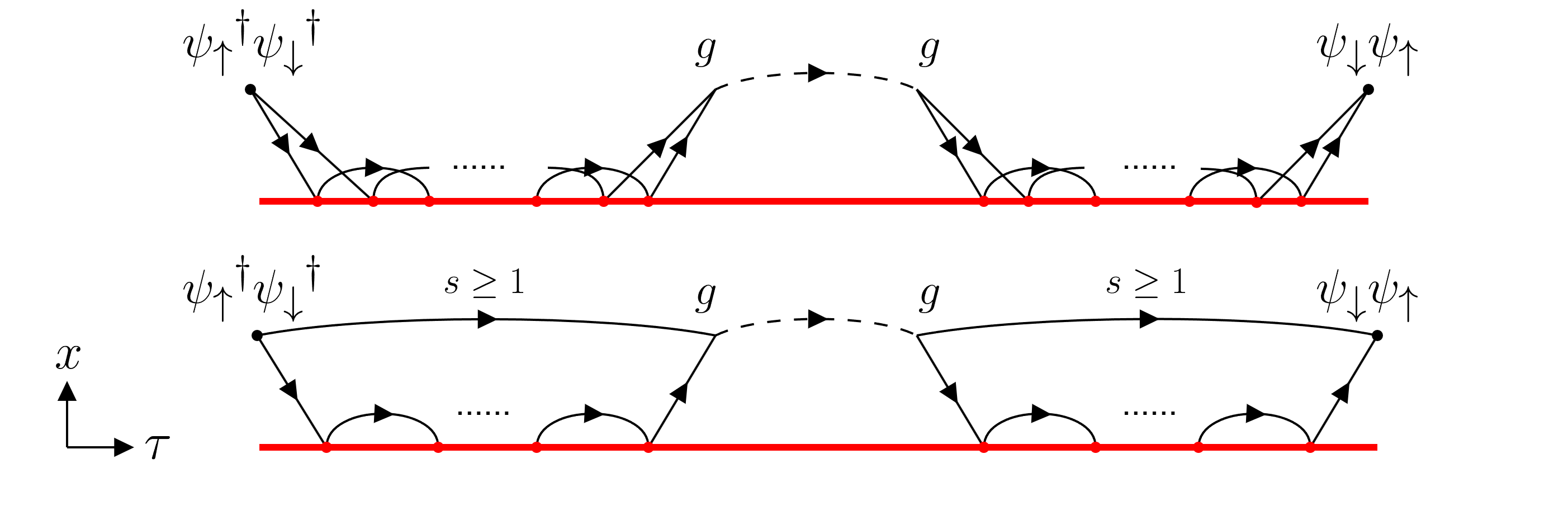}
  \caption[Feynman diagrams of the defect two-body operator for $d \lesssim 4$.]{\label{pic_d=4 defect 1-loop} Leading Feynman diagrams contributing to the anomalous dimension of the defect two-body operator for $d \lesssim 4$.}
\end{figure}

When $\epsilon=4-d\ll 1$, we instead use the description
\eqref{eq_d=4 bulk Largragian}, where the bulk coupling satisfies $|g|^2=O(\epsilon)$. The scaling dimensions of the two-body defect primaries can be computed using the
same approach as in \eqref{eq_sch 2bd defect dimension 1}, with the $O(|g|^2)$ loop Feynman diagrams shown in Figure \ref{pic_d=4 defect 1-loop}. We also refer the reader to the appendices of \cite{Raviv-Moshe:2024yzt} for a detailed treatment of the associated Feynman integrals. We find
\begin{equation}
\label{eq_sch 2bd defect dimension 2}
    \begin{aligned}
    \text{Unitary Fermi gas at }d=4-\epsilon~:~~{}&\Delta(\hat{\Phi}^{0})=\frac{19}{3}\epsilon+O\left(\epsilon^2\right)~,\\
    {}&\Delta(\hat{\Phi}^{s\geq 1})=2+s+O\left(\epsilon^2\right)~.
    \end{aligned}
\end{equation}
Under the state/operator correspondence, \eqref{eq_sch 2bd defect dimension 2} implies that the energy levels of spinning two-body states remain unperturbed at
$O(\epsilon)$. Qualitatively, this reflects the small wave-function overlap at large spatial dimensions.

\section{Giant superfluid vortices}
\label{sec_Giant superfluid vortices}

In the previous section, we focused on atomic quantum gases at zero density. Experimental setups, however, typically involve quantum gases at finite particle density confined in traps. For bosonic systems, the $U(1)$ particle-number symmetry is often spontaneously broken at finite density. This occurs, for example, in ${}^4\text{He}$ and in trapped alkali-metal gases, which enter a superfluid phase at low temperatures. See~\cite{fetter2001vortices, pethick2008bose, Schmitt:2014eka}, among many others, for reviews and references.

Vortices appear in the superfluid as the trap rotates. In 2-dimensional space, vortices are point-like defects in the particle condensate that carry angular momentum. As we increase the angular frequency of the trap, they form an Abrikosov lattice \cite{abo2001observation} and substantially change the superfluid dynamics.

Interestingly, superfluids exhibit new universalities at sufficiently high angular frequencies of the trap. It was predicted theoretically \cite{fischer2003vortex,aftalion2004giant} and observed experimentally \cite{guo2020supersonic} that a rapidly rotating superfluid forms a dynamical ring, which we investigate in the remainder of this section.

\subsection{Rapidly rotating superfluids}

We begin by reviewing the low-energy effective theory of Galilean invariant superfluids in 2-dimensional space. Physically, we consider a Bose–Einstein condensate confined in a trap. At low energies, we neglect fluctuations of the condensate density and assume that the only remaining degree of freedom is the phase of the condensate wave function. Denoted by $\varphi\sim \varphi+2\pi$, this phase is identified with the NGB of the $U(1)$ particle number symmetry.

To the lowest order in derivatives, the effective action of the condensate is written in terms of $\varphi$ via the combination~\cite{Son:2002zn, Son:2005rv}
\begin{equation}
\label{eq_functional argument}
    \textbf{X}\equiv D_t \varphi-\frac{(D_\mu\varphi)^2}{2m}~.
\end{equation}
where $m$ is both the mass of the microscopic constituents and the central extension of the Galilean algebra \eqref{eq_Galilean algebra 3}. The NGB field $\varphi$ couples to the background potential $\textbf{V}(x)$ via the covariant derivative $D_t\varphi=\partial_t \varphi-A_t$ and  $D_\mu\varphi=\partial_\mu \varphi-A_\mu$. In the static gauge, $A_t=\textbf{V}(x)$ and $A_\mu=0$. Accrodingly,
\begin{equation}
    \textbf{X}=\partial_t \varphi-\frac{(\partial_\mu\varphi)^2}{2m}-\textbf{V}(x)~.
\end{equation}

The low-energy effective action of the superfluid is given by a functional of $\textbf{X}$ with no additional derivatives,
\begin{equation}
 \label{eq_Eofstate}
S_\text{superfluid}=\int dtd^2x\, \textbf{P}(\textbf{X})~.
\end{equation}
Given a classical background for the NGB field, $ \textbf{P}(\textbf{X}_\text{cl})$ is identified with the thermodynamic pressure at chemical potential $\textbf{X}_\text{cl} $. The functional form of $P(X)$ generally depends on the UV details of the superfluid, and it could be complicated even for weakly coupled microscopic models \cite{Nicolis:2023pye}. In particular, for Schrödinger field theories in $d$-dimensional space,
\begin{equation}
    \textbf{P}(\textbf{X}) \propto \textbf{X}^{\frac{d+2}{2}}~.
\end{equation}
This is analogous to the effective actions of conformal superfluids \cite{Hellerman:2015nra, Monin:2016jmo, Cuomo:2017vzg, Cuomo:2022kio}, which capture universalities in the large charge sector of relativistic CFTs. In the following, we will often use the 2-dimensional Schrödinger superfluids with $\textbf{P}(\textbf{X}) \propto \textbf{X}^{2}$ as examples.

The superfluid action \eqref{eq_Eofstate} yield the equation of motion 
\begin{equation}
\label{eq_EOM}
 m\partial_t \textbf{P}'(\textbf{X}) -\partial^\mu \big(\textbf{P}'(\textbf{X})\partial_\mu\varphi\big)=0~.
\end{equation}
This equation admits classical solutions that take the form 
\begin{equation}
    \varphi_\text{cl}=\upmu t~~\Rightarrow ~~\textbf{X}_\text{cl}=\upmu -\textbf{V}(x)~,
\end{equation}
where $\upmu$ is the chemical potential. The parameter $\upmu$ controls the total particle number in the trap, with the local particle density given by $\textbf{P}'(\textbf{X})$. Note that it is not physically meaningful to allow $\textbf{P}'(\textbf{X})$ to attain negative values -- this restricts the domain of integration in~\eqref{eq_Eofstate} to where $\textbf{P}'(\textbf{X})\geq 0$. We further assume that $\textbf{P}'(\textbf{X})= 0$ for $\textbf{X}=0$. Physically, this means that the superfluid is confined to the region where $V(x)\leq \upmu$. For finite-size condensates, we require that no particles flow through the boundary of the superfluid. To the leading orders in derivatives, this condition reads
\cite{Hellerman:2020eff, Cuomo:2021cnb} 
\begin{equation}
   \textbf{P}'(\textbf{X})\, n^\perp_\mu\partial^\mu \varphi=0~,
\end{equation}
where $n^\perp_\mu$ is the unit vector transverse to the boundary.

When the trap is axisymmetric, the equation of motion \eqref{eq_EOM} also admits another set of solutions. We introduce polar coordinates $r\geq 0$ and $\theta\sim \theta+2\pi$, so that the axisymmetric trap $\textbf{V}(x)=\textbf{V}(r)$. We consider solutions with a single vortex at the center of the trap:
\begin{equation}
\label{eq_GV solution}
    \varphi_{\text{cl}}=\upmu t-J \theta~~\Rightarrow ~~\textbf{X}_\text{cl}=\upmu-\frac{J^2}{2mr^2}-\textbf{V}(r)~,
\end{equation}
where the vorticity $J\in\mathbb{Z}$ denotes the angular momentum carried by the superfluid. 

The core size of conventional vortices is set by a UV scale of the Bose--Einstein condensate, known as the healing length $\xi_{\text{h}}$. See \cite{Dalfovo:1999zz} and references therein. For small $J$, the vortex solution in \eqref{eq_GV solution} has a core size of order $ J \xi_{\text{h}}$ and can be analyzed using the method of \cite{Horn:2015zna}. The superfluid rotates more rapidly as $J$ increases. Consequently, the vortex core grows until its size becomes comparable to the size of the trap. In the limit $J\gg 1$, the superfluid is concentrated in a narrow annulus between the vortex core and the edge of the trap. We refer to vortices in this limit as giant vortices.

In the presence of a giant vortex, the superfluid system becomes effectively 1-dimensional. Do rapidly rotating superfluids exhibit universal low-energy dynamics? In the remainder of this section, we show that the dynamics of the superfluid circulating around the giant vortex are described by a Warped Conformal Field Theory (WCFT) \cite{Hofman:2011zj, Detournay:2012pc, Jensen:2017tnb}, up to perturbative corrections.

\subsection{Giant vortices in a hard-wall trap}

\begin{figure}[th]
\includegraphics[width=0.25\textwidth ]{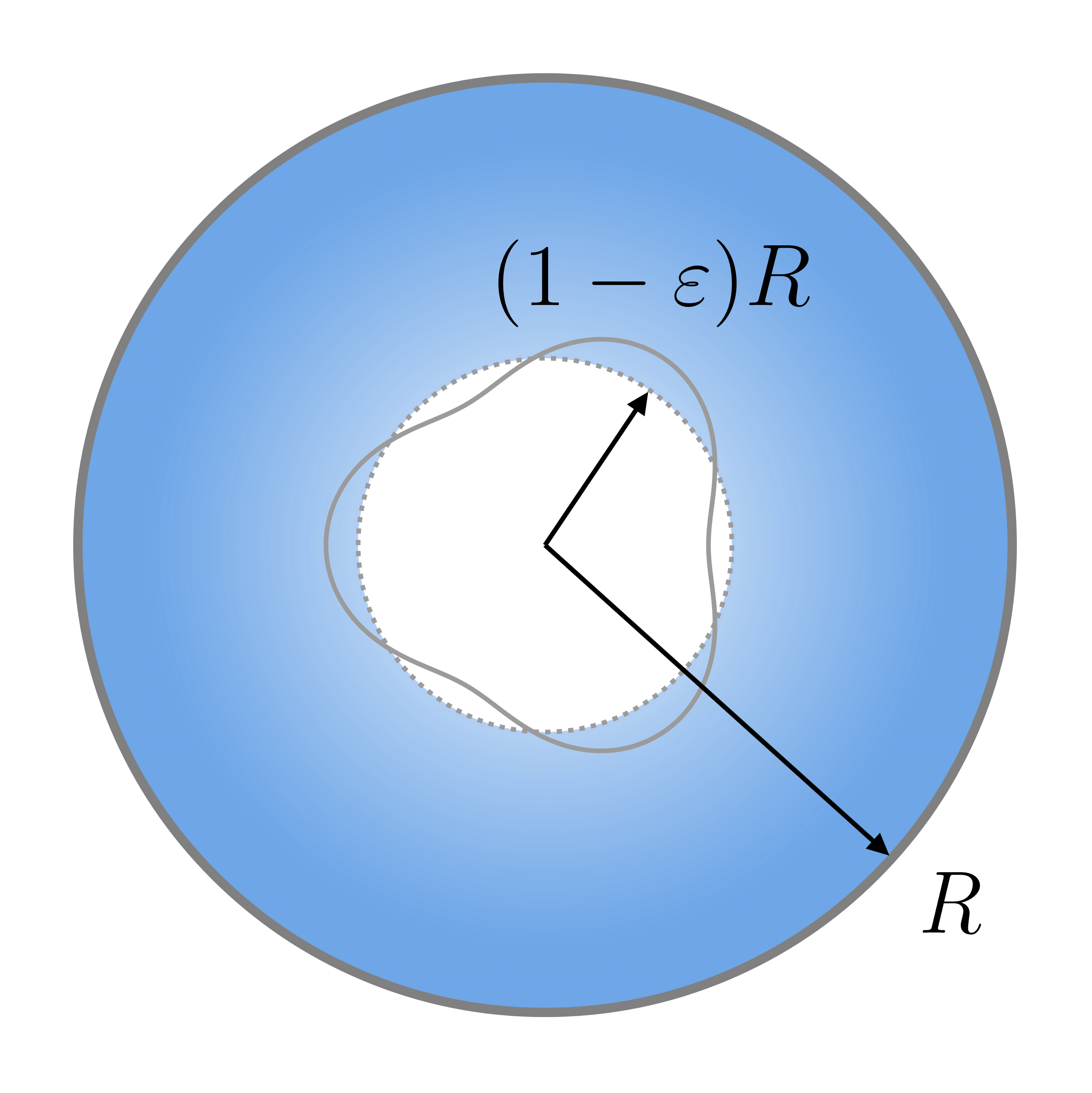}
\hspace{0.01\textwidth}
\includegraphics[width=0.73\textwidth ]{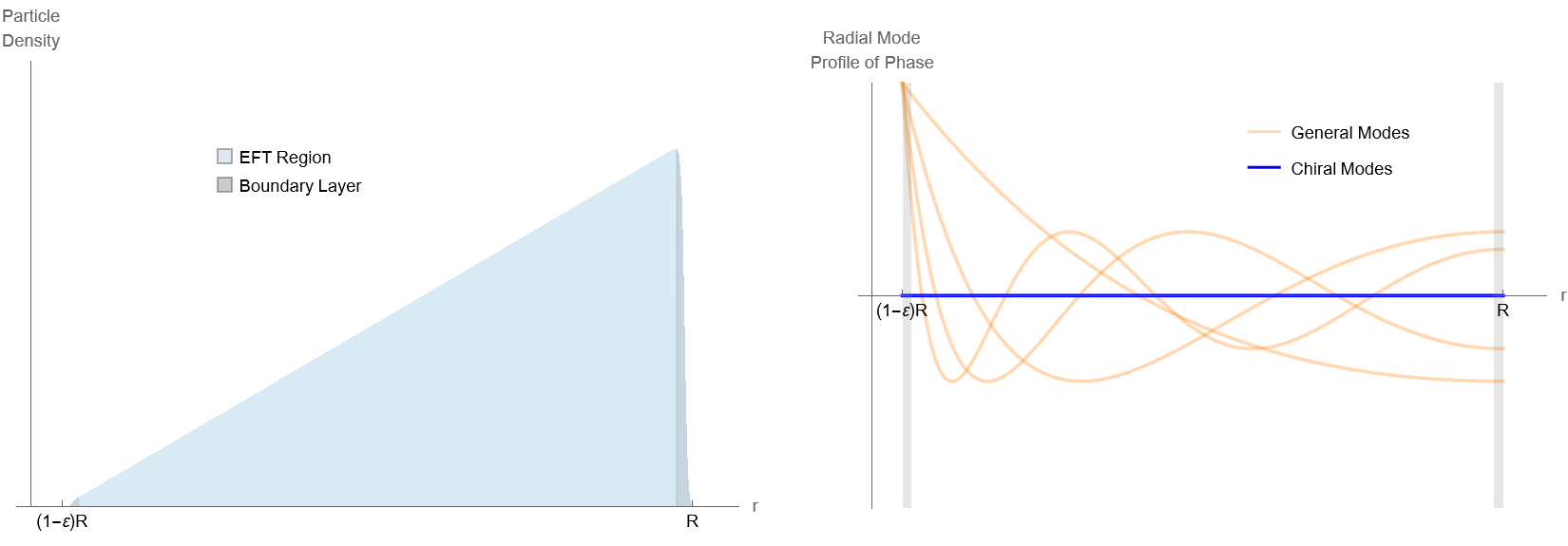}
\caption[Particle density and eigenmode profiles of a giant vortex in a hard-wall trap.]{\label{pic_GVinhardwalltrap}Particle density and eigenmode profiles of a giant vortex in a hard-wall trap. The figure shows a giant vortex in Schrödinger superfluids with $\textbf{P}(\textbf{X}) \propto \textbf{X}^{2}$.}
\end{figure}

Let us first consider superfluids confined in the region $r\leq R$ with $\textbf{V}(r)=0$, i.e., an axisymmetric hard-wall trap. We assume that the core of the giant vortex \eqref{eq_GV solution} occupies $r\leq (1-\varepsilon)R$, with the superfluid concentrated in the annular region $(1-\varepsilon)R\leq r \leq R$. See Figure \ref{pic_GVinhardwalltrap}. At the interior of the annulus $r=(1-\varepsilon)R$, we expect the particle density $\textbf{P}'(\textbf{X}_\text{cl})=0$, and accordingly $\textbf{X}_\text{cl}=0$. It then follows that
\begin{equation}
   \textbf{X}_\text{cl}=\frac{J^2}{2m}\big(\frac{1}{(1-\varepsilon)^2R^2}-\frac{1}{r^2}\big)~.
\end{equation}
For concreteness, we take the Schr\"odinger superfluids as an example, with $\textbf{P}(\textbf{X})=a \textbf{X}^2$. The total particle number confined in the trap is then given by
\begin{equation}
    N=\frac{2\pi aJ^2}{m}\int_{(1-\varepsilon)R}^R rdr \big(\frac{1}{(1-\varepsilon)^2R^2}-\frac{1}{r^2}\big)=\frac{\pi a J^2}{m}\big(\frac{1}{(1-\varepsilon)^2}-1+2\ln(1-\varepsilon)\big)~.
\end{equation}
Keeping the particle number $N$ fixed, we clearly see that $\varepsilon=O(1/J)\ll 1$ in the giant vortex limit.

We now analyze the spectrum of superfluid excitations in the limit $\varepsilon \ll 1$ while keeping the equation of state $\textbf{P}(\textbf{X})$ generic. Specifically, we expand $\varphi=\varphi_\text{cl}+\updelta \varphi$ and solve for the eigenmodes of the quantum fluctuation $\updelta \varphi$. These modes are labelled by the angular wave number $n\in \mathbb{Z}$ and the radial wave number ${n}' \in \mathbb{N}$, with the ansatz of the mode profile
\begin{equation}
\label{eq_fluc profile}
    \updelta \varphi(t,r,\theta)=e^{-i\omega t+in\theta}\updelta \varphi_{\tilde{n}}(r).
\end{equation}
We also introduce the angular velocity of the rotating superfluid near the edge of the hard-wall trap,
\begin{equation}
\label{eq_omega def}
    \varOmega=\frac{J}{mR^2}~.
\end{equation}
For modes with radial wave number $\tilde{n}\geq 1$, the profile \eqref{eq_fluc profile} depends nontrivially on the coordinate $r$, and the corresponding eigenfrequencies take the form \cite{Cuomo:2023vvd}
\begin{equation}\label{eq_hard_trap_ds1}
\omega_{n,\tilde{n}}=\varOmega\Big(\frac{k_{\tilde{n}}}{\sqrt{\varepsilon}}+n+O(\sqrt{\varepsilon})\Big)~,
\end{equation}
where $k_{\tilde{n}}>0$ are determined by the equation of state $\textbf{P}(\textbf{X})$. For Schr\"odinger superfluids, we further find that $k_{\tilde{n}}=j_{1,\tilde{n}}/2$, where $j_{1,\tilde{n}}$ denotes $\tilde{n}$-th zero of the Bessel function $J_1$. 

Modes with radial quantum number $\tilde{n}=0$ have profiles that are independent of the coordinate $r$. Their eigenfrequencies are
\begin{equation}\label{eq_hard_chiral_NLO}
\omega_{n,0}=\varOmega\Big(n+\sqrt{\varepsilon}\alpha_{\textbf{P}} |n|+O(\varepsilon)\Big)~.
\end{equation}
Here, the coefficient $\alpha_{\textbf{P}}$ is given by
\begin{equation}\label{eq_ap_hard}
\alpha_\textbf{P} \equiv \lim_{\varepsilon\to 0}\sqrt{\frac{\int  d^2x \,{\textbf{P}}'\big( \textbf{X}_\text{cl}\big)}{\upmu_{\text{eff}}\int  d^2x\, {\textbf{P}}''\big( \textbf{X}_\text{cl}\big)}}>0\,.
\end{equation}
where $\upmu_{\text{eff}}=\upmu-\frac{J^2}{2mR^2}$ denotes the effective chemical potential. For example, the Schr\"odinger superfluids with $\textbf{P}(\textbf{X})\propto \textbf{X}^2$ give $\alpha_\textbf{P}=\sqrt{2}/2$.

Let us now switch to a frame rotating with angular frequency $\varOmega$. The Hamiltonian in this frame is given by the differential operator
\begin{equation}
\label{eq_rot Hamiltonian}
    i(\partial_t+\varOmega\partial_\theta)~,
\end{equation}
with $\omega-n \varOmega$ denoting the excitation energy of the mode \eqref{eq_fluc profile}. Notably, \eqref{eq_rot Hamiltonian} takes the same form as the left-moving Virasoro zero mode of a relativistic CFT defined on a circle of circumference $2\pi/\varOmega$. In the rotating frame, the modes in \eqref{eq_hard_trap_ds1} have energies of the order $\varOmega/\sqrt{\varepsilon}$, and therefore become ultramassive in units of $\varOmega$ when $\varepsilon\ll 1$. By contrast, the modes in \eqref{eq_hard_chiral_NLO} have approximately zero energies of the order $\sqrt{\varepsilon}\varOmega$. This implies that the modes with $\tilde{n}=0$ can be effectively described by relativistic right-moving chiral waves. We therefore refer to them as chiral modes.

\subsection{Giant vortices in smooth traps}

\begin{figure}[th]
\includegraphics[width=0.25\textwidth ]{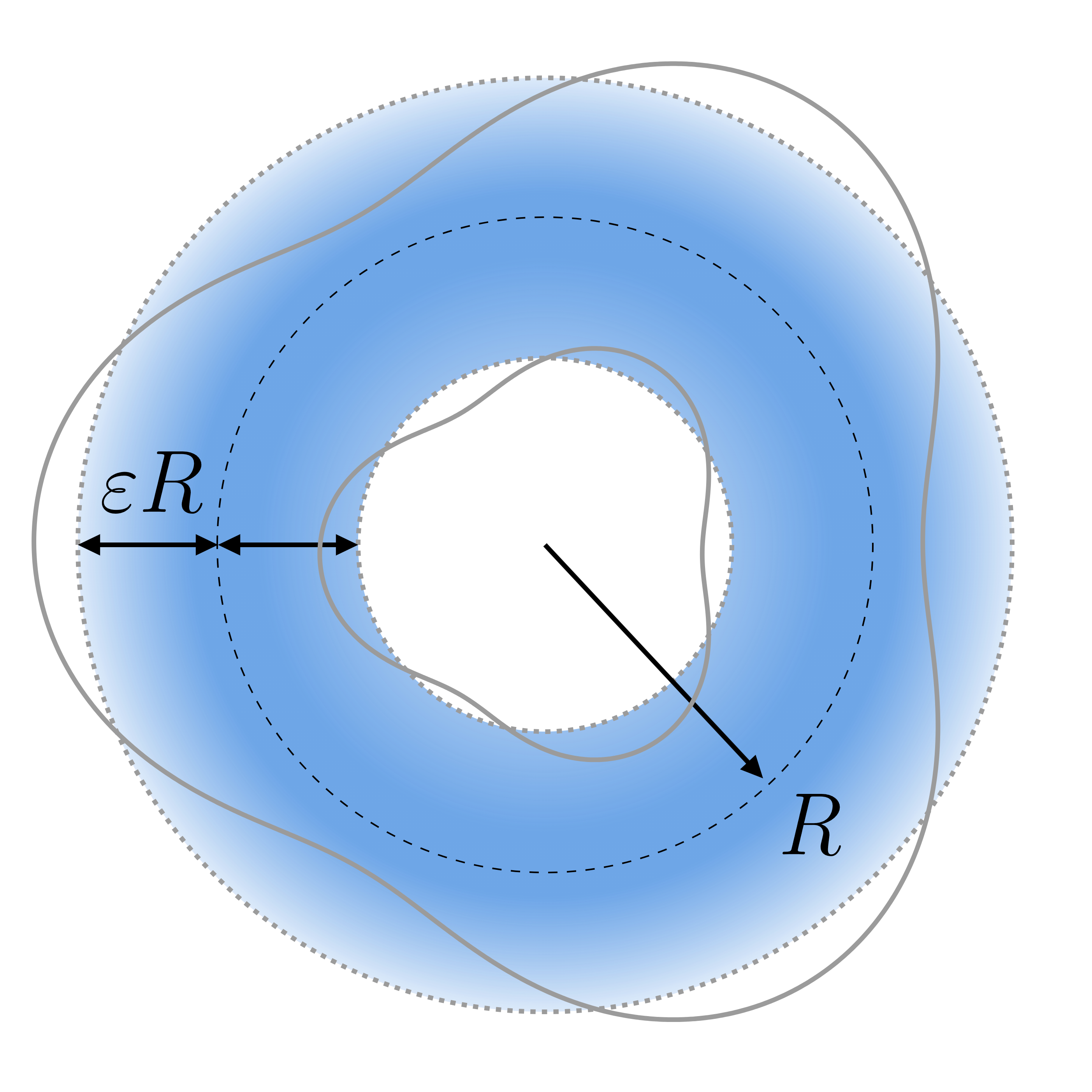}
\hspace{0.01\textwidth}
\includegraphics[width=0.73\textwidth ]{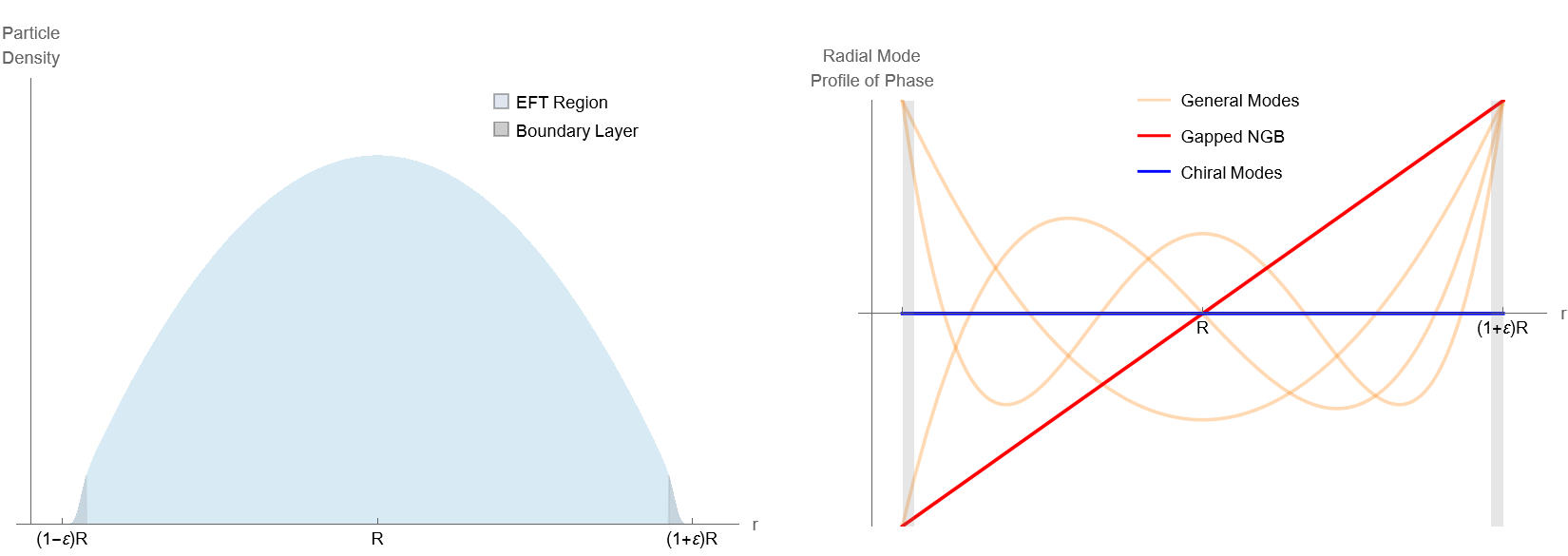}
\caption[Particle density and eigenmode profiles of a giant vortex in smooth traps.]{\label{pic_GVinsmoothtraps}Particle density and eigenmode profiles of a giant vortex in smooth traps. The figure shows a giant vortex in Schrödinger superfluids with $\textbf{P}(\textbf{X}) \propto \textbf{X}^{2}$.}
\end{figure}

In this section, we generalize the analysis to arbitrary smooth traps $\textbf{V}(r)$ and formulate the WCFT description of the chiral modes. The superfluid occupies the region where the particle density ${\textbf{P}}'(\textbf{X}_\text{cl})\geq0$, which we assume coincides with the region $\textbf{X}_\text{cl}\geq 0$. Let $R$ denote the radius at which $\textbf{X}_\text{cl}$ attains its maximum. It is then determined by
\begin{equation}
\label{eq_smooth_R_def}
\frac{J^2}{m R^3}-{\textbf{V}}'(R)=0~.
\end{equation}
We also assume that \eqref{eq_smooth_R_def} admits a unique solution, so that the superfluid concentrates in a single annulus. This condition holds for power-law traps $\textbf{V}(r) \propto r^v$, which we will use as examples below.

For convenience, we introduce the following parameters
\begin{equation}
   \upmu_{\text{eff}}=\mu-\frac{J^2}{2mR^2}-\textbf{V}(R)>0~,~~\varepsilon=\sqrt{\frac{2m \upmu_{\text{eff}}R^2}{3J^2+R^2{\textbf{V}}''(R)}}~.
\end{equation}
In particular, $\varepsilon$ denotes the width of the superfluid annulus in units of $R$, which we assume to be small in the giant vortex limit $J\gg 1$. See Figure \ref{pic_GVinsmoothtraps}. The expansion of $\textbf{X}_\text{cl}$ near its maximum then takes the form
\begin{equation}\label{eq_smooth_X_exp}
\textbf{X}_{\text{cl}}=\upmu_{\text{eff}}\Big(1-u^2+O(\varepsilon)\Big)~,
\end{equation}
where we have introduced a new coordinate $u$ through $r=(1+\varepsilon u)R$. To the leading order in $\varepsilon$, the superfluid occupies the region $-1\leq u\leq 1$, and $R$ coincides with the center of the annulus. We also use $\varOmega$ in \eqref{eq_omega def} to denote the angular velocity of the superfluid at $r=R$.

The superfluid excitations of giant vortices in smooth traps can be analyzed in close analogy with the hard-wall trap case. We refer the readers to the appendix in \cite{Cuomo:2023vvd} for further details. In the limit $\varepsilon\ll 1$, the dispersion relation involves the factor
\begin{equation}
\beta_\textbf{V}\equiv \sqrt{\frac{m\mu_\text{eff}}{ 2}}\frac{R}{\varepsilon J}~, 
\end{equation}
which roughly characterizes the steepness of the trap around $r=R$. For example, the power-law trap $\textbf{V}(r) \propto r^v$ yields $\beta_\textbf{V}=\sqrt{(v+2)/4}$.

For modes with nontrivial dependence in the radial direction, we find the eigenfrequencies
\begin{equation}
\label{eq_smooth general dispersion}
\omega_{n,\tilde{n}}=\varOmega\Big(n+\sqrt{2} \beta_\textbf{V} k_{\tilde{n}}+O(\varepsilon)\Big)~,
\end{equation}
where $\tilde{n}\geq 1$. While $k_{\tilde{n}}$ with $\tilde{n}\geq 2$ are generally determined by the equation of state $\textbf{P}(\textbf{X})$ as in \eqref{eq_hard_trap_ds1}, we note that $k_{1}=\sqrt{2}$ is universal. To the leading order in $\varepsilon$, these modes correspond to the extended symmetry generators of a locally harmonic trap \cite{Gibbons:2010fb}. For Schr\"odinger superfluids, we also find that $k_{\tilde{n}}=\sqrt{\tilde{n}(\tilde{n}+1)}$.

Interestingly, the modes with trivial radial dependence (i.e. $\tilde{n}=0$) are non-tachyonic only for $\beta_\textbf{V}>1$. This has a simple physical interpretation: the trap must be steep enough to balance the centrifugal force. For instance, a power-law trap $\textbf{V}(r)\propto r^v$ supports stable giant-vortex excitations only for $v>2$. For the stable modes, we find the eigenfrequencies
\begin{equation}\label{eq_chiral_NLO_smooth}
\omega_{n,0}=\varOmega\Big(n+\varepsilon \alpha_\textbf{P}\sqrt{2(\beta_{\textbf{V}}^2-1)}|n|+O(\varepsilon^2)\Big)~,
\end{equation}
where $\alpha_\textbf{P}$ is given in \eqref{eq_ap_hard}.

The dispersion relation \eqref{eq_chiral_NLO_smooth} implies that certain multi-phonon states are approximately degenerate. This occurs for any two Fock-space states, labeled by excitation numbers $n^a_1,n^a_2,\dots$ and $n^b_1,n^b_2,\dots$, that satisfy $\sum_i n^a_i=\sum_i n^b_i$ and $\sum_i |n^a_i|=\sum_i |n^b_i|$. Given two such states, we find that the degeneracy is lifted at order $O(\epsilon^3)$:
\begin{equation}
\label{eq_nonlinear lift deg}
    E(n^a_i)-E(n^b_i)=\varOmega \Big(\epsilon^3\gamma_{\textbf{P},\textbf{V}}\big(\sum_i |n^a_i|^3-\sum_i |n^b_i|^3\big)+O(\epsilon^4)\Big)~,
\end{equation}
where $ E(n^a_i)$ and $E(n^b_i)$ denote the energies of the corresponding multi-phonon states. Here, the coefficient $\gamma_{\textbf{P},\textbf{V}}$ depends on both the equation of state $\textbf{P}(\textbf{X})$ and the geometry of the trap $\textbf{V}(r)$. It is given by
\begin{equation}
\label{eq_def gammaPV}
\mathtoolsset{multlined-width=0.9\displaywidth}
\begin{multlined}
    \gamma_{\textbf{P},\textbf{V}}=-\frac{1}{\displaystyle\alpha_\textbf{P} \sqrt{2(\beta_\textbf{V}^2-1)}\int_{-1}^1d u\, {\textbf{P}}''(\upmu_\text{eff}(1-u^2))}\hfill\\
    \times\int_{-1}^1du\, \frac{\upmu_{\text{eff}}}{\textbf{P}'(\upmu_\text{eff}(1-u^2))}\Big[\int_{-1}^u \frac{d \tilde{u}}{\beta_\textbf{V}}\Big(\frac{\beta_\textbf{V}^2}{\upmu_\text{eff}}\textbf{P}'(\upmu_\text{eff}(1-\tilde{u}^2))\\
    \hfill-\Big(\alpha_\textbf{P}\sqrt{\beta_\textbf{V}^2-1}+\sqrt{2} \tilde{u}\Big)^2\textbf{P}''(\upmu_\text{eff}(1-\tilde{u}^2))\Big)\Big]^2<0~.
\end{multlined}
\end{equation}
Crucially, the sign $\gamma_{\textbf{P},\textbf{V}}<0$ implies that single-phonon states have lower energy than the corresponding multi-phonon states. For Schr\"odinger superfluids, \eqref{eq_def gammaPV} simplifies to
\begin{equation}
\gamma_{\textbf{P},\textbf{V}}=-\frac{\beta_{\textbf{V}} ^4+64\beta_{\textbf{V}} ^2 -56}{45\beta_{\textbf{V}}^2 \sqrt{3(\beta_{\textbf{V}} ^2-1)} }<0~.
\end{equation}

Finally, we discuss the effective description for the chiral modes of giant vortices. We introduce the light-cone coordinates 
\begin{equation}
    z^\pm=t\pm \Omega^{-1}\theta~,
\end{equation}
so that the dispersion relation \eqref{eq_chiral_NLO_smooth} can be reproduced from the effective action
\begin{equation}
\label{eq_WCFT}
    S_{\text{chiral}}\propto \int dz^+dz^-\Big((\partial_+\varphi)^2+\frac{\varepsilon^2}{2}\alpha_\textbf{P}^2(\beta_{\textbf{V}}^2-1) (\partial_+\varphi-\partial_-\varphi)^2+O(\epsilon^3)\Big)~.
\end{equation}
To the leading order in $\varepsilon$, this effective action describes a free WCFT whose conformal symmetry is generated by a right-moving Virasoro algebra and a left-moving $U(1)$ Kac-Moody algebra. Additionally, the leading order action \eqref{eq_WCFT} also admits a fractonic shift symmetry $\varphi\rightarrow \varphi+\updelta \varphi(x^-)$, which leads to infinite ground-state degeneracy in the rotating frame. These degeneracies are lifted by the $O(\varepsilon^2)$ marginal term in \eqref{eq_WCFT}. Moreover, the degeneracies among excited states are lifted by an $O(\varepsilon^4)$ irrelevant term,
\begin{equation}
    \int dz^+dz^- \big((\partial_+-\partial_-)^2\varphi\big)^2~,
\end{equation}
which corresponds to the correction in \eqref{eq_nonlinear lift deg}.

\include{sections/Conclusion}

\appendix

\include{sections/Appendix_1}

\renewcommand{\baselinestretch}{1}
\normalsize

\clearpage
\newpage
\phantomsection%
\addcontentsline{toc}{chapter}{\numberline{}{Bibliography}}%
\bibliography{BibFileThesis}
\bibliographystyle{JHEP.bst}

\clearpage
\newpage

\appendix
\include{sections/appendixSdkr}

\end{document}